\documentclass[12pt]{report}
\usepackage[utf8]{inputenc}
\usepackage[T1]{fontenc}
\usepackage[top=3cm, bottom=3cm, left=3.5cm, right=3cm]{geometry}
\usepackage{graphicx}
\usepackage[labelfont=bf]{caption}% titre de caption en gras
\usepackage[medium]{titlesec}
\usepackage[Sonny]{fncychap} 
\usepackage{caption}
\usepackage{subcaption}
\usepackage{shorttoc} %ajout du sommaire 
\usepackage{fancyhdr}
\usepackage[french]{babel}
\usepackage{amssymb}
\usepackage{amsmath}
\usepackage{txfonts}
\usepackage{mathdots}
\usepackage{lmodern}
\usepackage[toc,page]{appendix}% annexe
\usepackage{titlesec}
\usepackage{xcolor}
\definecolor{darkblue}{rgb}{0.0, 0.0, 0.55}
\usepackage{sectsty}
\chapterfont{\color{darkblue}}
\usepackage{microtype}
\usepackage{hyperref}
\hypersetup{
    colorlinks=true,
    linkcolor=blue,% Couleur des liens internes
    citecolor=blue, % Couleur des numéros de la biblio dans le corps
    filecolor=magenta,      
    urlcolor=cyan,% Couleur des url
    pdftitle={Sharelatex Example},
    bookmarks=true,
}
\usepackage[toc]{glossaries}
\makeglossaries

\begin{document}
%- - - - - - - - - - -- -- - --- - - - - - - - - - - -
% PAGE DE GARDE
%- - - - - - - - - - - - - - --  -- - - - - - - - - --

% - - - - - - - début de la page 
\thispagestyle{empty}

\begin{figure}[!htb]
    \centering
    \begin{tabular}{@{}c@{}c@{}c@{}}
        \includegraphics[width=0.2\textwidth]{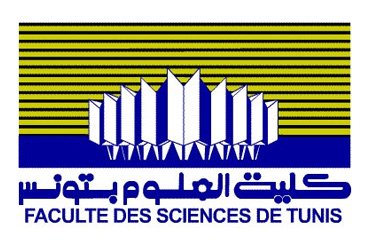}  &
        \includegraphics[width=0.2\textwidth]{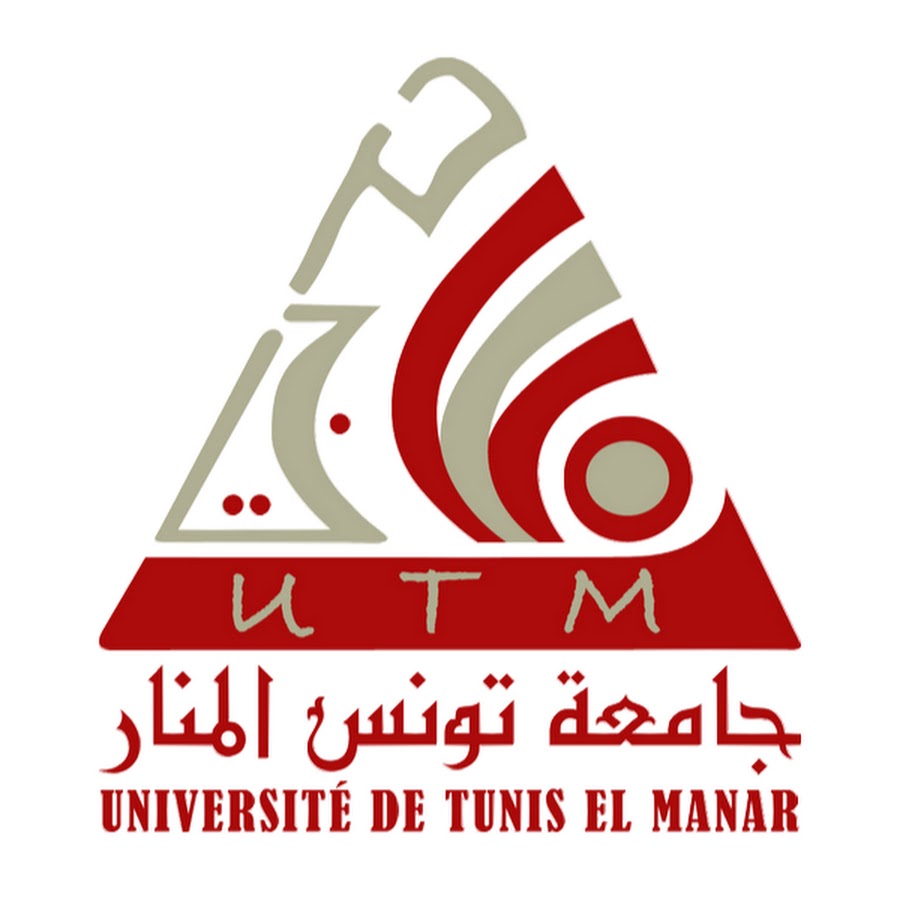} \\
    \end{tabular}
    \label{fig:hl_lhc}
\end{figure}

{

\begin{center}
{\bf{Université de Tunis EL Manar \\
Faculté des sciences de Tunis }}\\
\bf{Laboratoire de matériaux avancés et phénomènes quantiques}
\end{center}
\vspace*{1cm}

\begin{center}
\baselineskip=7mm
{\textcolor{black}{\Large {\bf Modélisation des performances des structures \\
lasers à base des chalcogénures de métaux de \\ transition: MoS2/WSe2 \\ }}}
\end{center}
\begin{center}

\vspace*{1cm}
Mémoire présenté par \\
\vspace*{0.5cm}
{\Large {\bf Khawla Jaffel}}

\vspace*{0.5cm}
Pour l'obtention du diplôme de mastère en physique de la matière condensée \ \\[1ex]

\vspace*{1cm}
Soutenu le 28 October 2017 devant le Jury composé de:
\begin{tabular}{ll}
    \multicolumn{2}{c}{}                        \\
    %\toprule
    Prof. Chaker Bouzidi (Examinateur)                        & CNRSM\\
    Prof. Nejmeddine Jaïdane (Président)                      & FST\\
    Prof. Sa\"{i}d Ridene (Encadreur)                         & FSB et FST\\ 
    & \\
    %\bottomrule
\end{tabular}
\end{center}
}

%----------------------------------------------------------------------------------------
%	ENTETE ET PIED DE PAGE
%---------------------------------------------------------------------------------------- 
 
\pagestyle{fancy}
\renewcommand{\chaptermark}[1]{\markboth{#1}{}}
\renewcommand{\sectionmark}[1]{\markright{\thesection\ #1}}
\fancyhf{} \fancyhead[LE,RO]{\normalfont\thepage}
\fancyhead[LO]{\itshape\rightmark}
\fancyhead[RE]{\itshape\leftmark}
\renewcommand{\headrulewidth}{0pt}
\addtolength{\headheight}{0.5pt}
\renewcommand{\footrulewidth}{0pt}
\fancypagestyle{plain}{ \fancyhead{}
\renewcommand{\headrulewidth}{0pt}}

%----------------------------------------------------------------------------------------
%	Remerciements
%---------------------------------------------------------------------------------------- 
 
\chapter*{Remerciements}
\thispagestyle{empty}
\baselineskip=6mm
Ce travail a été réalisé à la Faculté des Sciences de Tunis (FST) au sein du Laboratoire des Matériaux Avancés et Phénomènes Quantiques. J'exprime ma profonde gratitude au Prof. Habib Bouchriha et au Prof. Mohamed Mejatty, pour m'y avoir accueilli.\\

Je tiens particulièrement à remercier Prof. Sa\"{i}d Ridene, maître de conférences à la Faculté des Sciences de Bizerte (FSB), pour m'avoir soutenu tout au long de ce travail.\\
Je remercie chaleureusement Prof. Nejmeddine Jaïdane de la FST de s'être intéressé à ces travaux et d'avoir accepté de présider le jury de cette thèse.\\
Je remercie sincèrement Prof. Chaker Bouzidi, assistant au Centre National de Recherche en Sciences des Matériaux (CNRSM) de Borj Cédria, pour l'intérêt qu'il porte à mes travaux en acceptant d'être également membre du Jury.\\ Merci tout pour vos commentaires et suggestions qui ont contribué à améliorer cette thèse.
	
J'exprime ma sincère reconnaissance à tous mes collègues et les membres du laboratoire.\\
	
Enfin, les mots les plus simples étant les plus forts, j'adresse toute mon affection à ma famille. Merci papa, merci maman, merci mes sœurs et mon frère pour votre support et encouragements. 

%----------------------------------------------------------------------------------------
%	DEDICATION
%---------------------------------------------------------------------------------------- 
 
\cleardoublepage
	\thispagestyle{empty}
	%page blanche
	\begin{flushright}
	\large\em\null\vfill
		\large{\textit{\`A ma sœur jumelle "Khouloud"} }\vfill\vfill
	\end{flushright}
	\cleardoublepage
	%imprimé sur une page de droite
 
%----------------------------------------------------------------------------------------
%	TABLE DE MATIERES 
%---------------------------------------------------------------------------------------- 

\pagenumbering{roman}

\addcontentsline{toc}{chapter}{Table des matières}
\tableofcontents
\listoftables
\addcontentsline{toc}{chapter}{\listtablename}
\listoffigures
\addcontentsline{toc}{chapter}{\listfigurename}

%----------------------------------------------------------------------------------------
%	ACRONYMES
%---------------------------------------------------------------------------------------- 
 
\renewcommand*{\glstextformat}[1]{\textcolor{blue}{#1}}
\newacronym{AFM}{AFM}{force atomique moléculaire}
\newacronym{TMDCs}{TMDCs}{métaux de transition dichalcogènes}
\newacronym{ZB}{ZB}{zone de Brillouin}
\newacronym{MOSFET}{MOSFET}{Metal Oxide Semiconductor Field Effect Transistor}
\newacronym{CVD}{CVD}{Chemical Vapor Deposition}
\newacronym{CSD}{CSD}{Chemical Solution Deposition}
\newacronym{HEMT}{HEMT}{High Electrons Mobility Transistors}
\newacronym{BVO}{BVO}{Bande de valence Offset}
\newacronym{BCO}{BCO}{Bande de conduction Offset}
\newacronym{MOCVD}{MOCVD}{Metal-Organic Chemical Vapor Deposition}
\newacronym{PL}{PL}{photoluminescence}
\newacronym{QW}{QW}{Quantum Wells, puits quantique}
\newacronym{SLs}{SLs}{Super-Lattices, super-réseaux}
\newacronym{SO}{SO}{spin-orbite}
\newacronym{CSO}{CSO}{couplage spin-orbite}
\newacronym{RUI}{RUI}{Représentation Unitaire Irréductible}
\newacronym{EFA}{EFA}{approximation de la fonction enveloppe}
\newacronym{DFT}{DFT}{théorie de la densité fonctionnelle}
\newacronym{BC}{BC}{bande de conduction}
\newacronym{BV}{BV}{bande de valence}
\printglossary[title= Liste des acronymes, toctitle=Liste des acronymes]

%----------------------------------------------------------------------------------------
%	INTRODUCTION
%---------------------------------------------------------------------------------------- 

\shorttableofcontents{Introduction générale}{1}
\pagenumbering{arabic}

La recherche sur les matériaux bidimensionnels (2D), initiée en 2004-2005 avec l'étude du graphène, est considérablement diversifiée ces dernières années. Les semi-conducteurs bidimensionnelle des \acrlong{TMDCs}, suscitent l'intérêt de nombreux chercheurs car leurs caractères 2D s'accompagne de propriétés électroniques et optiques exceptionnelles et uniques. Ces composés appartiennent à la famille des matériaux lamellaires de formule générale $MX_2$, où M est un métal de transition de l’un des groupes IV, V ou VI du tableau périodique de Mendeleïev, et X est un chalcogénure. Ces matériaux présentent des propriétés, souvent liées à la nature de leur structure de bande électronique et phononique, qualitativement différentes de celles de leurs équivalents tridimensionnels (3D). Leur étude fine a révélé des effets remarquables, liés à des transitions de phases quantiques à deux dimensions, ou à l'optoélectronique dite de vallée, qui exploite une sélectivité en vecteur d'onde de l'émission et de l'absorption de photons. 
A l'inverse du graphène, MoS2, MoSe2,WS2 et WSe2 sont des semi-conducteurs qui possèdent une énergie de bande interdite variant de 1,2 à 1,9 eV (selon le nombre de couches). Cela permet de l'intégrer en tant qu'un élément actif dans des dispositifs optoélectroniques. De plus, son caractère 2D et sa robustesse mécanique permettent de réaliser des transitions de phases quantiques et des composants électroniques sur substrats flexibles. En effet il est possible aussi de réaliser des puits quantiques à base de $MoS_2$ tels que $MoS_2$/$MX_2$( $MX_2$= $MoSe_2$, $WS_2$ et $WSe_2$) tout dépend de l'offset des bandes entre les matériaux constitutifs des puits quantiques. Notons que depuis les années 80, le succès des puits quantiques et super-réseaux a motivé de multiples tentatives pour confiner les porteurs selon plus d'une direction spatiale et qui sont aujourd'hui d'un grand intérêt expérimental et théorique, vu leurs applications dans le domaine de l'optoélectronique. Les effets les plus spectaculaires se manifestant dans ces systèmes de basses dimensionnalités sont liés aux propriétés optiques. Le traitement de ces propriétés optiques, nécessite la connaissance du gain optique. Contrairement au cas des matériaux massifs à 3D, un gros travail à la fois conceptuel et numérique sera nécessaire pour obtenir le gain optique pour un puits quantique à 2D. 
Dans ce travail, il s'agit tout d'abord de se familiariser avec la physique des semi-conducteurs à 2D des \acrlong{TMDCs} et des puits quantiques $MoS_2$/$MX_2$ ; et ensuite de proposer un modèle théorique et numérique pour ces matériaux. La résolution du modèle proposé permettra de déterminer la structure de bande électronique et le gain optique  pour une zone active formée par des semi-conducteurs à 2D.

     Le premier chapitre présente un aperçu général sur les matériaux $MX_2$ ($MoS_2$, $MoSe_2$, $WS_2$ et $WSe_2$). Nous rappelons les principales propriétés électroniques et structurales de ces composés sous leurs diverses formes ainsi que les domaines de leurs applications. Nous introduisons brièvement les méthodes d'élaboration et de caractérisation de ces composés sous forme de couches minces.
     
Le deuxième chapitre présente le modèle théorique utilisé pour les semi-conducteurs à base des chalcogénures de métaux de transition $MX_2$. Dans un premier temps, un aperçu est donné sur les différentes méthodes de calcul de structure de bandes. En se basant sur des notions de base de la théorie de groupe, nous développons ensuite les hamiltoniens {\bf{k.p}} utilisés dans ce travail aux voisinages des points $\Gamma$ et K de la zone de Brillouin.

Le dernier chapitre comporte deux parties. La première partie est relative à l'étude de l'offset des bandes de ces matériaux afin de réaliser des puits quantiques à base de $MoS_2$. La deuxième partie de ce chapitre est relative à l'étude des performances de la zone active des structures lasers à base des puits quantiques de $MoS_2$/$WSe_2$, par le calcul de l'énergie de confinement et du gain optique. Nous mettrons en relief l'effet de la température et de la largeur du puits quantiques.

La conclusion présente un bilan des résultats obtenus et montre que la filière à base des chalcogénures de métaux de transition $MX_2$, peut permettre d'atteindre l'émission laser à température ambiante pour des longueurs d'ondes très utiles dans différents domaine de l'optoélectronique.

\addcontentsline{toc}{chapter}{Introduction générale}%ajout de l'introduction à la table de matière

%----------------------------------------------------------------------------------------
%	CHAPITRE 1
%---------------------------------------------------------------------------------------- 

\chapter{Généralités sur les chalcogénures de métaux de transition: MoS2, MoSe2, WS2 et WSe2}

Les \acrlong{TMDCs} $MoS_2$, $MoSe_2$, $WS_2$ et $WSe_2$ dont nous disposons dans ce travail sont devenus des piliers de l'optoélectronique moderne pour la réalisation des lasers, des cellules photovoltaïques et des photo-détecteurs de plus en plus performants. Ces matériaux appartiennent à la famille des matériaux lamellaires qui constituent des semi-conducteurs bidimensionnels.

 On commence tout d'abord par présenter un aperçu général sur les propriétés électroniques et structurales de ces composés sous leurs diverses formes ainsi que les domaines de leurs applications. Ensuite nous introduisons brièvement les méthodes d'élaboration et de caractérisation de ces composés sous forme de couches minces.     

\section{Structure cristalline }

Les \acrlong{TMDCs} ce qu'on appelle les \acrshort{TMDCs}, en particulières le disulfure de molybdène "{${MoS}_{2}$}" et le disélénure de tungstène  " {${WS}_{2}$}" existent dans la nature sous forme de cristaux naturels de molybdénite et de wolframite et aussi sous une forme planaire et sous la forme massif (bulk) \cite{D.Dophil}.

Les \acrshort{TMDCs} sont des semi-conducteurs constituent une famille de matériaux de formule générale {${MX}_{2}$} où M est un métal de transition et X est un chalcogéne, tel que leurs propriétés électroniques varient de métal à semi-conducteur, tout dépend de cation en présence exemple ; {${TiS}_{2}$} est un isolant, {${TaS}_{2}$} est un métal alors que le {${MoS}_{2}$} est un semi-conducteur \cite{S.Fang}.

%\begin{center}
%\includegraphics[scale=0.6]{figures/image2}
%\captionof{figure}{Tableau périodique de Mendeleïev sur lequel sont mis en évidence les éléments intervenant dans la composition de composés 2D de la famille des TMDCs.}
%\end{center}
Le tableau \ref{tab 1.1} ce dissous résume les différents matériaux à 2D de TMDCs les plus étudiés.
\begin{center}
{\renewcommand{\arraystretch}{1.5} %donne la distance entre les lignes%
{\setlength{\tabcolsep}{0.25cm} %donne la distance entre les collones%
\begin{tabular}{|p{0.3in}|p{2.1in}|p{2.1in}|p{1.2in}|} \hline 
           & \textbf{$S_2$} & \textbf{$Se_2$} & \textbf{$Te_2$} \\ \hline 
Nb & Métallique, Supraconducteur & Métallique, Supraconducteur & Métallique \\ \hline 
Tb & Métallique, Supraconducteur & Métallique, Supraconducteur & Métallique \\ \hline 
Mo & Semi-conducteur\newline SL~: 1.8 eV\newline Bulk~: 1.2 eV & Semi-conducteur\newline SL~: 1.5 eV\newline Bulk~: 1.1 eV & Semi-conducteur\newline SL~: 1.1 eV\newline Bulk~: 1 eV \\ \hline 
W & Semi-conducteur\newline SL~: 1.9-2.1 eV\newline Bulk~: 1.4 eV & Semi-conducteur\newline SL~: 1.7 eV\newline Bulk~: 1.2 eV & Semi-conducteur\newline SL~: 1.1 eV\newline  \\ \hline 
\end{tabular}}}\\
\captionof{table}{Caractère électronique de différents TMDCs qui peuvent sous forme 2D (SL = Single Layer)\protect\cite{M.Hugo}.}
\label{tab 1.1}
\end{center}
Dans ce mémoire, nous nous concentrons sur les monocouches semi-conductrices, telles que le disulfure de molybdène ($MoS_2$), le disulfure de tungstène ($WS_2$) ainsi que leurs voisins chimiques le disélénure de molybdène ($MoSe_2$) et le disélénure de tungstène ($WSe_2$), espérons qu'avec des électrons \textit{d} fortement corrélés ces matériaux pourrait montrer une physique distinctement nouvelle.\\

La figure ci-dessous \ref{fig 1.1} représente la structure massif du ${\mathrm{MoS}}_{\mathrm{2}}$ qui supposait \^{e}tre la m\^{e}me pour les autres semi-conducteurs TMDCs.

\begin{center}
\includegraphics[scale=0.9]{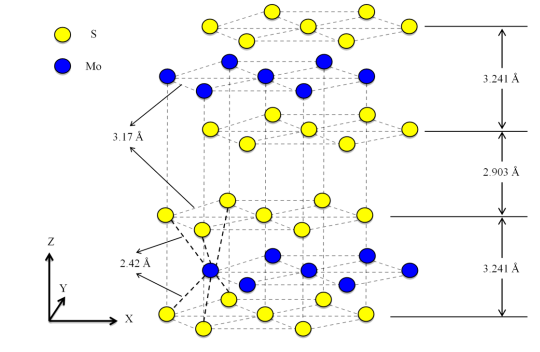}
\captionof{figure}{Structure cristalline de {${MoS}_{2}$} à l'état massif. Ce schéma montre les distances de liaison Mo-Mo, Mo-S, et S-S, il y'a une périodicité de la structure selon l'axe c de symétrie hexagonale \protect\cite{Liang}.}
\label{fig 1.1}
\end{center}

Les semi-conducteurs \acrshort{TMDCs} ont pratiquement tous la même structure, leurs paramètres de structures  tels que la distance interatomique, l'énergie de la bande interdite et le paramètre de maille changent en fonction de la nature de l'atome.
\begin{center}
{\renewcommand{\arraystretch}{1.5} %donne la distance entre les lignes%
{\setlength{\tabcolsep}{0.75cm} %donne la distance entre les collones%
\begin{tabular}{|p{0.6in}|p{0.5in}|p{0.5in}|p{0.5in}|p{0.5in}|} \hline 
Paramètre  &     $MoS_2$ &    $WS_2$         & $Mo{Se}_2$ &    $W{Se}_2$ \\ \hline 
$d_{M-X}(A{}^\circ )$  &     2.41 &    2.42 &   2.54 &     2.75 \\ \hline 
$d_{X-X}(A{}^\circ )$  &     3.13 &    3.14 &    3.34 &     3.35 \\ \hline 
   a (A${}^\circ$) &     3.18 &     3.18 &     3.32 &      3.32 \\ \hline 
$E_g\mathrm{(eV)}$ &     1.62 &     1.55 &     1.33 &      1.25 \\ \hline 
\end{tabular}}}\\
\captionof{table}{Paramètres de structures des TMDCs. \protect\cite{S.Fang,C.Gong}}
\end{center}

Les {${MX}_{2}$} sont des composés lamellaires dont la structure rappelle celle du graphite, ils se cristallisent dans une structure hexagonal et ils sont formés de trois couches atomiques, une couche de métal de transition (M=Mo, W…) entre deux couche de chalcogéne(X=S, Se, Te…). C'est pour cette raison, ils sont d'ailleurs considérés comme des matériaux à 2D, malgré que leur leur structure est formée de plusieurs  couches. Sachant qu'une couche de $MX_2$ est formée par un empilement d'atomes X-M-X . 
Au sein du feuillet, les atomes M sont en coordinence trigonale prismatique par rapport aux anions X ce qui montre le figure \ref{fig 1.3} ci-dissous.

\begin{center}
\includegraphics[scale=0.7]{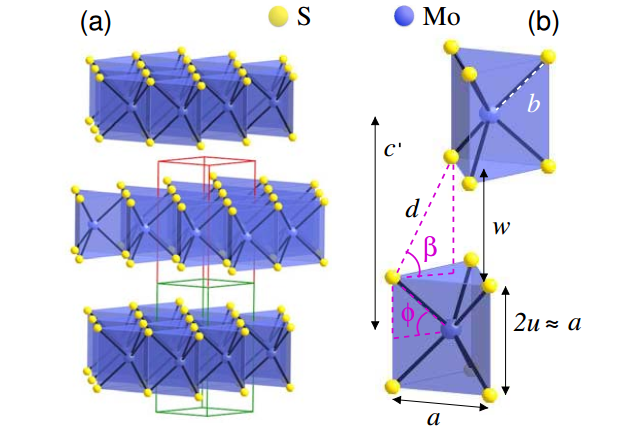}
\captionof{figure}{(a) Représentation tridimensionnelle de la structure atomique de $MX_2$.  Le rectangle vert représente une cellule unitaire d'une monocouche et double couches de $MX_2$ à l'état massif, tel que chaque couche est formé par des prismes trigonale. (b) Coordination trigonale prismatique de l'atome M de métal de transition par rapport aux anions X \protect\cite{Jose}.}
\label{fig 1.3}
\end{center}

Les liaisons types M-X à l'intérieur des couches sont fortes de natures covalentes, alors que les liaisons inter-couches sont beaucoup plus faibles de type \textit{Van Der Waals}, ceci explique la simple isolation dans le plan pratique d'une monocouche de ces matériaux par une exfoliation chimique ou mécanique.

\begin{center}
\includegraphics[scale=0.6]{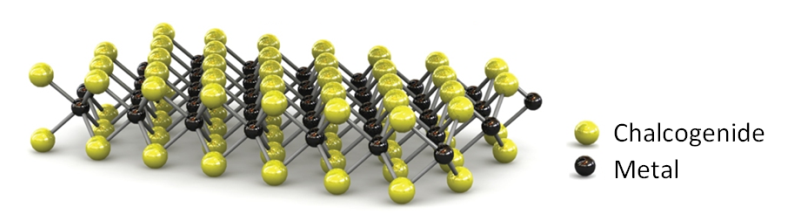}
\captionof{figure}{Structure bidimensionnelle d'une monocouche de TMDCs \protect\cite{B.Radisavljevic}. }
\end{center}

Les cristaux $MX_2$ à l'état massif montrent différents polytypes qui varient selon l'empilement et la coordination atomique, les polytypes sont soit 1T, 2H ou 3R. Dans le polytype 2H le feuillet supérieur est tourné de 60${}^\circ$ par rapport au feuillet précédent, les anions et les cations du feuillet supérieur se plaçant respectivement au-dessus des cations et des anions du feuillet précédent figure \ref{fig 1.5}. La périodicité de la structure selon l'axe (c)de symétrie hexagonale et le groupe d'espace associé est $D^4_{6h}$(P63/mmc).

Le polytype 3R est de symétrie rhombohédrale de groupe d'espace $C^5_{3v}$(R3m) tel que chaque feuillet garde la même orientation que le précédent mais translaté dans la direction [210] de 1/3 de la constante de réseau, les anions se placent dans ce cas au-dessus des interstices du feuillet précédent et les cations au-dessus des chalcogènes ainsi, la périodicité selon l'axe (c) \cite{Mingxiao}.
\begin{center}
\includegraphics[scale=0.5]{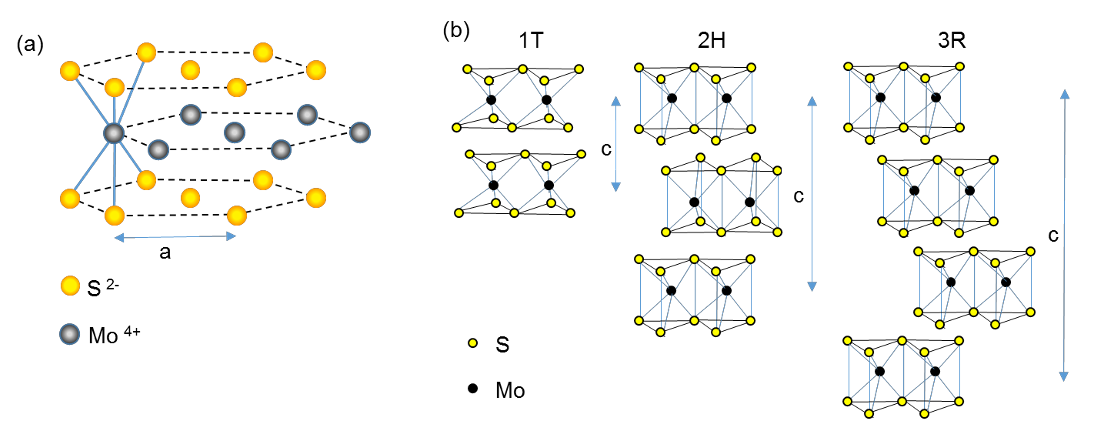}
\captionof{figure}{Différents polytypes des TMDCs de $MX_2$ à l'état massif \protect\cite{D.Costanzo}.}
\label{fig 1.5}
\end{center}

\section{Paramètre de maille }

Les vecteurs de base de réseau de Bravais dans la base orthonormé  $\left(O,{\overrightarrow{e}}_x,{\overrightarrow{e}}_y, {\overrightarrow{e}}_z\right)$ s'expriment :
\[\ {\overrightarrow{a}}_1=\frac{a_0}{2}\left(1,\ \sqrt{3},0\right)\quad \text{,} \quad {\overrightarrow{a}}_2=\frac{a_0}{2}(1,\ -\sqrt{3},0)\]

\begin{center}
\includegraphics[scale=0.7]{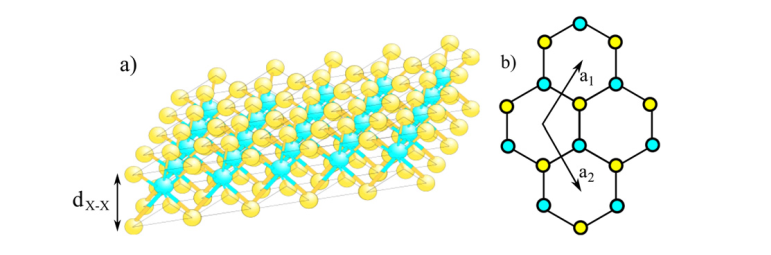}
\captionof{figure}{Structure cristalline d'une monocouche de $MX_2$. (a) Vue latérale. (b) vue de dessus montre les vecteurs de base $\vec{a}_1 \ et \ \vec{a}_2$ de réseau de Bravais dans la base (O, $\vec{e}_x , \vec{e}_y$)  \protect\cite{Andor}.}
\end{center}

Les vecteurs de réseau  ${\overrightarrow{a}}_1$ et ${\overrightarrow{a}}_{2\ }$ permettent de définir la translation qui décrit l'ensemble du cristal. La structure périodique du réseau réciproque permet de replier l'espace des k et de restreindre le vecteur d'onde à une cellule primitive, c'est la première \acrlong{ZB}\acrshort{ZB}.\\
La structure en nid d'abeilles des TMDCs a pour \acrshort{ZB} un hexagone avec deux vallées in-équivalents  ${\ K}^-$et${\ K}^+$  comme l'indique la figure \ref{fig 1.7} ci-dessous.\\
Notons les coordonnées des points de haute symétrie dans la base  (\textit{O},$\ {\overrightarrow{b}}_1,\ {\overrightarrow{b}}_2\ $)  de réseau réciproque sont :
\[\mathrm{\Gamma }\mathrm{=(0,0)} \quad \text{,} \quad   \mathrm{K=}\frac{4\pi }{3a_0}\left(1,0\right)  \quad \text{,} \quad   \mathrm{M=}\frac{4\pi }{3a_0}\left(0,\frac{\sqrt{3}}{2}\right)\] 

La principale différence entre les matériaux TMDCsmassif et monocouche est que la structure massif est centro-symétrique, alors que c'est ne pas le cas pour la monocouche.

\begin{center}
\includegraphics[scale=0.7]{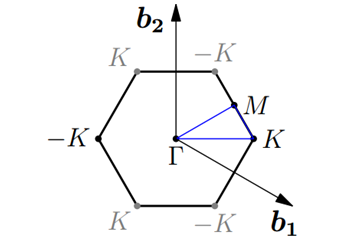}
\captionof{figure}{Représentation bidimensionnelle de la première \acrshort{ZB} des matériaux TMDCs. Les points $\Gamma$KM de haute symétrie formant la \acrshort{ZB} irréductible ainsi que les vecteurs de bases  $\vec{b}_1 \ \textit{et} \ \vec{b}_2$  de réseau réciproque sont indiqués\protect\cite{D.Costanzo}.}
\label{fig 1.7}
\end{center}

\section{Groupe ponctuel de symétrie}

La détermination de groupe ponctuel de symétrie d'une molécule nécessite l'identification de tous ses axes et ses plans de symétrie. Les éléments de symétrie pour les matériaux TMDCs sont:
\[\ \ \ \ \ \ \left\{\mathrm{E,}{\mathrm{\sigmaup }}_{\mathrm{h}},\left\{{\mathrm{C}}_{\mathrm{3}}\mathrm{,\ }{\mathrm{C}}^{\mathrm{-}\mathrm{1}}_{\mathrm{3}}\mathrm{\equiv }{\mathrm{C}}^{\mathrm{2}}_{\mathrm{3}}\right\}\mathrm{,\ }\left\{{\mathrm{S}}_{\mathrm{3}}\mathrm{=}{\mathrm{\sigmaup }}_{\mathrm{h}}.{\mathrm{C}}_{\mathrm{3}}\mathrm{,\ }{\mathrm{\ S}}^{\mathrm{2}}_{\mathrm{3}}\mathrm{=}{\mathrm{\sigmaup }}_{\mathrm{h}}.{\mathrm{C}}^{\mathrm{2}}_{\mathrm{3}}\right\}\mathrm{,\ }\left\{{\mathrm{C}}_{\mathrm{2}}\mathrm{\ }{\mathrm{,C'}}_{\mathrm{2}},{\mathrm{,C''}}_{\mathrm{2}}\right\}\mathrm{\ ,\ }\left\{{\mathrm{\sigmaup }}_{\mathrm{v}}{\mathrm{,}\mathrm{\sigmaup }\mathrm{'}}_{\mathrm{v}}{\mathrm{,}\mathrm{\sigmaup }\mathrm{''}}_{\mathrm{v}}\right\}\right\}\] 

Le groupe ponctuel de symétrie d'un matériau de type $MX_2$ à l'état massif est ${\mathrm{\ D}}_{\mathrm{3h}}\mathrm{(\ }\overline{\mathrm{6}}\mathrm{m2)}$. le tableau suivant donne la table de caractère de ce groupe.

\begin{center}
{\renewcommand{\arraystretch}{1.5} %donne la distance entre les lignes%
{\setlength{\tabcolsep}{0.45cm} %donne la distance entre les collones%
\begin{tabular}{|p{0.8in}|p{0.4in}|p{0.4in}|p{0.4in}|p{0.4in}|p{0.5in}|p{0.5in}|} \hline 
$D_{3h}=D_3\times {\sigma }_h$      $\ \ \ \ \ \ \ \left(\ \overline{6}m2\right)\ \ $ & E & ${\sigma }_h$ & 2$C_3$ & 2$S_3$ & ${3C'}_2$ & ${3\sigma }_v$ \\ \hline 
${A'}_1$ & 1 &  1 &  1   &  1 &  1 &  1 \\ \hline 
${A'}_2$ & 1 &  1 &  1 &  1 & -1 & -1 \\ \hline 
${A''}_1$ & 1 & -1 &  1 & -1 &  1 & -1 \\ \hline 

${A''}_2$ & 1 & -1 &  1 & -1 & -1 &  1 \\ \hline 
            $E'$ & 2 &  2 & -1 & -1 &  0 &  0 \\ \hline 
$E''$ & 2 & -2 & -1 &  1 &  0 &  0 \\ \hline 
\end{tabular}}}\\
\captionof{table}{Table de caractères présentant les différentes classes de symétrie de groupe \textbf{${\boldsymbol{D}}_{\boldsymbol{3}\boldsymbol{h}}\left(\boldsymbol{\ }\overline{\boldsymbol{6}}\boldsymbol{m}\boldsymbol{2}\right)$}, ainsi les représentations irréductibles et les caractères correspondant \protect\cite{Andor}.}
\end{center}

La table de caractère présentant le groupe de symétrie $C_{3h}(\overline{6})$ est donnée sur le tableau ci-dissous.    
\begin{center}
{\renewcommand{\arraystretch}{1.5} %donne la distance entre les lignes%
{\setlength{\tabcolsep}{0.5cm} %donne la distance entre les collones%
\begin{tabular}{|p{0.8in}|p{0.4in}|p{0.4in}|p{0.4in}|p{0.4in}|p{0.5in}|p{0.5in}|} \hline 
$C_{3h}(\overline{6})$ & E & $C_3$ & $C^2_3$ & ${\sigma }_h$ & $S_3$ & $({\sigma }_h{.C}^2_3)$ \\ \hline 
$A'$ &     1 &       1 &      1 &    1 &      1 &       1 \\ \hline 
$A''$ &    1 &       1 &      1 &   -1 &    -1 &      -1 \\ \hline 
${E'}_1$ &    1 & $\omega $ & $\omega ^2$ &    1 &    $\omega $  & $\omega ^2$ \\ \hline 
${E'}_2$ &    1 & $\omega ^2 $ & $\omega $ &    1 &    $\omega ^2$  & $\omega $ \\ \hline 
${E''}_1$ &    1 & $\omega $ & $\omega ^2$ &   -1 & $-\omega $  & $ -\omega ^2$ \\ \hline 
${E''}_2$ &    1 & $\omega ^2 $ & $\omega $ &   -1 & $-\omega ^2$  & $-\omega \ $  \\ \hline 
\end{tabular}}}\\
$\omega = e^{i2\pi /3}$
\captionof{table}{Table de caractères présentant les différentes classes de symétrie de groupe\textbf{$\boldsymbol{\ }{\boldsymbol{C}}_{\boldsymbol{3}\boldsymbol{h}}\boldsymbol{(}\overline{\boldsymbol{6}}\boldsymbol{)}$} , ainsi les représentations irréductibles et les caractères correspondants \protect\cite{Andor}.}
\end{center}

\section{Propriétés de transport}
Dans les matériaux TMDCs le transport de charge s'effectue suivant deux mécanismes: un transport au sein de la couche et un mécanisme par saut entre les couches qui permet au charge de passer d'une couche à l'autre. En effet le transport s'effectue pratiquement dans des bandes d'états délocalisés dues aux fortes liaisons covalentes entre atomes et à l'existence d'un ordre à grande distance. Le paramètre important qui décrit mieux les propriétés électriques des matériaux TMDCs que la conductivité est la mobilité. La mobilité intervient dans l'expression de la conductivité $ \gamma =nq \mu $.
Cette grandeur en ($cm^2$ $V^{-1}$ $s^{-1}$) est intrinsèque car la conductivité dépend de la concentration des porteurs libres. La mobilité des électrons dans le $MoS_2$ épais est de l'ordre de 200-500 $cm^2$ $V^{-1}$ $s^{-1}$ qui dépend du nombre des couches \cite{Fivaz}. Cette valeur de mobilité peut être conservée en réduisant l'épaisseur du $MoS_2$ à quelques couches.

\section{Propriétés mécaniques }

Une monocouche de métaux de transitions dichalcogènes $MX_2$ d'épaisseur quasi-atomique présente des propriétés intéressantes pour l'électronique flexible puisqu'ils sont à la fois semi-conducteurs et robustes mécaniquement. Des études sur les propriétés électriques d'une couche mince de $MoS_2$ ont montré qu'un courant peut être généré par effet piézoélectrique suite à une contrainte mécanique \protect\cite{Lei}. L'application d'une contrainte mécanique sur cette couche provoque l'apparition des deux charges positives et négatives qui se déplacent dans deux directions opposées, créant par la suite une différence de potentiel.

\begin{center}
\includegraphics[scale=0.9]{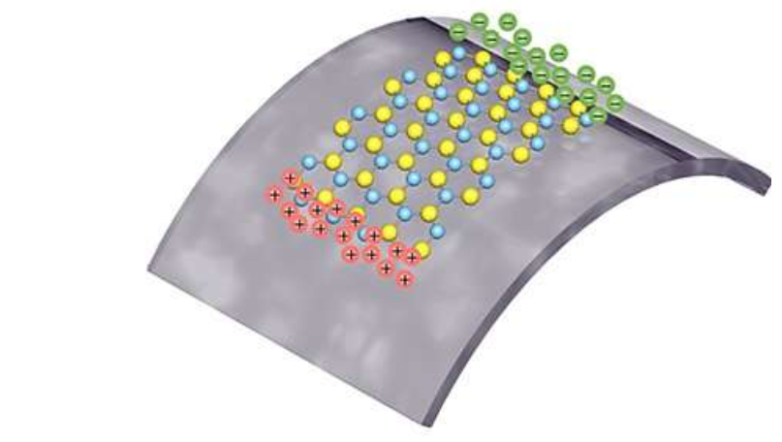}
\captionof{figure}{Effet piézoélectrique obtenu à la suite d'une contrainte mécanique.}
\end{center}

Ainsi, l'exploitation de l'effet piézoélectrique dans les matériaux à deux dimensions ouvre un nouveau champ d'application ou l'électronique devient flexible de l'ordre de l'atome, donc il est intéressant de savoir la déformation maximale que peuvent supporter les feuillets de $MoS_2$. La figure \ref{fig 1.10} montre la variation de F($\deltaup$) qui traduit la force appliquée au centre des feuillets de $MoS_2$ à l'aide d'une pointe \acrlong{AFM} en fonction de la déflection $\deltaup$ mesuré. Une quasi-linéarité pour les feuillets épais mais non-linéarité pour les feuillets les plus fins est marqué. La non linéarité 
traduit l'effet de flexibilité on peut alors mesurer à quelle force les feuillets se rompent et on peut déterminer la déformation maximale qu'ils peuvent subir un nombre des couches déterminé de $MoS_2$. Le $MoS_2$ peut subir une déformation de 6 à 11 \% sans se rompre \cite{Castellanos,Liu} et a titre de comparaison, le graphème peut \^{e}tre étiré jusqu'à 13\% de sa longueur \cite{Lee}.

\begin{center}
\includegraphics[scale=0.7]{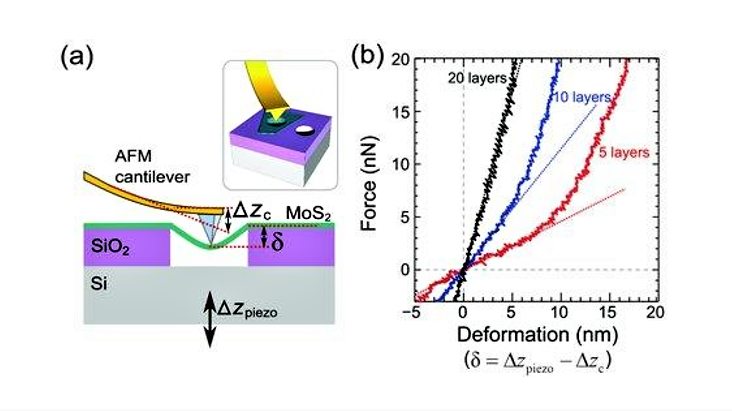}
\captionof{figure}{ (a) Schéma du dispositif expérimental mis en place pour mesurer la déformation des feuillets de $MoS_2$ suspendus. (b) Force appliquée sur le feuillet de $MoS_2$ en fonction de la déformation pour des feuillets de 5, 10 et 15 couches \protect\cite{Castellanos}.}
\label{fig 1.10}
\end{center}

\section{Propriétés optiques }

Les matériaux ${\mathrm{\ MoS}}_{\mathrm{2\ }},$ ${\mathrm{\ WS}}_{\mathrm{2}}$, ${\mathrm{\ MoSe}}_{\mathrm{2}}$  et ${\mathrm{\ WSe}}_{\mathrm{2}}$  sont des semi-conducteurs à gap direct ou indirect allant de 1.1 à 2.0 eV selon le nombre des couches  \cite{Christophe}.\textbf{ } Les transition excitoniques sont alors des transitions directes de 1.6 eV et indirectes de l'ordre de 2.0 eV . Ces hautes énergies de liaison des excitons impliquent que, m\^{e}me à température ambiante, les structures excitoniques sont visibles dans les spectres optiques. La figure \ref{fig 1.10} montre la variation de coefficient d'absorption $\alpha$ et de l'indice de réfraction n de $WS_2$ en fonction de l'énergie de photons à température ambiante. 

\begin{center}
\includegraphics[scale=0.8]{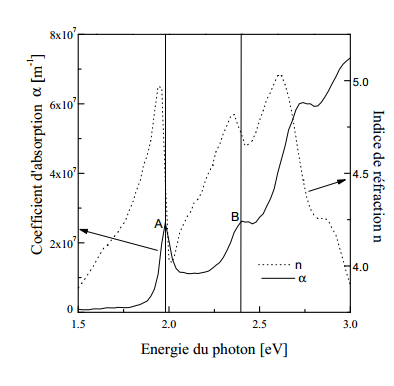}
\captionof{figure}{Coefficient d'absorption et indice de réfraction à température ambiante d'un monocristal $WS_2$ déterminée par ellipsométrie. Les pics A et B sont associés à l'absorption excitonique \protect\cite{Christophe}.}
\label{fig 1.11}
\end{center}

 En remarque qu'au-dessus de 1.92 eV, la valeur de $\alphaup$ est supérieure à \textbf{$\boldsymbol{\ }{\boldsymbol{10}}^{\boldsymbol{7}}{\boldsymbol{m}}^{\boldsymbol{-}\boldsymbol{1}}$}, ce qui signifie qu'au moins le 90\% de l'intensité lumineuse est absorbée sur 0.23 $\mu$m.

\section{Structure de bande}
Puisque les matériaux \acrshort{TMDCs} sont des semiconducteur à gap direct ou indirect selon le nombre de couche, il est évident qu'il existe des différences importantes entre un $MX_2$ à couche unique et bulk concernant la structure électronique tel que l'absence d'interaction entre couches entraîne de fortes modifications des propriétés électroniques et optiques dans tous les systèmes monocouches.
\begin{center}
\includegraphics[scale=0.7]{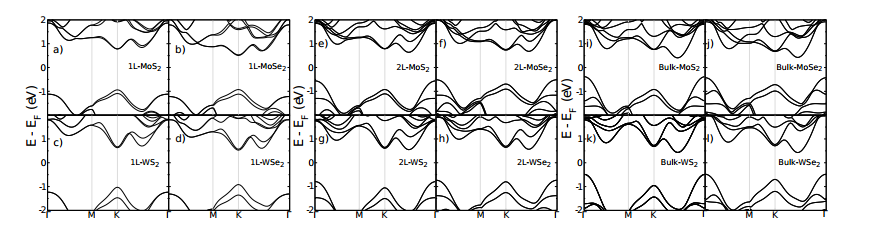}
\captionof{figure}{Structures des bandes calculées par DFT pour différentes \acrshort{TMDCs} et pour différentes nombre des couches (1L, 2L et bulk). Un changement remarquable de la structure électronique avec le nombre de couche pour les différents semi-conducteurs \acrshort{TMDCs} \protect\cite{Jose}. Le gap est direct au point K.}
\end{center}

Si nous concentrons notre étude sur le disulfure de molybdène $MoS_2$ la figure \ref{fig 1.13} ci dissous montre la dépendance de la largeur de la bande interdite avec le nombre de couches.

\begin{center}
\includegraphics[scale=0.8]{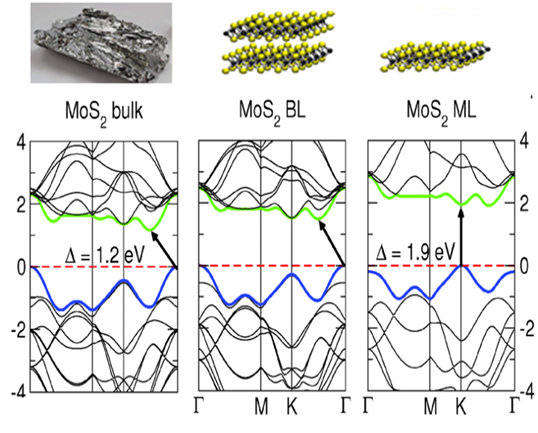}
\captionof{figure}{Structure de bandes de $MoS_2$, massif, deux couches et monocouche en allant de gauche vers la deroite \protect\cite{Kuc}.}
\label{fig 1.13}
\end{center}

Le disulfure de molybdène  possède une bande interdite direct quand il est sous forme monocouche et indirect à partir de deux couche. Le gap direct se situe au point de haute symétrie K, tandis dans le cas de gap indirect le maximum de la bande de valence se situe au point $\Gamma$ alors que le minimum de la bande de conduction se situe à mi-chemin entre le point  $\Gamma$ et K. Une diminution de nombre des couches de $MoS_2$ provoque une augmentation d'énergie du bas de bande de conduction à cause de confinement quantique ce qui traduit par une translation gap indirect/direct.

\section{Effet des interactions inter-couche sur l'intensité de photoluminescence }

Les propriétés optiques des \acrshort{TMDCs} dépendent très fortement du nombre de couches du feuillet de $MX_2$. La transition d'un semi-conducteur à gap indirect à un semi-conducteur à gap direct qui se fait par réduction de nombre des couches de $MX_2$ provoque une augmentation de rendement quantique de \acrlong{PL}(\acrshort{PL}).

Précisément, le $MoS_2$ est l'un des semi-conducteurs parmi les matériaux TMDCs prometteur d'être émergé dans divers applications électroniques (transistors) optoélectroniques (LED, cellules solaires, photodétecteurs, Laser).

La figure \ref{fig 1.14} ci dissous montre que les énergies des excitons A et B ont atteint un sommet de 1,82 et 1,98 eV respectivement pour une monocouche de MoS2.

\begin{center}
\includegraphics[scale=0.5]{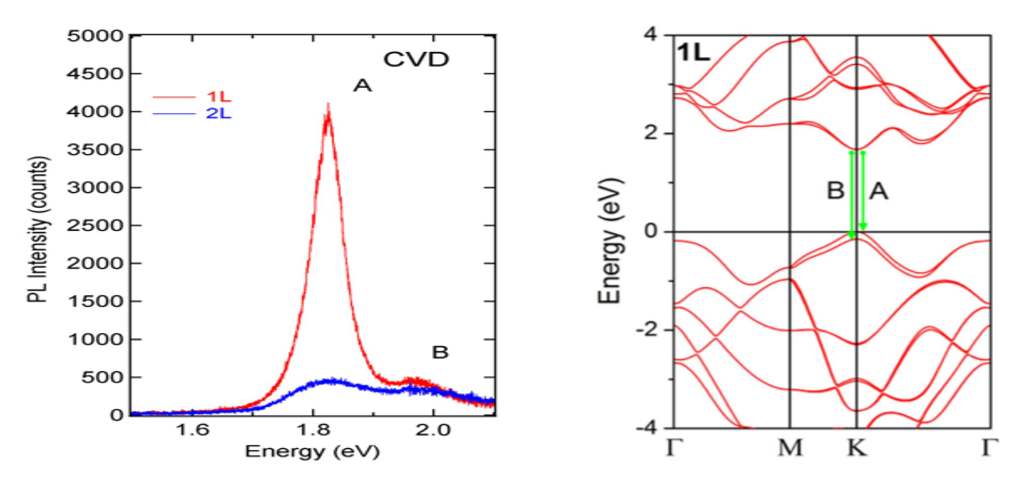}
\captionof{figure}{ Spectre de photoluminescence pour pour une couche et deux couche de $MoS_2$ \protect\cite{Jung}.}
\label{fig 1.14}
\end{center}

\section{Méthodes d'élaboration et de caractérisation des couches minces de MX2}
\subsection{Les techniques d'élaborations}
Une couche mince d'un matériau est un élément de ce même matériau dont l'épaisseur, qui est l'une des dimensions, a été vigoureusement réduite de façon à être exprimée usuellement en nanomètre. Les couches minces sont actuellement utilisées dans de nombreux domaines tels que l'optique (couches anti-reflets, miroires…), l'électronique (transistors, capteurs solaire…), la mécanique (couches résistant à l’érosion ou à l'usure, couches dures pour outils de coupe…), la chimie (couches anti-corrosion…) et la décoration (bijouterie, lunetterie…). Les couches minces peuvent être élaborés suivant deux procédés ; physique et chimique. La classification des techniques de déposition est présentée sur le tableau ci-dessous :

\begin{center}
{\renewcommand{\arraystretch}{1.5} %donne la distance entre les lignes%
{\setlength{\tabcolsep}{0.75cm} %donne la distance entre les collones%
\begin{tabular}{|p{1in}|p{1in}|p{1in}|p{1in}|} \hline 
\multicolumn{2}{c|}{Procédé physique (PVD)}      &     \multicolumn{2}{c|}{Procédé chimique}  \\ \hline 
En milieu vide poussé;    Evaporation sous vide  &     En milieu plasma;    Pulvérisations cathodiques &    En milieu de gaz réactif; Dépôt chimique en phase vapeur &   En milieu liquide; spray , Déposition par bain chimique, sol-gel \\ \hline 
\end{tabular}}}\\
\captionof{table}{Classification des principaux techniques d'élaboration des couches minces.}
\end{center}
Les dépôts physiques en phase vapeur consistent à utiliser des vapeurs du matériau à déposer pour réaliser un dépôt sur un substrat. Le transport des vapeurs de la source au substrat nécessite un vide assez poussé de $(10^{-5} à  10^{-10} Pa)$ afin d'éviter la formation de  poudre ou toute forme de pollution.\\
	{\bf Dépôt par évaporation sous vide}
Cette technique consiste à évaporer le matériau à déposer en le portant à une température suffisante. Dés que la température de liquéfaction est dépassée, il se trouve que la pression du matériau est sensiblement supérieure à celle résiduelle dans l'enceinte. Alors des atomes du matériau s'échappent et se propagent en ligne droite jusqu'à ce qu'ils rencontrent un obstacle.
Cette rencontre peut être le fait soit d'une surface solide (substrat, paroi de l'enceinte) soit d'un atome ou d'une molécule se déplaçant dans l'espace. Dans le cas de rencontre d'une surface, il y aura séjour de l'atome sur la surface avec échange d'énergie et si la surface est sensiblement plus froide que l'atome il y'a condensation définitive.\\
	{\bf Techniques de dépôt par voi chimique}
Les techniques de dépôt chimique en milieu de gaz réactif \acrlong{CVD} (\acrshort{CVD})  ou en milieu liquide \acrshort{CSD} (\acrlong{CSD})  permettent de réaliser des dépôts à partir de précurseurs qui réagissent chimiquement pour former un film solide déposé sur un substrat. Cette technique est une méthode dans laquelle les constituants d'une phase gazeuse réagissent pour former un film solide déposé sur un substrat. Les composés volatils du matériau à déposer sont éventuellement dilués dans un gaz porteur et introduits dans une enceinte où sont placés les substrats. Le film est obtenu par réaction chimique entre la phase vapeur au niveau du substrat chauffé. La réaction chimique détermine la nature, le type et les espèces présentes.\\
Le principe de la sulfurisation de MoS2 et le WS2 par la techniques \acrshort{CVD}est que la couche mince de MoS2 est obtenue par sulfurisation de poudre de MoO3 pure à (99.99 pour-cent) en utilisant la technique de la \acrshort{CVD} à haute température.

\begin{center}
\includegraphics[scale=0.5]{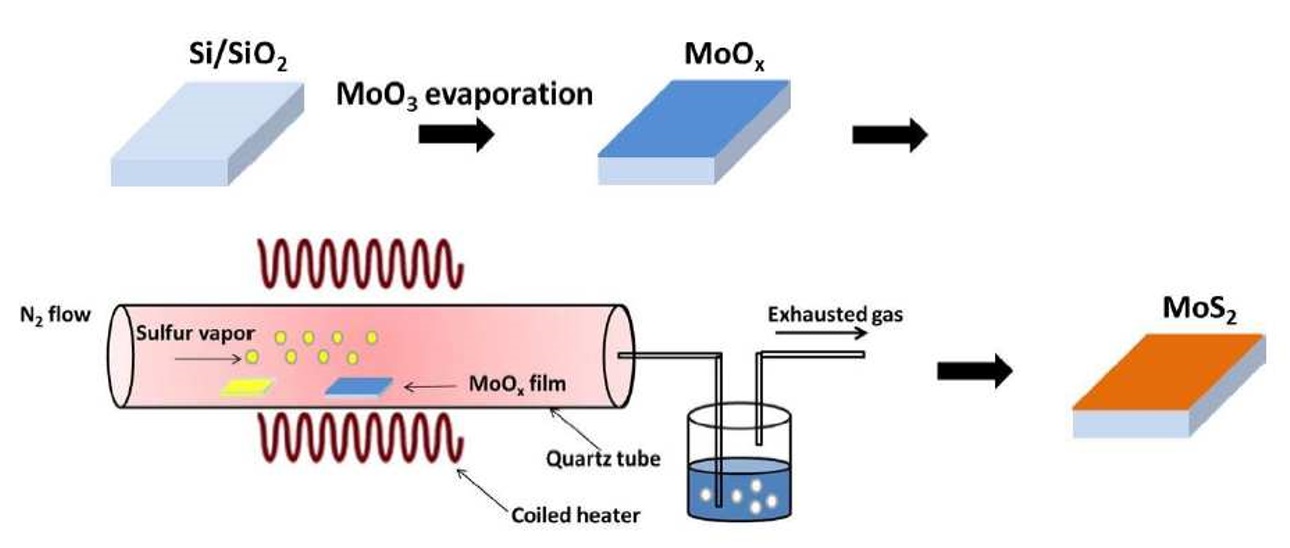}
\captionof{figure}{Illustration d'un schéma de synthèses d'une couche mince par la sulfurisations de $MoO_x$ via la méthode \acrshort{CVD}.}
\end{center}
La figure \ref{fig 1.8} ci-dissous montre un échantillon d'une monocouche de $MoS_2$, obtenu par cette technique. 
\begin{center}
\includegraphics[scale=0.6]{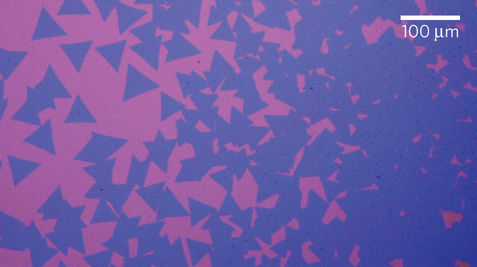}
\captionof{figure}{Une monocouche de $MoS_2$.\protect\cite{Arend}}
\label{fig 1.8}
\end{center}

\subsection{Les techniques de caractérisations }

L'un de principale méthodes, pour la détermination de la structure cristalline est la diffraction des rayons X. Lorsqu'un faisceau de rayon X frappe un cristal (solide ordonné), sous un angle $\theta$ (figure), l'interaction des rayons X avec la matière entraîne une diffusion cohérente laquelle est caractérisée par le fait que le champ électromagnétique des rayons X incidents fait vibrer les électrons des atomes du cristal. Chaque atome devient alors une source de rayons de même longueur d'onde que les rayons X incidents.
\begin{center}
\includegraphics[scale=0.9]{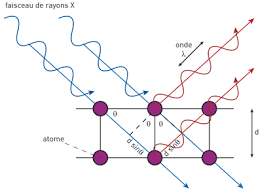}
\captionof{figure}{Géométrie de diffraction des rayons X.}
\end{center}
Les ondes diffractées à partir de différents plans d'atomes s'interfèrent entre elles et donnent un diagramme de diffraction. Les atomes qui sont arrangés d'une manière périodique donnent des figures de diffraction avec des maxima d'interférence fins dits pics de diffraction.Les pics de diffraction sont reliés aux dimensions de la maille élémentaire. Un pic de diffraction apparaît si la loi de Bragg est vérifiée. Cette loi relie la distance dhkl entre les plans cristallins parallèles, la longueur d'onde des rayons X et l'angle $\theta$.\\

La spectroscopie Raman est aussi une technique de caractérisation structurale très utilisées aussi pour l'analyse des solides et des couches minces.\\
En fin, on utilise la microscopie à force atomique (\acrshort{AFM}) pour mesurer l'épaisseur de la couche. L'étude par spectroscopie de photoélectrons (XPS) permet de confirmer la structure et d'établir la stoechiométrie du $MoS_2$.Les domaines seront également caractérisés optiquement par spectroscopie Raman et photoluminescence. une mesure en spectroscopie d'émission des photoélectrons dans l'espace réciproque destinée à révéler la structure de bandes du matériau a été entreprise.

\section{Application des matériaux TMDCs }
Les matériaux TMDCS peuvent conduire à de multiples applications: diodes électroluminescentes, transistor à effet de champ, cellules solaires ou photovoltaïque, lasers... Ces différentes systèmes dépendent notamment   du caractère radiatif ou non de leur désexcitation. Les points suivants ont fait de ces matériaux des candidats à très fort potentiels pour le développement de l'électronique moderne :\\
$\bullet$ Une mise en ouvre facile\\
$\bullet$ Faible coût de fabrication\\
$\bullet$ Flexibilité des dispositifs \\
$\bullet$ Légèreté des dispositifs
\subsection{Applications dans les dispositifs optoélectroniques}
Les possibilités d'applications des matériaux TMDCs comparés aux semiconducteurs habituels sont de ce fait multiples et variées, notamment dans les transistors à effet de champ. Les TMDCs monocouches ont déjà été implémentés dans des transistors à effet de champ \cite{RadiB.R}, des dispositifs logiques \cite{RadiB.W} et des structures optoélectroniques \cite{Britnell}. En particulier, le $MoS_2$ a suscité un intérêt considérable dans diverses applications \cite{Q.H,X.Huang}.\\

\textbf{Transistors à base de MoS2}\\
Il est possible d'étudier le $MoS_2$ pour fabriquer des diodes électroniques et des transistors à effet de champ de haute qualité. En effet il est possible de contrôler ces propriétés électriques et optiques de ces dispositifs puisque la bande d'énergie interdite est contrôlable via les nombres des couches.

\begin{center}
\includegraphics[scale=0.8]{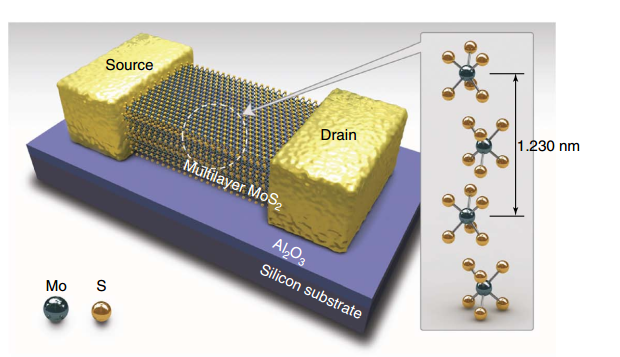}
\captionof{figure}{Transistor à base d'une monocouche de $MoS_2$ .}
\end{center}

\subsection{Cellules solaires ou photovoltaïques à base des matériaux TMDCs }

Les matériaux \acrshort{TMDCs} avec un haut coefficient d'absorption et une bande interdite permettant l'absorption d'une grande partie du spectre solaire semblent être efficace comme éléments actifs dans des cellules photovoltaïques et ils ont déjà démontré leur efficacité dans des cellules photo-électrochimiques \cite{D.Dophil}. Le $MoS_2$ et $WS_2$ suscitent un intérêt particulier car il y a tout d'abord le facteur prix ; les cellules au silicium monocristallin sont onéreuses à cause de l'exigence de qualité et de la quantité importante de matière nécessaire. La faible absorption optique du Si cristallin due à la nature indirecte de sa bande interdite implique la nécessité d'une épaisseur de 200 à 400 $\mu$m  afin d'absorber la quasi-totalié de la lumière \cite{M.Grätzel}. Deuxièmement le silicium nécessite une grande pureté afin d'assurer un temps de vie et par conséquent une longueur de diffusion suffisante pour que les porteurs générés à près de 200 $\mu$m de la zone de déplétion puissent y arriver avant de se recombiner.  Alors que les matériaux à haut coefficient d'absorption optique comme le $MoS_2$ ou le $WS_2$ absorbent la lumière sur une distance de l'ordre du $\mu$m ce qui rend possible la réalisation des dispositifs minces et presque sans pureté. De plus, lorsque les semi-conducteurs conventionnels (Si, GaAs) ont besoin d'un traitement de passivation préalable pour échapper à la photocorrosion, les monocristaux de $MoS_2$ ou de $WS_2$ sont remarquablement stables chimiquement plutôt ils résistent activement à la photocorrosion dans des cellules électrochimiques \cite{H.Tributsch}. 

\section*{Conclusion} 

Dans ce chapitre nous avons vu que les \acrlong{TMDCs} attirent beaucoup d'intérêt autant théorique que expérimental vue propriétés optiques et électriques exceptionnelles. Le $MoS_2$ a également un faible coefficient de frottement et de robustesse, ce qui lui permet de construire une perspective passionnante pour plusieurs domaines d'applications. Ils important de créer des hétérostructures à base des matériaux TMDCs afin d'amiliorer les performances des ces matériaux. Ce à nous conduit à construire un modèle théorique pour étudier les propriétés électroniques de ces matériaux, c'est l'objet du deuxième chapitre.

%----------------------------------------------------------------------------------------
%	CHAPITRE 2
%---------------------------------------------------------------------------------------- 
\chapter{Modèle théorique pour les semi-conducteurs à base des TMDCs}

\baselineskip=7mm
L'accumulation de données expérimentales et l’implémentation des \acrlong{TMDCs} monocouches dans des transistors à effet de champ \cite{RadiB.R}\cite{Lembke D,Das S}, des dispositifs logiques \cite{Wang H,Radisav l jević B} et dans des structures optoélectroniques\cite{Britnell} \cite{Pospischil A,Jo S} nécessitent des modèles théoriques pour étudier  leurs propriétés électroniques. La structure de bande est l'un des concepts les plus importants en physique de l'état solide. Nous rappelons que la relation de dispersion est celui qui traduit la variation de  l'énergie en fonction du vecteur d'onde obtenue à partir de la résolution de l'équation de Schrödinger dans un potentiel périodique formé par les ions supposés fixes.

 Le but ce chapitre est de développer le modèle théorique des TMDCs au point K et dans d’autres points d’intérêt de la zone hexagonal de Brillouin. Nous introduisons dans un premier temps les différentes méthodes de calcul de structure de bandes. Nous développons ensuite les hamiltoniens k.p utilisés dans ce travail aux points $\Gamma$ et K.

\section{Position du problème : Effet à N corps}
On considère un système physique constitué de N' noyaux et N électrons.   
L'équation de départ est celui de Schrödinger\footnote{Le comportement des noyaux et des électrons est gouverné par la mécanique quantique. On utilise alors les méthodes quantiques pour laquelle l'énergie est calculée par la résolution de l'équation de Schrödinger.} qui décrit par : 
\begin{equation}
\hat{H}\psi =E\psi 
\end{equation}

où \^H est l'hamiltonien non relativiste. 
Nous somme face à un problème à (N+N') particules chargée, difficile à le résoudre vu qu'on ne sait résoudre qu'un problème à deux corps et qui se complique encore plus par les nombres d'interactions possibles au sein de système et les degrés de liberté de chaque électron et chaque noyau, puisque les électrons sont repérés par le variable (r) contenant les coordonnées spatiales ainsi que le spin alors que les noyaux sont repérés par le variable (R).
\begin{equation}
\hat{H}\psi \left(r_1.\ .\ .\ .\ .\ .r_N,R_1.\ .\ .\ .\ .\ .\ .\ R_{N'}\right)=E\psi (r_1.\ .\ .\ .\ .\ .r_N,R_1.\ .\ .\ .\ .\ .\ .\ R_{N'}) 
\end{equation}

L'hamiltonien \^H est un opérateur décrivant toutes les combinaisons possibles d'interaction :
\begin{equation}
\hat{H}={\hat{T}}_e+{\hat{T}}_{noy}+{\hat{V}}_{e-noy}+{\hat{V}}_{noy-noy} 
\end{equation}

Un nombre des simplifications est alors nécessaire pour simplifier l'équation de Schrödinger, la première approximation est celui de \textit{Born Oppenheimer} qui consiste à découpler les degrés de liberté ioniques à celle électroniques du fait de la différence de masse entre l'électron et le noyau, où les noyaux sont considérés fixes par rapport aux électrons, ainsi la fonction d'onde s'écrit sous la forme d'un produit :
\begin{equation}
\psi \cong {\psi }_e(r,R)*{\psi }_{noy}(R) 
\end{equation}

Par conséquent l'équation de Schrödinger électronique et dans le cadre de cette approximation s'écrit :
\begin{equation} 
\left[\widehat{T_e}+{\hat{V}}_{e-e}+{\hat{V}}_{e-noy}\right]{\psi }^e\left(x_1.\ .\ .\ .\ .\ .x_N\right)=E{\psi }^e(x_1.\ .\ .\ .\ .\ .x_N) 
\end{equation} 

Une autre approximation nous permet de simplifie énormément le calcul est celui si nous considérerons les électrons indépendants, cela nous permet d'exprimer N fonctions d'onde à une seul variable plutôt que une fonction à N variable.\\
En définitive, dans le cadre ainsi précisé l'équation aux valeurs propres à résoudre est celle associée au Hamiltonien d'un unique électron dans un potentiel périodique.
\begin{equation} 
{\rm {\mathcal H}}=\frac{\vec{P}^{2} }{2m_{0} } +V\left(\vec{r}\right) 
\end{equation} 

où V est périodique possède la périodicité de cristal:$\ V(\overrightarrow{r}+\overrightarrow{R}$)=$V(\overrightarrow{r}$) pour tout vecteur $\vec{R}$ du réseau de Bravais.\\
Vu que nos matériaux sont de type cristallin cela nous permet d'appliquer alors le théorème de Bloch, on associe donc pour chaque vecteur d'onde k un indice n de bande de la structure électronique et les fonctions de Bloch périodiques deviennent des solutions pour l'équation de Schrödinger : 
\begin{equation} 
\psi_{n,k}(r)={U}_{n,k}(r)e^{ikr}
\end{equation} 

où {${U}_{n,k}$}(r) ayant la périodicité de la structure cristalline, n désigne l'indice de la bande et k le vecteur d'onde appartenant à la première zone de Brillouin.\\
Le but est de savoir résoudre l'équation de Schrödinger ci-dessous en utilisant la théorie k.p :        
\begin{equation} 
\label{eq:2.7} 
\left[\frac{-{\hslash }^2}{2m_0}{\overrightarrow{\mathrm{\nabla }}}^2+V(r)\right]{\psi \ }_{n,k}(r)=E_n\left(k\right)\ {\psi \ }_{n,k}(r) \quad \textbf{avec} \quad \ \ \overrightarrow{P}=-i\hslash \ \overrightarrow{\mathrm{\nabla }}\ \ \
\end{equation}

\section{Détermination théorique des structures de bandes}

Ils existent plusieurs méthodes pour résoudre l'équation \ref{eq:2.7} aux valeurs propres afin de déterminer les bandes d'énergie, tel que les méthodes ab initio ou les méthodes semi-empiriques. Bien que certains de ces méthodes divergent dans leur approche pour traiter la physique et notamment les interactions, mais tous convergent sur leur schéma numérique puisque ils sont tous basées sur des calculs mono-électroniques de fonctions d'onde.

\subsection{Les méthodes ab-initio}

Le principe de ces méthodes repose sur le principe variationnelle \textit{Rayleigh-Ritz} non perturbatif   et qui s'appuie sur l'extraction de l'état fondamental correspondant à l'énergie la plus basse du système, un calcul variationnelle alors nous permet de trouver la fonction d'onde associé à ce minimum d'énergie, où la valeur moyenne de l'hamiltonien d'un système dans n'importe qu'elle état $\left|\psi >\right.$ est toujours supérieur à l'énergie fondamental $E_{min}$ . 
\begin{equation} 
{\ \ \ \ \ \ \ \ E}_{min}=min\frac{\left\langle \psi \mathrel{\left|\vphantom{\psi  \mathcal{H} \psi }\right.\kern-\nulldelimiterspace}\mathcal{H}\mathrel{\left|\vphantom{\psi  \mathcal{H} \psi }\right.\kern-\nulldelimiterspace}\psi \right\rangle }{\left\langle \psi \mathrel{\left|\vphantom{\psi  \psi }\right.\kern-\nulldelimiterspace}\psi \right\rangle }\ \ \ \ \ \ \ \ \ \ \ \ \ \ \ \ \  
\end{equation} 

Par la suite toute la difficulté réside dans l'expression de l'Hamiltonien $\ \mathcal{H}$.

\subsection{Les méthodes semi-empiriques}

Lorsque les méthodes ab-initio exigent des puissances numériques qui limitent leur utilisation, les méthodes empiriques entrent en jeu pour contrer cette limitation, ces méthodes sont basées sur des hamiltoniens modèles utilisant des paramètres ajustables afin de simplifier le problème, dans laquelle chaque électron est dans un état  $\left|\left.n,\ k\right\rangle \right.\ $  associé à une fonction d'onde {${\psi}_{n,k}$}. La structure cristalline peut alors être déterminé dans toute la premier \acrlong{ZB} (méthode pseudo-potentiels ou liaison forte) soit au voisinage de vecteur d'onde spécifique (méthode k.p ou l'approximation de la masse effective).

\section{Théorie k.p}

La théorie k.p est une méthode semi-empirique profondément liée aux expériences. En revanche, malgré la complexité des hétéro-structures bi- et unidimensionnelles, elles permettent d'expliquer certains résultats expérimentaux, qui sans cela resteraient une suite de nombres sans signification. L'intérêt important de la théorie k.p est qu'elle est considérée comme une méthode analytique et elle entraîne une compréhension en profondeur des propriétés des bandes et fonctions d'onde au voisinage d'un certain nombre de points de la zone de Brillouin. La détermination des relations de dispersion au voisinage des extrema des bandes via la théorie k.p aboutira à la détermination de la masse effective, les transitions inter-bandes, gain optique, gain modal, courant de seuil,...). On est ainsi capable de décrire, sans difficulté importante les performances optiques des semi-conducteurs à basses dimensions telle par exemple les puits, les fils et les boites quantiques. 

\subsection{Historique de la théorie k.p }

La théorie est née dans les années cinquante, les premiers articles \cite{G.Dresselhaus},\cite{J.M.Luttinger} décrivent en détails le sommet de la bande de valence, via des perturbations du second ordre appliquées un état dégénéré, le couplage de la bande de valence et de la bande de conduction est pour la première fois décrit en détail par Kane\cite{E.O.Kane} et celui qui relie la masse effective, la bande interdite et le couplage \acrlong{SO}. Par la suite les deux théories de Luttinger-Kohn et de Kane seront combinées par Pidgeon et Brown\cite{ C.R.Pidgeon}, qui définiront les paramètres de type Luttinger et construiront un hamiltonien 8 bandes. 
 La théorie à ce point est bien comprise, et la méthode k.p permet de décrire la bande de valence et la première bande de conduction des semi-conducteurs à bande interdite directe au voisinage du centre de la \acrlong{ZB}.

L'ajout des bandes permet détendre de domaine de validité de la méthode ou offre une meilleure précision mais il faut prises en compte l'influence des bande. Pfeffer et Zawadski\cite{P. Pfeffer} passent alors à un modèle à 14 bandes pour prendre en compte le « splitting » de spin de la bande de conduction ainsi Nicolas Cavassilas et al. ont développé un hamiltonien 20 bandes \cite{ Boujdaria}, en simulant l'influence des bandes d à l'aide de bandes fictives $s^*$ et de paramètres de type Luttinger aussi en bande de conduction et ils ont obtenu une description précise de la structure de bande de -1 eV à 3.5 eV. Mais finalement, nous verrons que passer à 30 bandes en introduisant l'interaction \acrshort{SO} dans l'hamiltonien de Cardona et Pollak permet d'étendre le domaine de validité de cette méthode de 6 eV à +5 eV sur toute la \acrshort{ZB} sans multiplier le nombre de paramètres ajustables.

\subsection{Structure de bande et caractère orbital }
Certaines propriétés des semi-conducteurs sont plus faciles à comprendre si l'on connaît la contribution des orbitales atomiques pour chaque bande à un point k donné par exemple la différence de la composition atomique orbitale peut expliquer le gap entre la valeur de spin-splitting du \acrlong{BC} et du \acrlong{BV} au point K \cite{Xiao D}.

\begin{center}
\includegraphics[scale=1.0]{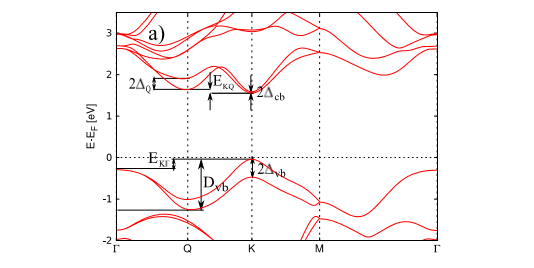}
\captionof{figure}{Dispersion dans l'axe de $\Gamma$-K-M-$\Gamma$ dans le \acrshort{ZB} de la structure de bande d'une monocouche de TMDCs. SOC est prise en compte ainsi les différentes différences d'énergie de la bande sont également indiquées\protect\cite{Andor}.}
\end{center}

Dans notre cas les fonctions orbitales de base sont les trois orbitales  ${p(p}_x,\ p_y,\ p_z)$ vient de l'atome de chalcogènes X alors que cinq autres orbitales \textit{d} viennent de l'atome M de métal de transition ${d(d}_{3z^2-r^2},\ d_{xz},d_{yz},\ d_{x^2-y^2}\ ,\ d_{xy}).\ $

\[\left|p_{y} >=\right. \frac{i}{\sqrt{2} } \left[\left|1,1>+\left|1,-1>\right. \right. \right]\] 
\[\left|p_{z} >=\right. \left|1,0>\right. \] 
\[\left|d_{3z^2-r^2} \right. >=\left|2,0\right. >\] 
\[\left|d_{xz} \right. >=\frac{-1}{\sqrt{2} } \left[\left|2,1>-\left|2,-1>\right. \right. \right]\] 
\[\left|d_{yz} \right. >=\frac{i}{\sqrt{2} } \left[\left|2,1>+\left|2,-1>\right. \right. \right]\] 
\[\left|d_{x^2-y^2} \right. >=\frac{1}{\sqrt{2} } \left[\left|2,2>+\left|2,-2>\right. \right. \right]\] 
\[\left|d_{xy} \right. >=\frac{-i}{\sqrt{2} } \left[\left|2,2>-\left|2,-2>\right. \right. \right]\] 

La figure ci-dessous \ref{fig 2.1 } montre la contribution des orbitales atomiques individuelles d et p respectivement pour les atomes métalliques et les atomes de chalcogène à une bande donnée pour une monocouche de \acrshort{TMDCs} type 2H-$MX_2$.

\begin{center}
\includegraphics[scale=0.75]{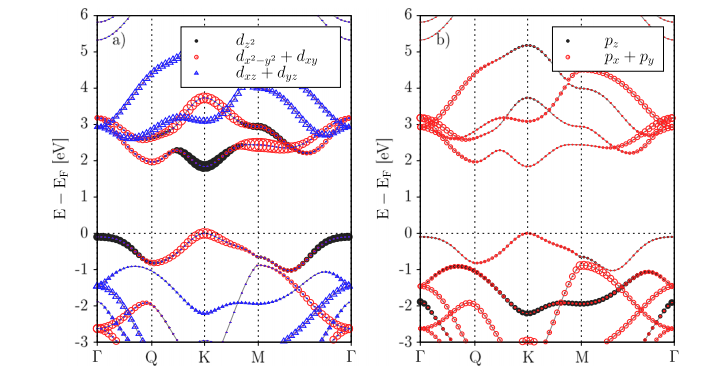}
\captionof{figure}{La contribution des orbitales atomiques dans les bandes d'énergie pour une monocouche d'un semi-conducteur TMDCs type $MX_2$. (a) Les orbitales d de l'atome M de métal de transition. (b) Les orbitales p de l'atome de chalcogène X. L'épaisseur de la bande représente le poids de l'orbital. La couplage spin-orbite ne pas tenir en compte\protect\cite{Andor}.}
\label{fig 2.1 }
\end{center}

En comparant les figures \ref{fig 2.1 } (a) et (b), nous constatons que plus d'un type d'orbital atomique contribue à la fois au \acrshort{BC} et au \acrshort{BV} et que le poids des orbitales atomiques change à travers le \acrlong{ZB}. En concentrons sur le caractère orbital des extremums de \acrshort{BV} et de \acrshort{BC} au point K, on peut constater que la  minimum de \acrshort{BC} est gouvernée par les orbitales $d_{x^2 - y^2}$ dû aux atomes M de métal de transition et par les orbitales $p_x$  , $d_{z^2}$ dû aux atomes X de chalcogène, alors que le maxima de \acrshort{BV} possède les orbitales $d_{x^2-y^2}$  ,$d_{xy}$  des atomes M et les orbitales $p_x$ et $p_y$  des atomes X.\\
Une classification des fonctions orbitales selon les représentations irréductibles de groupe $C_{3h}$ permet la restriction de nombre des fonctions orbitales qui identifie une bande donnée à sa \acrlong{RUI}. Les propriétés de symétrie des fonctions d'ondes de Bloch aux bords de zone sont résumées dans le tableau ci-dessous en indiquant la combinaison appropriée d'orbitales atomiques et les signatures associées de $C_{3h}$  le groupe ponctuel de symétrie à ces points  de la \acrshort{ZB}. Il est possible alors de savoir la nature de la \acrshort{RUI} gouvernée pour chaque bande. 

\begin{center}
{\renewcommand{\arraystretch}{2} %donne la distance entre les lignes%
{\setlength{\tabcolsep}{0.5cm} %donne la distance entre les collones%
\begin{tabular}{|p{0.2in}|p{0.2in}|p{0.2in}|p{2.2in}|p{1.3in}|} \hline 
RUI & $C_3$  & ${\sigma }_h$  &                   M atome &             X atome \\ \hline 
$A'$  & 1 & 1 & $\frac{1}{\sqrt{2}}(d_{x^2-y^2}\pm id_{xy})$~; $\frac{1}{\sqrt{2}}(p_x\mp ip_y)$~ & $\frac{1}{\sqrt{2}}(p_x\pm ip_y)$ (b) \\ \hline 
$A''$  & 1 & -1 & $\frac{1}{\sqrt{2}}(d_{xz}\pm id_{yz})$ & $\frac{1}{\sqrt{2}}(p_x\pm ip_y)$  (ab) \\ \hline 
${E'}_1$  & ${\omega }^{\pm 1}$  & 1 & $d_{3z^2-r^2}$ & $\frac{1}{\sqrt{2}}(p_x\mp ip_y)$~ (b) \\ \hline 
${E'}_2$  & ${\omega }^{\mp 1}$  & 1 & $\frac{1}{\sqrt{2}}(d_{x^2-y^2}\pm id_{xy})$~; $\ \frac{1}{\sqrt{2}}(p_x\pm ip_y)$ & $p_z,\ s\ \ $(b) \\ \hline 
${E''}_1$  & ${\omega }^{\pm 1}$  & -1 & $p_z$ & $\frac{1}{\sqrt{2}}(p_x\pm ip_y)$  (ab) \\ \hline 
${E''}_2$  & ${\omega }^{\mp 1}$  & -1 & $\frac{1}{\sqrt{2}}(d_{xz}\pm id_{yz})$ & $p_z\ \left(b\right)\ ,\ s\ \ $(ab) \\ \hline
\end{tabular} }}
\captionof{table}{Classification des fonctions de Bloch au point $K^\pm$ de la zone de Brillouin. Le signe $\pm$ se réfère aux deux vallées  $K^\pm$ . (a) indique un orbital liant alors que (b) indique un orbital anti-liant.}
\end{center}

On peut conclure alors que la bande de valence possède la représentation irréductible A' alors que la bande de conduction possède la représentation $E'_1$.

Le tableau ci-dissous \ref{tab 2.3} montre les fonctions de base c'est à dire les représentations irréductibles de groupe $C_{3h}$ au point $K^+$, ainsi la bande à laquelle dépend une fonction de base donnée.Notons que les fonctions de base au point $K^-$ sont les complexes conjuguées de celui au point $K^+$.

\begin{center}
{\renewcommand{\arraystretch}{1.5} %donne la distance entre les lignes%
{\setlength{\tabcolsep}{0.75cm} %donne la distance entre les collones%
\begin{tabular}{|p{0.6in}|p{2.3in}|p{0.8in}|} \hline 
      RUI & Fonction de base & Bande  \\ \hline 
$\boldsymbol{A}'$ & ${\psi }^M_{2,-2} \ ,\  \frac{1}{\sqrt{2}}({\psi }^{X1}_{1,-1}+{\psi }^{X2}_{1,-1})$ & vb \\ \hline 
$\boldsymbol{A}''$ & ${\psi }^M_{2,1} \ ,\  \frac{1}{\sqrt{2}}({\psi }^{X1}_{1,-1}-{\psi }^{X2}_{1,-1})$ & cb+1 \\ \hline 
${\boldsymbol{E}'}_{\boldsymbol{1}}$ & $\left|\left.{\psi }^M_{2,0}\right\rangle \right. \ ,\  \frac{1}{\sqrt{2}}({\psi }^{X1}_{1,1}+{\psi }^{X2}_{1,1})$ & cb \\ \hline 
${\boldsymbol{E}'}_{\boldsymbol{2}}$ & ${\psi }^M_{2,2} \ ,\  \frac{1}{\sqrt{2}}({\psi }^{X1}_{1,0}-{\psi }^{X2}_{1,0})$ & vb-3\newline cb+2 \\ \hline 
${\boldsymbol{E}''}_{\boldsymbol{1}}$ & $\left|\left.{\psi }^M_{1,0}\right\rangle \right.\  \ ,\  \frac{1}{\sqrt{2}}({\psi }^{X1}_{1,1}-{\psi }^{X2}_{1,1})$ & vb-2 \\ \hline 
${\boldsymbol{E}''}_{\boldsymbol{2}}$ & ${\psi }^M_{2,-1}\ ,\  \frac{1}{\sqrt{2}}({\psi }^{X1}_{1,0}+{\psi }^{X2}_{1,0})$ & vb-1 \\ \hline 
\end{tabular}}}
\captionof{table}{Les fonctions de base des repr\'{e}sentations irréductibles de groupe  \textbf{${\boldsymbol{C}}_{\boldsymbol{3}\boldsymbol{h}}$} au point $K^+$ \protect\cite{Andor}.}
\label{tab 2.3}
\end{center}

\subsection{Théorie k.p pour un semi-conducteur à gap direct au point $\Gamma$}
On part de l'équation de Schrödinger indépendante du temps à une dimension : 
\begin{equation}  
\left\{ \begin{array}{c}
\left.-\frac{{\hslash }^2}{2m_0}\frac{{\partial }^2}{\partial x^2}+V\left(x\right)\right\}{\psi }_{n,k_x}=E_{n,k_x}{\psi }_{n,k_x}
\end{array}
\right. 
\end{equation} 

Où $E_{n,k_x}$ était une valeur propre du système associé à la fonction d'onde ${\psi }_{n,k_x}$\textit{ }et \textit{V} le potentiel périodique à une dimension vu par un électron dans le cristal alors que $m_0$ est la masse de l'électron libre.\\

En remplacent ${\psi }_{n,k_x}$ par $U_{n{,k}_x}.e^{ik_x.x}$ on trouve~que les fonctions $U_{n{,k}_x}$doivent vérifier l'équation aux valeurs propres suivante :

\begin{equation} 
\label{eq 2.10} 
\left\{\frac{{{\hslash }^2k}^2_x}{2m_0}+\frac{\hslash }{m_0}k_x.p_x+\frac{{\ P}^2_x}{2m_0}+V\left(x\right)\right\}U_{n{,k}_x}\left(x\right)=E_{n,k_x}U_{n{,k}_x}\left(x\right)
\end{equation}
 
Ainsi on peut réécrire l'équation \ref{eq 2.10} sous la forme~:~
\begin{equation} 
\left\{ \begin{array}{c}
\left.\frac{{(\ P}_x+\hslash k_x)^2}{2m_0}+V\left(x\right)\right\}U_{n{,k}_x}\left(x\right)=E_{n,k_x}U_{n{,k}_x}\left(x\right)
\end{array}
\right. 
\end{equation} 
Pour que la méthode k.p soit plus générale on écrit~:
\begin{equation} 
\left\{ \begin{array}{c}
\left.\frac{{\hslash }^2k^2}{2m_0}+\frac{\hslash }{m_0}k.p+\frac{P^2}{2m_0}+V\left(r\right)\right\}U_{n{,k}_x}(r)= \end{array}
\right.\left\{ \begin{array}{c}
\left.\frac{(p+\hslash k)^2}{2m_0}+V\left(r\right)\right\}U_{n,k}(r)=E_{n,k}U_{n,k}(r) \end{array}
\right. 
\end{equation} 
Ainsi, pour k=0 l'équation de Schr\"{o}dinger prend une forme particulière simple, ayant la symétrie du potentiel cristallin V(r)~:  

\begin{equation} 
\left\{ \begin{array}{c}
\left.\frac{P^2}{2m_0}+V\left(r\right)\right\}U_{n,0}(r)=E_{n,0}U_{n,0}(r) 
\end{array}
\right. 
\end{equation} 

Rappelons que la méthode k.p suppose que l'on connait les valeurs propres $E_{n,0}$ solutions de l'équation  soit à partir d'une théorie soit à partir de l'expérience.

On limite tout d'abord le calcul à des niveaux d'énergie non dégénéré, et a des valeurs de k voisines de k=0. O\`{u} en traite le terme $\frac{\hslash }{m_0}k.p$ comme étant une perturbation de l'hamiltonien${\ \ \hat{H}}_0=\frac{{\hslash }^2k^2}{2m_0}+\frac{p^2}{2m_0}+V\left(r\right)$. 

L'hamiltonien total s'écrit~alors comme suit :

\begin{equation} 
{\ \hat{H}=\hat{H}}_0+{\hat{H}}_{k.p}=\left\{\frac{p^2}{2m_0}+\frac{{\hslash }^2k^2}{2m_0}+V\left(r\right)\right\}+\frac{\hslash }{m_0}k.p
\end{equation}

Tel que l'énergie propre de${\hat{H}}_0$ est~:  $E_{n,k}(r)=E_{n,0}+\ \frac{{\hslash }^2k^2}{2m_0}~$  associée à la fonction propre $U_{n,0}\left(r\right).$ 

L'application de théorie de perturbation nous permet de trouver les niveaux d'énergie de l'hamiltonien $\hat{H}$ connaissant le spectre de ${\ \hat{H}}_0.$

Sachant que l'opérateur $\hat{p}$ est un opérateur impaire ce qui limite son couplage que avec des états de parité différente, ainsi les éléments de matrices diagonaux $\left\langle n,0\mathrel{\left|\vphantom{n,0 p n,0}\right.\kern-\nulldelimiterspace}p\mathrel{\left|\vphantom{n,0 p n,0}\right.\kern-\nulldelimiterspace}n,0\right\rangle $ s'annulent vu que le forme intégrale donne~:
\begin{equation} 
\int^{+\infty }_{-\infty }{{U_{n,0}(\overrightarrow{r})}^*}\left(-i\hslash \frac{\partial }{\partial r}\right)U_{n,0}(\overrightarrow{r})d\overrightarrow{r}  = 0                                          
\end{equation} 
La correction du premier ordre s'annule par la suite~:
\begin{equation} 
E^{(1)}_{n,k}=\left.<n\right|~{\hat{H}}_{k.p}\left|n\right.\mathrm{>=}\frac{\hslash }{m_0}k\left\langle n,0\mathrel{\left|\vphantom{n,0 p n,0}\right.\kern-\nulldelimiterspace}p\mathrel{\left|\vphantom{n,0 p n,0}\right.\kern-\nulldelimiterspace}n,0\right\rangle =0 
\end{equation} 
Il ne reste donc que la correction du second ordre. Les~ énergies à cet ordre s'écrit~:
\begin{equation}
E^{(2)}_{n,k}={\hslash}^2\sum_{n\neq n'} \frac{\left\langle  n', 0\vert\frac{k_p}{m_0}\vert n,0\right\rangle  \left\langle  n, 0\vert\frac{k_p}{m_0}\vert n',0\right\rangle }{E_{n,0} - E_{n',0}}
\end{equation}
Si on considère alors des petites valeurs de k voisines de k=0, pour laquelle la théorie de perturbation stationnaire reste valable ou on traite l'opérateur ${\hat{H}}_{k.p}=\frac{\hslash }{m_0}k.\hat{p}$  comme une perturbation de ${\hat{H}}_0\ $  . L'énergie total pour un développement en k jusqu'à l'ordre 2 de système s'écrit alors :    
\begin{equation}
E_{n,k}=E_{n,0} + \frac{\hslash^2 k^2}{2m_0}\hslash^2\sum_{n\neq n'} \frac{\left\langle  n', 0\vert\frac{k_p}{m_0}\vert n,0\right\rangle  \left\langle  n, 0\vert\frac{k_p}{m_0}\vert n',0\right\rangle }{E_{n,0} - E_{n',0}}
\end{equation}
Et donc,
\begin{equation}
E_{n,k}=E_{n,0} +\sum_{\alpha\beta}{\frac{\hslash^2}{2m_0}\left\lbrace k_\alpha k_\beta \delta_{\alpha\beta} +\frac{2}{m_0} \sum_{n\neq n'}\frac{\left\langle  n', 0\vert\frac{P_\alpha}{m_0}\vert n,0\right\rangle  \left\langle  n, 0\vert\frac{P_\beta}{m_0}\vert n',0\right\rangle }{E_{n,0} - E_{n',0}}\right\rbrace} 
\end{equation}
Ce qui est marqué que la théorie de la méthode k.p fait simplement varier la masse effective des bandes, tel qu'on peut écrit l'énergie comme suit ;
\begin{equation}
E_{n,k} = E_{n,0} + \sum_{\alpha \beta} \frac{\hslash^2}{2m_0}\left( \frac{m_0}{m^*}\right)_{\alpha \beta} k_\alpha k_\beta
\end{equation}
Avec,
\begin{equation}
\Rightarrow\left(\frac{m_0}{m^*}\right)_{\alpha\beta} = \delta_{\alpha\beta} + \frac{2}{m_0} \sum_{n\neq n'}{\frac{\left\langle  n', 0\vert\frac{P_\alpha}{m_0}\vert n,0\right\rangle  \left\langle  n, 0\vert\frac{P_\beta}{m_0}\vert n',0\right\rangle }{E_{n,0} - E_{n',0}}}
\label{eq 2.21}
\end{equation}

On peut conclure alors que le modèle de la masse effective peut être considéré comme une théorie k.p à une bande et l'influence des autres bandes étant traité comme perturbation.
Pour le cas isotrope où $\alpha = \beta$, la relation de dispersion \ref{eq 2.21} adopte la forme simple :
\begin{equation}
E_{n,k} - E_{n,0} = \frac{\hslash^2}{2m^*} k^2
\end{equation}

\renewcommand{\labelitemi}{$\bullet$}
\begin{itemize}
\item \textbf{Cas d'hamiltonien à deux bandes :}
\end{itemize}
Prenons l'exemple de bandes de conduction et de valence d'énergie respectives $E_c$ et $E_v$ mais plus on considère de bandes, plus l'ordre du développement limité est grand, plus le domaine de validité de la méthode s'accroît.  
La matrice représentant l'hamiltonien \^H dans la base $\left\lbrace |U_v > , |U_c >\right\rbrace $ s'écrit donc :
\begin{equation}
H = \left( \begin{array}{c}
\ E_v + \frac{\hslash^2 k^2}{2m_0}\ \ \ \ \ \ \ \ \ \ \ \ \frac{\hslash}{m_0} k \left\langle U_v |p|U_c\right\rangle\\
\ \frac{\hslash}{m_0} k \left\langle U_c |p|U_v\right\rangle\ \ \ \ \ \ \ \ \ \ \ \ E_c + \frac{\hslash^2 k^2}{2m_0} \end{array}
\ \right)
\end{equation}

En posant que $\left\langle U_c |p|U_v\right\rangle = \tilde{p}$ : on obtient les relations de dispersion pour la bande de valence et de conduction au voisinage de k=0 s'écrit comme suit : 
\begin{equation}
E_v(k) = E_v + \frac{\hslash^2 k^2}{2m_0} + \frac{\hslash^2}{m_0^2}\frac{k^2 \tilde{p}^2}{E_v-E_c} = E_v + \frac{\hslash^2 k^2}{2m_0} \left( 1 - \frac{2 \tilde{p}^2}{m_0} \frac{1}{E_g}\right)= E_v + \frac{\hslash^2 k^2}{2m_v^*}
\end{equation}

\begin{equation}
E_c(k) = E_c + \frac{\hslash^2 k^2}{2m_0} + \frac{\hslash}{m_0}\frac{k^2 \tilde{p}^2}{E_v-E_c} = E_c + \frac{\hslash^2 k^2}{2m_0} \left( 1 + \frac{2 \tilde{p}^2}{m_0} \frac{1}{E_g}\right)= E_c + \frac{\hslash^2 k^2}{2m_c^*}
\end{equation}

Remarquons que si on connait les énergies $E_c$ (k) et $E_v$ (k)on peut connait directement les masses effectives : 
\begin{equation}
\frac{m_0}{m^*_c} = 1+ \frac{2 \tilde{p}^2}{m_0} \frac{1}{E_g}
\end{equation}
\begin{equation}
\frac{m_0}{m^*_v} = 1- \frac{2 \tilde{p}^2}{m_0} \frac{1}{E_g}
\end{equation}

Soit : $\frac{1}{m_c^*} = - \frac{1}{m_v^*} + \frac{1}{m_0}$ mais dans la plupart des cas, dans les régions des faibles k, ces deux bandes sont l'un miroir de l'autre tel que $m_c^* \cong - m_v^*$.

\begin{center}
{\renewcommand{\arraystretch}{1.5} %donne la distance entre les lignes%
\begin{tabular}{|p{1.8in}|p{0.9in}|p{0.9in}|p{0.6in}|p{0.9in}|} \hline 
 & ${MoS}_2$\textit{} & ${MoSe}_2$\textit{} & ${WS}_2$\textit{} & ${WSe}_2$\textit{} \\ \hline 
$m^{(\mathrm{\Gamma })}_{vb}/m_0\ $(HSE, LDA) & -2.60 & -3.94 & -2.18 & -2.87 \\ \hline 
Exp [\textit{me}V] & $-2.4\pm 0.3$ & $-3.9\pm 0.3$ & - & - \\ \hline 
$E_{K\mathrm{\Gamma }}$[\textit{me}V] & $-140$ & $-380$ & - & $-590\pm 40$ \\ \hline 
$n_{\mathrm{\Gamma }}\left[{10}^{12}{cm}^{-2}\right]$(HSE,LDA) & 15.8 & 130 & 36.86 & 81.4 \\ \hline 
\end{tabular}}
\captionof{table}{Les masses effective de trou $m_{vb}^{(\Gamma)}$ au point $\Gamma$ pour différentes TMDCs-bulk donnée à partire DFT, Différence d'énergie ainsi la densité de trou $n_\Gamma$ est donnée à ce point d'haute symétrie de la BZ\protect\cite{Zhang Yi et al,Jin W et al}.}
\end{center}

Notons que, pour que la théorie de perturbation soit une bonne approximation il faut que les éléments de matrices de la perturbation $\frac{\hslash}{m_0}k.p$ entres les fonctions propres $U_{n,0}$ et$U_{n',0}$ soient tels que :
\begin{equation}
\left|{\frac{\left\langle  n', 0\left| \frac{\hslash}{m_0} k.p \right| n,0\right\rangle }{E_{n,0} - E_{n',0}}} \right|< 1
\end{equation}

Ce que signifie dans notre cas~: 
\begin{equation}
\frac{1}{m_c^*}\backsim \frac{2}{m_0^2} \frac{\left|\tilde{p}\right|^2}{E_g}
\end{equation}
\begin{equation}
\frac{\hslash k}{m_0} < \frac{E_g}{\left|\tilde{p}\right|}
\end{equation}
En combinant ces deux relations on obtient :
\begin{equation}
\frac{\hslash^2 k^2}{2m_c^*} < E_g
\end{equation}

Ainsi pour que la théorie de perturbation ci-dessus donne une bonne approximation, il faut que l'énergie dans la bande cinétique de la bande $E_c$ reste petite devant la largeur de la bande interdite.

\subsection{Théorie k.p pour un semi-conducteur à gap indirect au point $K^{\pm}$ de la zone de Brillouin}

Suivant la méthode de Kane\cite{E.O.Kane}, nous effectuons maintenant un calcul perturbatif afin d'obtenir les courbes de dispersion $E_n\left(\overrightarrow{k}\right)\ $ au voisinage des points $\ K^{\pm }$. Selon cette théorie, l'hamiltonien à un électron s'écrit :
\begin{equation} 
\hat{{\rm {\mathcal H}}}=\frac{\hat{p}^2}{2m_{0} } +V\left(\vec{r}\right) 
\end{equation} 

Et l'équation de Schrödinger qui devrait être satisfaite par ${\ U}_{n,\ \overrightarrow{k}}\left(\overrightarrow{r}\right)\ $ est donnée par :
\begin{equation} 
{\widehat{\ h}}_{\overrightarrow{k}}U_{n,\ \overrightarrow{k}}\left(\overrightarrow{r}\right)=E_{n,\ \overrightarrow{k}}{\ U}_{n,\ \overrightarrow{k}}\left(\overrightarrow{r}\right) 
\end{equation} 

O\`{u}  ${\hat{h}}_{\overrightarrow{k}}=\frac{1}{2m_0}{\left(\hat{p}+\hslash \overrightarrow{k}\ \right)}^2+V(\overrightarrow{r})$\\

Nous supposons comme d'habitude que si la solution est connue à un point donné ${\ \ \overrightarrow{k}}_0$ alors la solution \`{a} n'importe quel point $\overrightarrow{k}$ près de ${\overrightarrow{k}}_0$ peut \^{e}tre obtenue à partir des solutions dans ${\overrightarrow{k}}_0\ $puisque ils constituent une base complète de fonctions périodiques.
\begin{equation} 
{\hat{h}}_{{\overrightarrow{k}}_0}=\frac{1}{2m_0}{\left(\hat{p}+\hslash {\overrightarrow{k}}_0\ \right)}^2+V(\overrightarrow{r}) 
\end{equation} 

Ainsi, on peut écrire notre hamiltonien ${\hat{h}}_{\overrightarrow{k}}\ $de fa\c{c}on o\`{u} on peut apparaitre l'Hamiltonien ${\hat{h}}_{{\overrightarrow{k}}_0}$dont les solutions sont connues perturbé par un opérateur $\hat{W}\left(\overrightarrow{k}\right)$:
\[{\hat{h}}_{\overrightarrow{k}}={\hat{h}}_{{\overrightarrow{k}}_0}+\hat{W}\left(\overrightarrow{k}\right)\] 
\begin{equation}
\begin{split}
\left\{\frac{1}{2m_0}{\left(\hat{p}+\hslash \overrightarrow{k}\ \right)}^2+V\left(\overrightarrow{r}\right)+\frac{1}{2m_0}{\left(\hat{p}+\hslash {\overrightarrow{k}}_0\ \right)}^2-\frac{1}{2m_0}{\left(\hat{p}+\hslash {\overrightarrow{k}}_0\ \right)}^2\right\}U_{n,\ \overrightarrow{k}}\left(\overrightarrow{r}\right) \\
= E_{n,\ \overrightarrow{k}}{\ U}_{n,\ \overrightarrow{k}}\left(\overrightarrow{r}\right)
\end{split} 
\end{equation}
On obtient~:
\begin{equation} 
\hat{W}\left(\overrightarrow{k}\right)={\hat{h}}_{\overrightarrow{k}}-{\hat{h}}_{{\overrightarrow{k}}_0}=\frac{\hslash }{m_0}\left(\overrightarrow{k}-{\overrightarrow{k}}_0\right)\hat{p}+\frac{\hslash ^2}{2m_0}(k^2-k^2_0) 
\end{equation} 
 Les solutions sont écrites dans la base~ $\left\{\left|U_{n,{\overrightarrow{k}}_0\ }>\right.\ \ ,n\in {\mathbb{N}}^*\right\}$ puisque les solutions constitue une base propre pour de l'hamiltonien non perturbé${\ \hat{h}}_{{\overrightarrow{k}}_0}$, ainsi les éléments de matrice de $\hat{W}\left(\overrightarrow{k}\right)$ dans cette base peut \^{e}tre écrit~de la manière suivante~:
\begin{equation} 
\left\langle U_{{\overrightarrow{k}}_0,n'\ }\mathrel{\left|\vphantom{U_{{\overrightarrow{k}}_0,n'\ } \hat{W}\left(\overrightarrow{k}\right) U_{n,{\overrightarrow{k}}_0\ }}\right.\kern-\nulldelimiterspace}\hat{W}\left(\overrightarrow{k}\right)\mathrel{\left|\vphantom{U_{{\overrightarrow{k}}_0,n'\ } \hat{W}\left(\overrightarrow{k}\right) U_{n,{\overrightarrow{k}}_0\ }}\right.\kern-\nulldelimiterspace}U_{n,{\overrightarrow{k}}_0\ }\right\rangle =\frac{\hslash }{m_0}\left(\overrightarrow{k}-{\overrightarrow{k}}_0\right)\left\langle U_{{\overrightarrow{k}}_0,n'\ }\mathrel{\left|\vphantom{U_{{\overrightarrow{k}}_0,n'\ } \hat{p} U_{n,{\overrightarrow{k}}_0\ }}\right.\kern-\nulldelimiterspace}\hat{p}\mathrel{\left|\vphantom{U_{{\overrightarrow{k}}_0,n'\ } \hat{p} U_{n,{\overrightarrow{k}}_0\ }}\right.\kern-\nulldelimiterspace}U_{n,{\overrightarrow{k}}_0\ }\right\rangle +\frac{\hslash^2}{2m_0}(k^2-k^2_0){\delta }_{n',n} 
\end{equation} 

A ce stade, la théorie de perturbation stationnaire (annexe. A) nous permet d'écrire l'énergie de premier ordre $E^{(1)}_{n,\ {\overrightarrow{k}}_0}(\overrightarrow{k})$ correction de l'énergie $E_{n,{\overrightarrow{k}}_0\ }$et qui est associé à l'état propre $|U_{n,{\overrightarrow{k}}_0} >$ :

\[E^{(1)}_{n,\overrightarrow{k}_0} (\overrightarrow{k}) = \left\langle U_{\overrightarrow{k}_0,n} \left| \hat{W} (\overrightarrow{k})\right| U_{n,\overrightarrow{k}_0}  \right\rangle  = \frac{\hslash}{m_0} (\overrightarrow{k} - \overrightarrow{k}_0)\left\langle U_{\overrightarrow{k}_0,n} \left| \hat{p}  \right| U_{n,\overrightarrow{k}_0}  \right\rangle \frac{\hslash^2}{2m_0} (\overrightarrow{k}^2 - \overrightarrow{k}^2_0) \]

Posons que :${\widetilde{\overrightarrow{p}}}^{ {\overrightarrow{k}}_0}_{\ n,n}={\widetilde{\overrightarrow{p}}}_{n,\ m}({\overrightarrow{k}}_0)=$ $\left\langle U_{{\overrightarrow{k}}_0,n\ }\mathrel{\left|\vphantom{U_{{\overrightarrow{k}}_0,n\ } \hat{p} U_{n,{\overrightarrow{k}}_0\ }}\right.\kern-\nulldelimiterspace}\hat{p}\mathrel{\left|\vphantom{U_{{\overrightarrow{k}}_0,n\ } \hat{p} U_{n,{\overrightarrow{k}}_0\ }}\right.\kern-\nulldelimiterspace}U_{n,{\overrightarrow{k}}_0\ }\right\rangle $  
l'énergie de premier ordre s'écrit~:
\begin{equation}
E^{\left(1\right)}_{n,\ {\overrightarrow{k}}_0}\left(\overrightarrow{k}\right)=\frac{\hslash }{m_0}\left(\overrightarrow{k}-{\overrightarrow{k}}_0\right){\widetilde{\overrightarrow{p}}}^{\ \ {\overrightarrow{k}}_0}_{\ n,n}+\frac{\hslash ^2}{2m_0}(k^2-k^2_0) 
\end{equation} 

Nous évaluons maintenant la correction du second ordre à l'énergie pour un état orbital non dégénéré $E_{n,{\overrightarrow{k}}_0\ }$ on a :
\begin{equation} 
E^{\left(2\right)}_{n,\ {\overrightarrow{k}}_0}\left(\overrightarrow{k}\right)={\left(\frac{\hslash }{m_0}\right)}^2{\left(\overrightarrow{k}-{\overrightarrow{k}}_0\right)}^2\sum_{n\neq m}{\frac{{\left|\left\langle U_{n,{\overrightarrow{k}}_0\ }\mathrel{\left|\vphantom{U_{n,{\overrightarrow{k}}_0\ } \hat{W}\left(\overrightarrow{k}\right) U_{{\overrightarrow{k}}_0,m\ }}\right.\kern-\nulldelimiterspace}\hat{W}\left(\overrightarrow{k}\right)\mathrel{\left|\vphantom{U_{n,{\overrightarrow{k}}_0\ } \hat{W}\left(\overrightarrow{k}\right) U_{{\overrightarrow{k}}_0,m\ }}\right.\kern-\nulldelimiterspace}U_{{\overrightarrow{k}}_0,m\ }\right\rangle \right|}^2}{E_{n,{\overrightarrow{k}}_0\ }\mathrm{\ -}E_{m,{\overrightarrow{k}}_0\ }\mathrm{\ \ }}} 
\end{equation}
Soit 
\begin{equation}
\begin{split}
{\overrightarrow{p}}_{n,\ m}({\overrightarrow{k}}_0)= \left\langle {\psi }_{{\overrightarrow{k}}_0,n\ }\mathrel{\left|\vphantom{{\psi }_{{\overrightarrow{k}}_0,n\ } \hat{p} {\psi }_{m,{\overrightarrow{k}}_0\ }}\right.\kern-\nulldelimiterspace}\hat{p}\mathrel{\left|\vphantom{{\psi }_{{\overrightarrow{k}}_0,n\ } \hat{p} {\psi }_{m,{\overrightarrow{k}}_0\ }}\right.\kern-\nulldelimiterspace}{\psi }_{m,{\overrightarrow{k}}_0\ }\right\rangle \\
= \left\langle U_{{\overrightarrow{k}}_0,n\ }\mathrel{\left|\vphantom{U_{{\overrightarrow{k}}_0,n\ } \hat{p} U_{m,{\overrightarrow{k}}_0\ }}\right.\kern-\nulldelimiterspace}\hat{p}\mathrel{\left|\vphantom{U_{{\overrightarrow{k}}_0,n\ } \hat{p} U_{m,{\overrightarrow{k}}_0\ }}\right.\kern-\nulldelimiterspace}U_{m,{\overrightarrow{k}}_0\ }\right\rangle +\hslash {\overrightarrow{k}}_0{\delta }_{n,m}={\widetilde{\overrightarrow{p}}}^{ {\overrightarrow{k}}_0}_{n,n} + \hslash {\overrightarrow{k}}_0 {\delta }_{n,m}
\end{split}
\end{equation}
Ainsi, pour 
\begin{equation}
{\overrightarrow{k}}_0={\overrightarrow{K}}^{\pm }\ \ \ :\ \ {\overrightarrow{p}}_{n,\ m}({\overrightarrow{K}}^{\pm })={\widetilde{\overrightarrow{p}}}^{{ \overrightarrow{K}}^{\pm }}_{n,m}+\hslash {\overrightarrow{K}}^{\pm }{\delta }_{n,m}
\end{equation}
Finalement, le développement de l'énergie totale où en limite la correction pour le second ordre au voisinage de $\vec{K}_0$ s'écrit : 
\begin{equation}
\begin{split}
E_{n,{\overrightarrow{k}}_0\ }\left(\overrightarrow{k}\right)=E_{n,{\overrightarrow{k}}_0\ }+\frac{\hslash }{2m_0}\left(\overrightarrow{k}-{\overrightarrow{k}}_0\right){\widetilde{\overrightarrow{p}}}^{ {\overrightarrow{k}}_0}_{\ n,n}+ \frac{\hslash }{2m_0}\left(k^2-k^2_0\right)+ \\
{\left(\frac{\hslash }{m_0}\right)}^2{\left(\overrightarrow{k}-{\overrightarrow{k}}_0\right)}^2\sum_{n\neq m}{\frac{{\left|{\widetilde{\overrightarrow{p}}}^{ {\overrightarrow{k}}_0}_{n,m}\right|}^2}{E_{n,{\overrightarrow{k}}_0\ }\mathrm{\ -}E_{m,{\overrightarrow{k}}_0\ }\mathrm{\ \ }}}+\dots
\end{split}
\end{equation} 

Où on peut définir l'énergie et la vitesse de Kane comme suit :
\begin{equation} 
{\varepsilon }^{\ \ {\overrightarrow{k}}_0}_{\ n,m}=\frac{2{\left|{\widetilde{\overrightarrow{p}}}^{\ \ {\overrightarrow{k}}_0}_{\ n,m}\right|}^2}{m_0}\qquad ;\qquad {\tilde{v}}_{n,m}=\left|\frac{{\widetilde{\overrightarrow{p}}}^{\ \ {\overrightarrow{k}}_0}_{\ n,m}}{m_0}\right|\ \  
\end{equation}
Les valeurs propres solution de l'hamiltonien 
${\hat{h}}_{\overrightarrow{k}}$ s'écrit alors~: $\forall \ \overrightarrow{k}\in \left]\overrightarrow{k}-{\overrightarrow{k}}_0,\overrightarrow{k}+{\overrightarrow{k}}_0\right[$~
\begin{equation}
\begin{split}
E_{n,{\overrightarrow{k}}_0}\left(\overrightarrow{k}\right)=E_{n,{\overrightarrow{k}}_0}+\frac{\hslash }{2m_0}\left(\overrightarrow{k}-{\overrightarrow{k}}_0\right){\widetilde{\overrightarrow{p}}}^{ {\overrightarrow{k}}_0}_{n,n}+\frac{\hslash }{2m_0}\left(k^2-k^2_0\right)\\
+\frac{{\hslash }^2}{2m_0}{\left(\overrightarrow{k}-{\overrightarrow{k}}_0\right)}^2\sum_{n\neq m}{\frac{{\varepsilon }^{{\overrightarrow{k}}_0}_{n,m}}{E_{n,{\overrightarrow{k}}_0\ }\mathrm{\ -}E_{m,{\overrightarrow{k}}_0\ }\mathrm{\ \ }}}+\dots
\end{split} 
\end{equation}
Pour ${\overrightarrow{k}}_0={\overrightarrow{K}}^{\pm }\ \ $et si on limite le calcul seulement \`{a} la bande de valence et de conduction, tel que$\ \left(n,m\right)=\left\{c,v\right\}\ $on obtient~:

\begin{equation} 
{\overrightarrow{\ p}}_{n,\ m}\left({\overrightarrow{K}}^{\pm }\right)= \left\langle {\psi }_{{\overrightarrow{K}}^{\pm },n\ }\mathrel{\left|\vphantom{{\psi }_{{\overrightarrow{K}}^{\pm },n\ } \hat{p} {\psi }_{m,{\overrightarrow{K}}^{\pm }\ }}\right.\kern-\nulldelimiterspace}\hat{p}\mathrel{\left|\vphantom{{\psi}_{{\overrightarrow{K}}^{\pm },n\ } \hat{p} {\psi }_{m,{\overrightarrow{K}}^{\pm }\ }}\right.\kern-\nulldelimiterspace}{\psi }_{m,{\overrightarrow{K}}^{\pm }\ }\right\rangle ={\widetilde{\overrightarrow{p}}}^{\ \ {\overrightarrow{K}}^{\pm }}_{\ n,n}+\hslash {\overrightarrow{K}}^{\pm }{\delta }_{n,m}\ \  
\end{equation} 
 
\textbf{\underbar{Pour n = m : }}
\[{\overrightarrow{\ p}}_{n,\ n}\left({\overrightarrow{K}}^{\pm }\right)= \left\langle {\psi }_{{\overrightarrow{K}}^{\pm },c\ }\mathrel{\left|\vphantom{{\psi }_{{\overrightarrow{K}}^{\pm },c\ } \hat{p} {\psi }_{c,{\overrightarrow{K}}^{\pm }\ }}\right.\kern-\nulldelimiterspace}\hat{p}\mathrel{\left|\vphantom{{\psi }_{{\overrightarrow{K}}^{\pm },c\ } \hat{p} {\psi }_{c,{\overrightarrow{K}}^{\pm }\ }}\right.\kern-\nulldelimiterspace}{\psi }_{c,{\overrightarrow{K}}^{\pm }\ }\right\rangle =\left\langle {\psi }_{{\overrightarrow{K}}^{\pm },v\ }\mathrel{\left|\vphantom{{\psi }_{{\overrightarrow{K}}^{\pm },v\ } \hat{p} {\psi }_{v,{\overrightarrow{K}}^{\pm }\ }}\right.\kern-\nulldelimiterspace}\hat{p}\mathrel{\left|\vphantom{{\psi }_{{\overrightarrow{K}}^{\pm },v\ } \hat{p} {\psi }_{v,{\overrightarrow{K}}^{\pm }\ }}\right.\kern-\nulldelimiterspace}{\psi }_{v,{\overrightarrow{K}}^{\pm }\ }\right\rangle =0\] 

\begin{equation} 
\Rightarrow {\widetilde{\overrightarrow{p}}}^{\ \ {\overrightarrow{K}}^{\pm }}_{\ n,n}=-\hslash {\overrightarrow{K}}^{\pm } 
\end{equation} 
L'énergie s'écrit alors~:
\begin{equation}
\begin{split}
E_{n,{\overrightarrow{K}}^{\pm }\ }\left(\overrightarrow{k}\right)=E_{n,{\overrightarrow{K}}^{\pm }\ }+\frac{\hslash }{2m_0}\left(\overrightarrow{k}-{\overrightarrow{K}}^{\pm }\right)\left(-\hslash {\overrightarrow{K}}^{\pm }\right)+\frac{\hslash }{2m_0}\left(k^2-{\left(K^{\pm }\right)}^2\right)\\
+\frac{{\hslash }^2}{2m_0}{\left(\overrightarrow{k}-{\overrightarrow{K}}^{\pm }\right)}^2\sum_{n\neq m}{\frac{{\varepsilon }^{\ \ {\overrightarrow{K}}^{\pm }}_{\ n,m}}{E_{n,{\overrightarrow{K}}^{\pm }\ }\mathrm{\ -}E_{m,{\overrightarrow{K}}^{\pm }\ }\mathrm{\ \ }}}
\end{split}
\end{equation}

Par développement de deux premier termes de l'expression de de l'énergie, cela donne :
\[\frac{\hslash }{m_0}\left(\overrightarrow{k}-{\overrightarrow{K}}^{\pm }\right)\left(-\hslash {\overrightarrow{K}}^{\pm }\right)+\frac{\hslash }{2m_0}\left(k^2-{\left(K^{\pm }\right)}^2\right)=\frac{{\hslash }^2}{2m_0}{\left(\overrightarrow{k}-{\overrightarrow{K}}^{\pm }\right)}^2\] 
On obtient finalement :
\begin{equation}
E_{n,{\overrightarrow{K}}^{\pm }\ }\left(\overrightarrow{k}\right)=E_{n,{\overrightarrow{K}}^{\pm }\ }+\frac{\hslash ^2}{{2m}_0}{\left(\overrightarrow{k}-{\overrightarrow{K}}^{\pm }\right)}^2\left\{1+\sum_{n\neq m}{\frac{{\varepsilon }^{\ \ {\overrightarrow{K}}^{\pm }}_{\ n,m}}{E_{n,{\overrightarrow{K}}^{\pm }\ }\mathrm{\ -}E_{m,{\overrightarrow{K}}^{\pm }\ }\mathrm{\ \ }}}\right\} 
\end{equation}
 
A ce stade on peut introduit la notion de masse effective~:
\begin{equation}
\frac{m_0}{m^*_n}=\left\{1+\sum_{n\neq m}{\frac{{\varepsilon }^{\ \ {\overrightarrow{K}}^{\pm }}_{\ n,m}}{E_{n,{\overrightarrow{K}}^{\pm }\ }\mathrm{\ -}E_{m,{\overrightarrow{K}}^{\pm }\ }\mathrm{\ \ }}}\right\}\quad ;\quad \left(n,m\right)=\left\{c,\ v\right\} 
\end{equation}

Et l'énergie au point $K^{\pm }$ de la zone de Brillouin, suite à un calcul perturbatif de la théorie k.p est donné par~:  
\begin{equation} 
E_{n,{\overrightarrow{K}}^{\pm }\ }\left(\overrightarrow{k}\right)=E_{n,{\overrightarrow{K}}^{\pm }\ }+\frac{\hslash ^2}{{2m}^*_n}{\left(\overrightarrow{k}-{\overrightarrow{K}}^{\pm }\right)}^2 
\end{equation} 
Une structure à deux bandes, $n\in \left\{c,\ v\right\}$ nous donne la masse effective de la bande de conduction ainsi pour celle de la bande de valence~:
\begin{equation}
\frac{m_0}{m^*_c}=\left\{1+\frac{{\varepsilon }^{ {\overrightarrow{K}}^{\pm }}_{c,v}}{E_{c,{\overrightarrow{K}}^{\pm }\ }\mathrm{\ -}E_{v,{\overrightarrow{K}}^{\pm }\ }\mathrm{\ \ }}\right\}=1+\frac{2{\left|{\widetilde{\overrightarrow{p}}}^{{\overrightarrow{K}}^{\pm }}_{c,v}\right|}^2}{m_0E_{c,v\ }\mathrm{\ \ \ }}
\end{equation} 
\begin{equation}
\frac{m_0}{m^*_v}=\left\{1+\frac{{\varepsilon }^{ {\overrightarrow{K}}^{\pm }}_{v,c}}{E_{v,{\overrightarrow{K}}^{\pm }\ }\mathrm{\ -}E_{c,{\overrightarrow{K}}^{\pm }\ }\mathrm{\ \ }}\right\}=1-\frac{2{\left|{\widetilde{\overrightarrow{p}}}^{ \ {\overrightarrow{K}}^{\pm }}_{c,v}\right|}^2}{\mathrm{\ }m_0 E_{c,v\ }\mathrm{\ }}
\end{equation} 
Rappelons que pour tous les \acrshort{TMDCs} le gap direct est au point ${\overrightarrow{K}}^+$ de la zone de Brillouin :$\ E_{c,{\overrightarrow{K}}^+\ }\mathrm{\ -}E_{v,{\overrightarrow{K}}^+\ }=E_{g,{\overrightarrow{K}}^+\ }~$\textbf{ }

La masse de trou est donné par :
\begin{equation}
\frac{m_0}{m^*_{vh}}=-1+\frac{2{\left|{\widetilde{\overrightarrow{p}}}^{ {\overrightarrow{K}}^{\pm }}_{\ c,v}\right|}^2}{\mathrm{\ }m_0E_{c,v\ }\mathrm{\ }}
\end{equation}
Ceci montre que la masse effective de trou est différent du bas de la bande de conduction au top de bande de valence par la méthode k.p.\\
En outre, les calculs ab initio prédisent alors différentes masses effectives pour les électrons et les trous, ce qui évidemment brise la symétrie électron-trou.\\
On définit alors la masse moyenne da la courbure des courbes de dispersion de la conduction et du trou au gap :
\begin{equation}
\frac{m_0}{m^*_{c,v}}=\frac{m_0}{2}(\frac{1}{m^*_c}+\frac{1}{m^*_h}) 
\end{equation}
Une dernière expression o\`{u} en introduit la vitesse de Kane~est donné par : 
\begin{equation}
\frac{m_0}{m^*_{c,h}}=\frac{2{\left|{\widetilde{\overrightarrow{p}}}^{ {\overrightarrow{K}}^{\pm }}_{c,v}\right|}^2}{\mathrm{\ }m_0E_{c,v\ }\mathrm{\ }}\quad \Longrightarrow \quad E_{c,v\ }=2m^*_{c,h}{\left|\frac{{\widetilde{\overrightarrow{p}}}^{ {\overrightarrow{K}}^{\pm }}_{c,v}}{m_0}\right|}^2\equiv 2m^*_{c,h}\ {{\tilde{v}}_{c,v}}^2
\end{equation}

\begin{center}
{\renewcommand{\arraystretch}{1.5} %donne la distance entre les lignes%
\begin{tabular}{|p{1.0in}|p{1.0in}|p{1.0in}|p{1.0in}|} \hline 
\textit{} & $E_g$\textit{(e}V\textit{)} & $m^*_e/m_0$\textit{} & $m^*_h/m_0$\textit{} \\ \hline 
${MoS}_2$\textit{} & 1.80 & 0.56 & 0.64 \\ \hline 
${MoSe}_2$\textit{} & 1.51 & 0.62 & 0.72 \\ \hline 
${WS}_2$\textit{} & 1.93 & 0.33 & 0.43 \\ \hline 
${WSe}_2$\textit{} & 1.62 & 0.35 & 0.46 \\ \hline 
\end{tabular}}
\captionof{table}{Les valeurs expérimentales des masses effectives et des énergies de bande interdite pour une monocouche de TMDCs\protect\cite{Jiwon C}.}
\end{center}

Si en fait translater la matrice de perturbation dans la base d'amplitude de Bloch$\ U_{{\overrightarrow{K}}^+}$ o\`{u}$\ \overrightarrow{q}=\overrightarrow{k}-{\overrightarrow{K}}^+\ $on écrit alors~:
\begin{equation}
{\left\langle {\hat{h}}_{\overrightarrow{k}}\right\rangle }_{{\ \ U}_{{\overrightarrow{K}}^+}}={\left\langle {\hat{h}}_{{\overrightarrow{K}}^+}\right\rangle }_{{\ \ U}_{{\overrightarrow{K}}^+}}+{\left\langle \hat{W}(\overrightarrow{k})\right\rangle }_{{\ \ U}_{{\overrightarrow{K}}^+}} 
\end{equation}
\begin{equation}
\begin{split}
\hat{W}\left(\overrightarrow{k}\right)={\hat{h}}_{\overrightarrow{k}}-{\hat{h}}_{{\overrightarrow{K}}^+}=\frac{\hslash }{m_0}\left(\overrightarrow{k}-{\overrightarrow{K}}^+\right)\hat{p}+\frac{{\hslash }^2}{2m_0}\left(k^2-{(K}^+)^2\right)\\
=\frac{\hslash }{m_0}\overrightarrow{q}.\hat{p}+\frac{{\hslash }^2}{2m_0}\left(k^2-{(K}^+)^2\right)
\end{split}
\end{equation}

\textbf{\underbar{Pour $n\neq m$  }}
\begin{equation}
{\left\langle \hat{W}(\overrightarrow{k})\right\rangle }_{{\  U}_{{\overrightarrow{K}}^+}}=\frac{\hslash }{m_0}\overrightarrow{q}\left\langle {\ U}_{{\overrightarrow{K}}^+,m}\mathrel{\left|\vphantom{{\ \ U}_{{\overrightarrow{K}}^+,m} \hat{p} {\ U}_{{n,\overrightarrow{K}}^+}}\right.\kern-\nulldelimiterspace}\hat{p}\mathrel{\left|\vphantom{{\ U}_{{\overrightarrow{K}}^+,m} \hat{p} {\ U}_{{n,\overrightarrow{K}}^+}}\right.\kern-\nulldelimiterspace}{\ U}_{{n,\overrightarrow{K}}^+}\right\rangle +\frac{{\hslash }^2}{2m_0}\left(k^2-{(K}^+)^2\right)\left\langle {\ U}_{{\overrightarrow{K}}^+,m}\mathrel{\left|\vphantom{{\ \ U}_{{\overrightarrow{K}}^+,m} {\ U}_{{n,\overrightarrow{K}}^+}}\right.\kern-\nulldelimiterspace}{\ U}_{{n,\overrightarrow{K}}^+}\right\rangle
\end{equation}
Les éléments de matrice de l'opérateur$\ \hat{W}(\overrightarrow{k})$, ce sont les éléments de matrice de l'Hamiltonien $\overrightarrow{q}.\overrightarrow{p}$:
\begin{equation}
{\left\langle \hat{W}(\overrightarrow{k})\right\rangle }_{{\ \ U}_{{\overrightarrow{K}}^+}}=\frac{\hslash }{m_0}\overrightarrow{q}.{\widetilde{\overrightarrow{\ p}}}^{ {\overrightarrow{K}}^+}_{\ m,n}
\end{equation} 

Une écriture peut \^{e}tre encore simplifiée due aux propriétés de symétrie au point ${\overrightarrow{K}}^+$ en utilisant l'expression~: 
${\widetilde{\overrightarrow{\ p}}}^{ {\overrightarrow{K}}^+}_{\ m,n}={\overrightarrow{\ p}}^{\ \ \ }_{\ m,n}\left({\overrightarrow{K}}^+\right)=\left\langle {\ \ U}_{{\overrightarrow{K}}^+,m}\mathrel{\left|\vphantom{{\ \ U}_{{\overrightarrow{K}}^+,m} \hat{p} {\ \ U}_{{n,\overrightarrow{K}}^+}}\right.\kern-\nulldelimiterspace}\hat{p}\mathrel{\left|\vphantom{{\ \ U}_{{\overrightarrow{K}}^+,m} \hat{p} {\ \ U}_{{n,\overrightarrow{K}}^+}}\right.\kern-\nulldelimiterspace}{\ \ U}_{{n,\overrightarrow{K}}^+}\right\rangle $
\[{\left\langle \hat{W}(\overrightarrow{k})\right\rangle }_{{\ \ U}_{{\overrightarrow{K}}^+}}={\left\langle {\widehat{\mathcal{H}}}_{k.p}\right\rangle }_{{\ \ U}_{{\overrightarrow{K}}^+}}\]
 
Rappelons que~:    $\overrightarrow{q}.\overrightarrow{p}=\frac{1}{2}\left(q_+{\hat{p}}_-+q_-{\hat{p}}_+\right)$    $\ \ \ \ \ ;\ \ \ \ \ \ \left\{ \begin{array}{c}
q_{\pm }=q_x\pm iq_y \\ 
{\hat{p}}_{\pm }={\hat{p}}_x\pm i{\hat{p}}_y \end{array}
\right.$ \\
L'Hamiltonien effective décrivant les états à proximité des bords de la zone de Brillouin $\ K^{\pm }\ $est de la forme.
\begin{equation}
{\widehat{\mathcal{H}}}_{k.p}=\frac{\hslash }{2m_e}\left(q_+{\hat{p}}_-+q_-{\hat{p}}_+\right)={\widehat{\mathcal{H}}}^-_{k.p}+{\widehat{\mathcal{H}}}^+_{k.p}
\end{equation} 
\begin{equation}
\begin{split}
{\left\langle {\widehat{\mathcal{H}}}_{k.p}\right\rangle }_{{\ \ U}_{{\overrightarrow{K}}^+}}=\frac{\hslash }{2m_e}\overrightarrow{q}.{\overrightarrow{p}}_{\ m,n}\left({\overrightarrow{K}}^+\right)=\frac{\hslash }{2m_e}\left(q_+{\hat{p}}_{m,n,-}+q_-{\hat{p}}_{m,n,+}\right) \\
=\frac{\hslash }{2m_e}\left\{q_+\left\langle {\psi }_{{\overrightarrow{K}}^+,m}\mathrel{\left|\vphantom{{\psi }_{{\overrightarrow{K}}^+,m} {\hat{p}}_- {\psi }_{n,{\overrightarrow{K}}^+}}\right.\kern-\nulldelimiterspace}{\hat{p}}_-\mathrel{\left|\vphantom{{\psi }_{{\overrightarrow{K}}^+,m} {\hat{p}}_- {\psi }_{n,{\overrightarrow{K}}^+}}\right.\kern-\nulldelimiterspace}{\psi }_{n,{\overrightarrow{K}}^+}\right\rangle +q_-\left\langle {\psi }_{{\overrightarrow{K}}^+,m}\mathrel{\left|\vphantom{{\psi }_{{\overrightarrow{K}}^+,m} {\hat{p}}_+ {\psi }_{n,{\overrightarrow{K}}^+}}\right.\kern-\nulldelimiterspace}{\hat{p}}_+\mathrel{\left|\vphantom{{\psi }_{{\overrightarrow{K}}^+,m} {\hat{p}}_+ {\psi }_{n,{\overrightarrow{K}}^+}}\right.\kern-\nulldelimiterspace}{\psi }_{n,{\overrightarrow{K}}^+}\right\rangle \right\}
\end{split}
\end{equation}
Si on limite notre cas pour deux bande ou $\ \left(n,m\right)=\left\{v,c\right\}$ \\
Vu que l'Hamiltonien est hermétique on a alors~:\\

\renewcommand{\labelitemi}{$\bullet$}
\begin{itemize}

\item  ${\left\langle {\psi }_{{\overrightarrow{K}}^+,c}\mathrel{\left|\vphantom{{\psi }_{{\overrightarrow{K}}^+,c} {\hat{p}}_- {\psi }_{v,{\overrightarrow{K}}^+}}\right.\kern-\nulldelimiterspace}{\hat{p}}_-\mathrel{\left|\vphantom{{\psi }_{{\overrightarrow{K}}^+,c} {\hat{p}}_- {\psi }_{v,{\overrightarrow{K}}^+}}\right.\kern-\nulldelimiterspace}{\psi }_{v,{\overrightarrow{K}}^+}\right\rangle }^*=\left\langle {\psi }_{{\overrightarrow{K}}^+,v}\mathrel{\left|\vphantom{{\psi }_{{\overrightarrow{K}}^+,v} {({\hat{p}}_-)}^{\dagger } {\psi }_{c,{\overrightarrow{K}}^+}}\right.\kern-\nulldelimiterspace}{({\hat{p}}_-)}^{\dagger }\mathrel{\left|\vphantom{{\psi }_{{\overrightarrow{K}}^+,v} {({\hat{p}}_-)}^{\dagger } {\psi }_{c,{\overrightarrow{K}}^+}}\right.\kern-\nulldelimiterspace}{\psi }_{c,{\overrightarrow{K}}^+}\right\rangle =\left\langle {\psi }_{{\overrightarrow{K}}^+,v}\mathrel{\left|\vphantom{{\psi }_{{\overrightarrow{K}}^+,v} {\hat{p}}_+ {\psi }_{c,{\overrightarrow{K}}^+}}\right.\kern-\nulldelimiterspace}{\hat{p}}_+\mathrel{\left|\vphantom{{\psi }_{{\overrightarrow{K}}^+,v} {\hat{p}}_+ {\psi }_{c,{\overrightarrow{K}}^+}}\right.\kern-\nulldelimiterspace}{\psi }_{c,{\overrightarrow{K}}^+}\right\rangle $

\item  ${\left\langle {\psi }_{{\overrightarrow{K}}^+,v}\mathrel{\left|\vphantom{{\psi }_{{\overrightarrow{K}}^+,v} {\hat{p}}_- {\psi }_{c,{\overrightarrow{K}}^+}}\right.\kern-\nulldelimiterspace}{\hat{p}}_-\mathrel{\left|\vphantom{{\psi }_{{\overrightarrow{K}}^+,v} {\hat{p}}_- {\psi }_{c,{\overrightarrow{K}}^+}}\right.\kern-\nulldelimiterspace}{\psi }_{c,{\overrightarrow{K}}^+}\right\rangle }^*=\left\langle {\psi }_{{\overrightarrow{K}}^+,c}\mathrel{\left|\vphantom{{\psi }_{{\overrightarrow{K}}^+,c} {({\hat{p}}_-)}^{\dagger } {\psi }_{v,{\overrightarrow{K}}^+}}\right.\kern-\nulldelimiterspace}{({\hat{p}}_-)}^{\dagger }\mathrel{\left|\vphantom{{\psi }_{{\overrightarrow{K}}^+,c} {({\hat{p}}_-)}^{\dagger } {\psi }_{v,{\overrightarrow{K}}^+}}\right.\kern-\nulldelimiterspace}{\psi }_{v,{\overrightarrow{K}}^+}\right\rangle =\left\langle {\psi }_{{\overrightarrow{K}}^+,c}\mathrel{\left|\vphantom{{\psi }_{{\overrightarrow{K}}^+,c} {\hat{p}}_+ {\psi }_{v,{\overrightarrow{K}}^+}}\right.\kern-\nulldelimiterspace}{\hat{p}}_+\mathrel{\left|\vphantom{{\psi }_{{\overrightarrow{K}}^+,c} {\hat{p}}_+ {\psi }_{v,{\overrightarrow{K}}^+}}\right.\kern-\nulldelimiterspace}{\psi }_{v,{\overrightarrow{K}}^+}\right\rangle $ 

\item $\left\langle {\psi }_{{\overrightarrow{K}}^+,c}\mathrel{\left|\vphantom{{\psi }_{{\overrightarrow{K}}^+,c} {\hat{p}}_- {\psi }_{v,{\overrightarrow{K}}^+}}\right.\kern-\nulldelimiterspace}{\hat{p}}_-\mathrel{\left|\vphantom{{\psi }_{{\overrightarrow{K}}^+,c} {\hat{p}}_- {\psi }_{v,{\overrightarrow{K}}^+}}\right.\kern-\nulldelimiterspace}{\psi }_{v,{\overrightarrow{K}}^+}\right\rangle =\left\langle {\psi }_{{\overrightarrow{K}}^+,v}\mathrel{\left|\vphantom{{\psi }_{{\overrightarrow{K}}^+,v} {\hat{p}}_+ {\psi }_{c,{\overrightarrow{K}}^+}}\right.\kern-\nulldelimiterspace}{\hat{p}}_+\mathrel{\left|\vphantom{{\psi }_{{\overrightarrow{K}}^+,v} {\hat{p}}_+ {\psi }_{c,{\overrightarrow{K}}^+}}\right.\kern-\nulldelimiterspace}{\psi }_{c,{\overrightarrow{K}}^+}\right\rangle =0$ 

\item $\left\langle {\psi }_{{\overrightarrow{K}}^+,v}\mathrel{\left|\vphantom{{\psi }_{{\overrightarrow{K}}^+,v} {\hat{p}}_- {\psi }_{c,{\overrightarrow{K}}^+}}\right.\kern-\nulldelimiterspace}{\hat{p}}_-\mathrel{\left|\vphantom{{\psi }_{{\overrightarrow{K}}^+,v} {\hat{p}}_- {\psi }_{c,{\overrightarrow{K}}^+}}\right.\kern-\nulldelimiterspace}{\psi }_{c,{\overrightarrow{K}}^+}\right\rangle =\left\langle {\psi }_{{\overrightarrow{K}}^+,c}\mathrel{\left|\vphantom{{\psi }_{{\overrightarrow{K}}^+,c} {\hat{p}}_+ {\psi }_{v,{\overrightarrow{K}}^+}}\right.\kern-\nulldelimiterspace}{\hat{p}}_+\mathrel{\left|\vphantom{{\psi }_{{\overrightarrow{K}}^+,c} {\hat{p}}_+ {\psi }_{v,{\overrightarrow{K}}^+}}\right.\kern-\nulldelimiterspace}{\psi }_{v,{\overrightarrow{K}}^+}\right\rangle  = 0$
\end{itemize}

Les deux Hamiltonien effectives s'écrit dans la base $\left\{\left.\left|{\psi }_{v,{\overrightarrow{K}}^+}\right.\right\rangle ,\left.\left|{\psi }_{c,{\overrightarrow{K}}^+}\right.\right\rangle \right\}$ sous la forme~:
\[{\widehat{\mathcal{H}}}^-_{k.p}=\left(\genfrac{}{}{0pt}{}{{\frac{1}{2}E}_{v,{\overrightarrow{K}}^+}\ \ \ \ \ \ \ \ \ \ \ \ \ \ \ \frac{\hslash }{2m_0}q_+{\hat{p}}_{v,c,-}}{0\ \ \ \ \ \ \ \ \ \ \ \ \ \ \ \ \ \ \ \ \ \ \ \ {\frac{1}{2}E}_{c,{\overrightarrow{K}}^+}}\right)\ \ \ \ \ \ ;\ \ \ \ {\ \ \widehat{\mathcal{H}}}^+_{k.p}=\left(\genfrac{}{}{0pt}{}{{\frac{1}{2}E}_{v,{\overrightarrow{K}}^+}\ \ \ \ \ \ \ \ \ \ \ \ \ \ \ \ \ \ \ \ \ \ \ 0\ \ \ \ \ \ \ \ }{\frac{\hslash }{2m_0}q_-{\hat{p}}_{c,v,+}\ \ \ \ \ \ \ \ \ \ \ \ \ \ \ \ {\frac{1}{2}E}_{c,{\overrightarrow{K}}^+}}\right)\] 

Ainsi l'Hamiltonien total de la théorie k.p au point $K^+$ de la zone de Brillouin s'écrit~:
\[{\widehat{\mathcal{H}}}^-_{k.p}=\left(\genfrac{}{}{0pt}{}{E_{v,{\overrightarrow{K}}^+}\ \ \ \ \ \ \ \ \ \ \ \ \ \ \ \ \ \ \frac{\hslash }{2m_0}q_+{\hat{p}}_{v,c,-}}{\frac{\hslash }{2m_0}q_-{\hat{p}}_{c,v,+}\ \ \ \ \ \ \ \ \ \ \ \ \ \ \ \ \ E_{c,{\overrightarrow{K}}^+}}\right)=\left(\genfrac{}{}{0pt}{}{E_{v,{\overrightarrow{K}}^+}\ \ \ \ \ \ \ \ \ \ \ \ \ \hslash q_+v_K}{\hslash q_-v_K\ \ \ \ \ \ \ \ \ \ \ E_{c,{\overrightarrow{K}}^+}}\right)\ \ \ \ \ \ \] 

On obtient finalement~:
\begin{equation}
{\left\langle {\hat{h}}_{\overrightarrow{q}}\right\rangle }_{{\ U}_{{\overrightarrow{K}}^+}}=\left(\frac{{E_{v,{\overrightarrow{K}}^+}+E}_{c,{\overrightarrow{K}}^+}}{2}+\frac{\hslash ^2}{2m_0}q^2\right)+\left(\genfrac{}{}{0pt}{}{-\frac{E_{g,{\overrightarrow{K}}^+}}{2}\ \ \ \ \ \ \ \ \ \ \ \ \ \hslash v_Kq_+\ \ }{\hslash v_Kq_-\ \ \ \ \ \ \ \ \ \ \ \ \ \ \ \frac{E_{g,{\overrightarrow{K}}^+}}{2}}\right)
\end{equation} 
\begin{equation}
E_{c,{\overrightarrow{K}}^+}\left(\overrightarrow{q}\right)=\left(\frac{{E_{v,{\overrightarrow{K}}^+}+E}_{c,{\overrightarrow{K}}^+}}{2}+\frac{\hslash ^2}{2m_0}q^2\right)+\sqrt{{(\frac{E_{g,{\overrightarrow{K}}^+}}{2})}^2+(\hslash v_Kq)^2}
\end{equation}
\begin{equation} 
E_{v,{\overrightarrow{K}}^+}\left(\overrightarrow{q}\right)=\left(\frac{{E_{v,{\overrightarrow{K}}^+}+E}_{c,{\overrightarrow{K}}^+}}{2}+\frac{\hslash ^2}{2m_0}q^2\right)-\sqrt{{(\frac{E_{g,{\overrightarrow{K}}^+}}{2})}^2+(\hslash v_Kq)^2}
\end{equation} 

Ainsi l'Hamiltonien au point $K^-$ de la zone de Brillouin~calculées dans la base $\ \left\{\left.\left|{\psi }_{v,{\overrightarrow{K}}^+}\right.\right\rangle ,\left.\left|{\psi }_{c,{\overrightarrow{K}}^+}\right.\right\rangle \right\}$ est obtenu directement par application de l'opérateur d'inversion de temps sur ${\left\langle {\hat{h}}_{\overrightarrow{q}}\right\rangle }_{{\ U}_{{\overrightarrow{K}}^+}}.$ Rappelons que$\ {\left\langle {\hat{h}}_{\overrightarrow{q}}\right\rangle }_{{\ U}_{{\overrightarrow{K}}^-}}={\left\langle {\hat{h}}_{-\overrightarrow{q}}\right\rangle }^*_{{\ U}_{{\overrightarrow{K}}^+}}$ , ce qui nous donne~:
\[{\left\langle {\hat{h}}_{\overrightarrow{q}}\right\rangle }_{{\ U}_{{\overrightarrow{K}}^-}}=\left(\frac{{E_{v,{\overrightarrow{K}}^-}+E}_{c,{\overrightarrow{K}}^-}}{2}+\frac{\hslash ^2}{2m_0}q^2\right)+\left( \begin{array}{cc}
{\frac{1}{2}E}_{g,{\overrightarrow{K}}^-} & -\hslash v_Kq_-\ \  \\ 
-\hslash v_Kq_+\ \ \  & -{\frac{1}{2}E}_{g,{\overrightarrow{K}}^-} \end{array}
\right)\] 

Les éléments de matrice ${\widehat{\mathcal{H}}}^{K^+}_{k.p}$ calculés au point $K^+$ de la \acrlong{ZB} sont présentés dans le matrice ci-dissous, o\`{u} les éléments diagonaux sont les énergies de bord de bande. Les éléments de matrice au point $K^-$peuvent \^{e}tre obtenus avec les substitutions~:  ${\gamma }_i\longrightarrow {\gamma }^*_i$ et  $q_{\pm }\longrightarrow {-q}_{\mp }$ . \\
Notons que l'Hamiltonien ${\widehat{\mathcal{H}}}^{K^+}_{k.p} \text{est écrite dans la base:}
\lbrace|{\psi }^{A'}_{vb},\ s>,\quad |{\psi }^{E'_1}_{cb},\ s> ,$ \\
$ \quad |{\psi }^{E'_2}_{vb-3},s>,\quad |{\psi }^{E'_2}_{cb+2},s>,\quad |{\psi }^{E^{''}_1}_{vb-2},s>,\quad |{\psi }^{E^{''}_2}_{vb-1},s>,\quad |{\psi }^{A''}_{cb+1},s> \rbrace$ \\

\begin{center}
\includegraphics[scale=0.5]{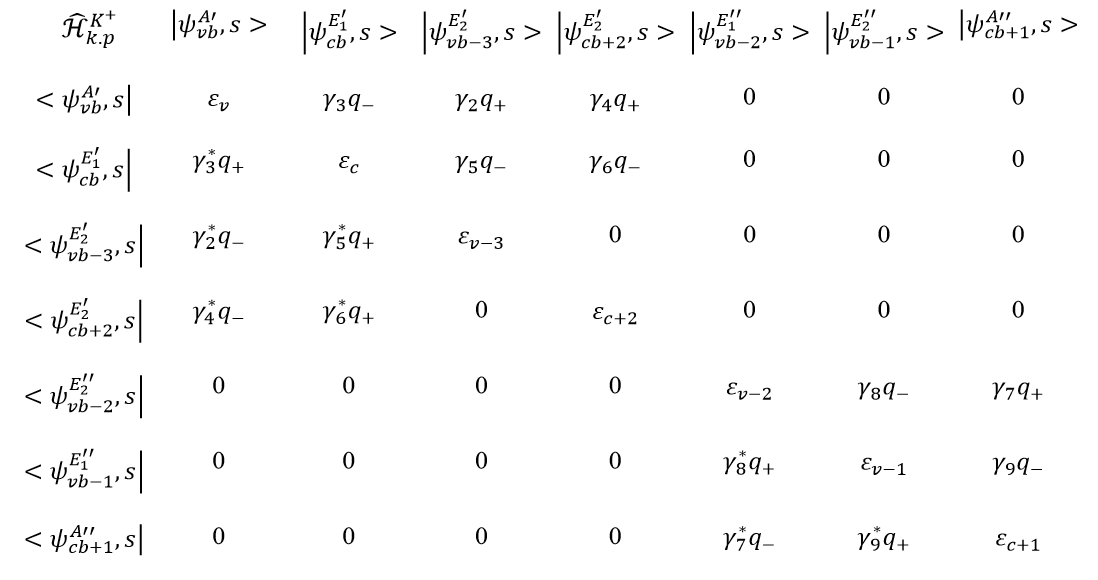} 
\end{center}
$$ \gamma_2 = \frac{\hslash}{2m_e} \left\langle {\psi_{vb}^{A'}| \hat{p}-|\psi_{cb}^{E'_2} }\right\rangle \quad \text{;} \quad \gamma_3 = \frac{\hslash}{2m_e} \left\langle \psi_{vb}^{A'}| \hat{p} + |\psi_{cb}^{E'_1} \right\rangle \quad \text{;} \quad \gamma_4 = \frac{\hslash}{2m_e} \left\langle \psi_{vb}^{A'}| \hat{p} -|\psi_{cb+2}^{E'_2} \right\rangle ;$$
$$\gamma_5 = \frac{\hslash}{2m_e} \left\langle {\psi_{cb}^{E'_1}| \hat{p}+|\psi_{vb-3}^{E'_2} }\right\rangle \quad \text{;} \quad \gamma_6 = \frac{\hslash}{2m_e} \left\langle \psi_{cb}^{E'_1}| \hat{p} + |\psi_{cb+2}^{E'_2} \right\rangle \quad \text{;} \quad \gamma_7 = \frac{\hslash}{2m_e} \left\langle \psi_{vb-2}^{E''_2}| \hat{p} -|\psi_{cb+1}^{A''} \right\rangle ; $$
\[\gamma_8 = \frac{\hslash}{2m_e} \left\langle {\psi_{vb-2}^{E''_1}| \hat{p}+|\psi_{vb-1}^{E''_2} }\right\rangle \quad \text{;} \quad \gamma_9 = \frac{\hslash}{2m_e} \left\langle \psi_{vb-1}^{E''_2}| \hat{p} + |\psi_{cb+1}^{A''} \right\rangle \]
Généralement ces éléments de matrices seront déduits numériquement. 
\section{Interaction spin-orbite }
Nous discutons maintenant la physique à faible énergie autour des points ${K}^{\pm }$. En particulière on veut répondre à la question ; comment la structure électronique de ces matériaux 2D va changer lorsque l'interaction \acrlong{SO} est incluse. En se concentrant sur le fait qu'il existe une différence entre les matériaux $MoX_2$ et $WX_2$ concernant le signe de la constante \acrshort{SO} dans la \acrlong{BC}.

L'interaction \acrshort{SO} est un effet relativiste lié au mouvement des électrons par lequel un électron en mouvement dans un champ électrique voit un champ magnétique effectif  $B_{so}$, qui agit sur son spin. Ce champ est celui qui fait splitté les niveaux d'énergie des atomes, donnant naissance à leur structure fine.

Le \acrlong{CSO} rend le degré de liberté de spin réagir à son environnement orbital, dans les solides cela génère des phénomènes aussi fascinants que le spin-splitting des électrons dans les systèmes d'inversion-asymétrie même à zéro champ magnétique . Cependant, les effets du \acrshort{CSO} ont souvent été négligés conduisant à, par exemple, la conclusion que les masses efficaces des \acrlong{BV} à spin-split sont les mêmes mais des preuves expérimentales récentes montrent que ce n'est pas le cas \cite{Zhang Yi et al}.

Sans \acrshort{SO} les états des électrons de bande de valence sont de type d, alors que si on tient compte de l'interaction spin orbite on obtient des états électroniques avec un moment cinétique total j=5/2 et j=3/2 écarté par un gap $\ {\Delta }_{SO}$ . En outre, la quantification de la taille dans ces systèmes donne lieu à de nombreux phénomènes complètement nouveaux qui n'existent pas dans les semi-conducteurs tridimensionnels. Une compréhension détaillée des phénomènes liés aux effets d'interaction spin-orbite dans les systèmes 2D est importante tant dans la recherche fondamentale que dans les applications des systèmes 2D dans les dispositifs électroniques aussi spintronique.

\subsection{Effet de couplage spin-orbite sur la structure de bande au point $\Gamma$  de la zone de Brillouin }

Le couplage spin-orbite à deux effets principaux sur la structure de bande des semi-conducteurs de TMDCs-bulk au point $\Gamma$ de la \acrlong{ZB} illustrés dans la figure \ref{fig 2.5} ci dissous.
\renewcommand{\labelitemi}{$\bullet$}
\begin{itemize}
  \item 	Conduit à un splitting des états dégénérée ; les représentations irréductibles E' et $E^{''}$ qui sont de dimension 2. (pour le E' de $MoS_2$ le splitting est très petit d'être vu sur l'échelle de figure).
  \item 	Près de point $\Gamma$, il n'y a plus de croisement entre les bandes E' et $E^{''}$.
\end{itemize}

\begin{center}
\includegraphics[scale=0.7]{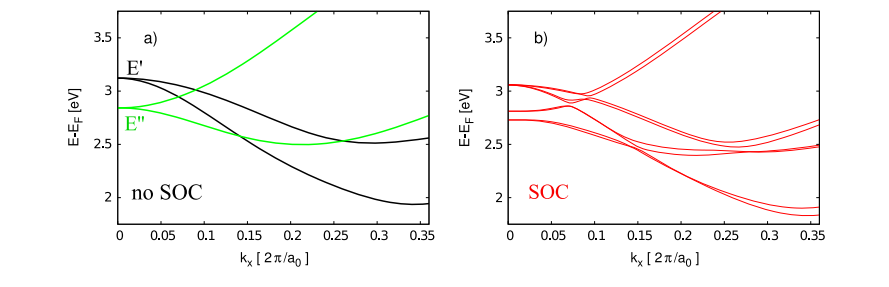}
\captionof{figure}{(a) Les courbes de dispersion des bandes CB et CB+1 tout au long la direction  $\Gamma-K$ calculé par la DFT pour le semi-conducteur  $MoS_2$-bulk, les courbes en noir sont les bandes de symétrie E' alors que les courbes vert sont de symétrie  $E^{''}$. (b) en fait tenir compte de couplage spin-orbite \protect\cite{Andor}.}
\label{fig 2.5}
\end{center}
 
\subsection{Effet de couplage spin-orbite sur la structure de bande dans les deux vallées $K^\pm$}.

L'un des phénomènes qui a d'abord suscité un vif intérêt pour les \acrshort{TMDCs} monocouches était le prononcé de l'effet spin-orbite sur la bande de valence autour des points  $K^+$ et  $K^-$ \cite{Zhu Z Y,Kadantsev E S}. Le terme d'interaction spin-orbite conduit à un splitting et polarisation de spin de la bande de valence, ainsi de la bande de conduction tel que le clivage s'effectue sur une échelle d'énergie de plusieurs centaines de MeV.

En effet le couplage spin orbite affecte également la bande de conduction. Cela a d'abord été négligé, principalement parce que dans $MoS_2$, qui est le plus largement étudié des TMDCs, c'est en effet un petit effet et on a supposé que la situation serait similaire dans les autres. 
\begin{center}
\includegraphics[scale=0.7]{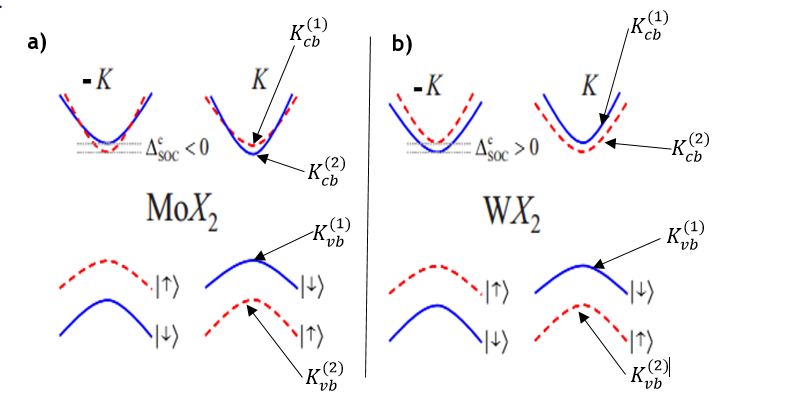}
\captionof{figure}{Schéma montre le splitting de spin de la bande de conduction et de valence dans les vallées $K^\pm$   pour le $MoX_2$ (a) et pour le $WX_2$ (b). Les courbures pointillées en rouge indiquent un état de spin up, alors que les courbures en bleu indiquent un état de spin down. Le spin splitting de la bande de conduction possède un changement de signe entre $MoX_2$ et $WX_2$.Pour $MoX_2$ le spin splitting de la bande de conduction montre un chauvauchement\cite{Gui-Bin Liu}}
\end{center}
On introduit la notation $K_{vb}^{(1)}$ ($K_{vb}^{(2)}$) pour l'énergie supérieure (inférieur) pour le spin splitting de bande de valence et similairement pour la bande de conduction.

Le figure ci-dessus simplifie le comportement de $MoX_2$ et $WX_2$ au niveau de la bande de conduction ainsi de la bande de valence si on tient compte de l'effet spin-orbite dans le calcule de la structure de bande électronique. La bande de conduction de $MoX_2$ montre un spin splitting avec $\left\langle S_z \right\rangle >0$ \quad $(\left\langle S_z \right\rangle >0)$ pour la bande d'énergie supérieur respectivement pour la bande d'énergie inférieur, alors le cas est inversé pour le $WX_2$. De plus la bande avec la masse effectives la plus légère pour $MoX_2$ est la plus faible en énergie, conduisant à un chevauchement des deux bandes de spin-split à proximité des points K et $-K$ alors que pour $WX_2$ la bande la plus légère de spin-split est plus élevée en énergie et donc il n'y a pas de chevauchement de bande\cite{Liu G B,Kormányos A}.

Et comme conséquence de la polarisation du spin des bandes les transitions optiques possibles pour les bandes d'énergie inférieur sont~; $K^{1}_{vb}\to K^{2}_{cb}$ pour MoX2 et $K^{1}_{vb}\to K^{1}_{cb}$ pour WX2.

\begin{center}
\includegraphics[scale=0.6]{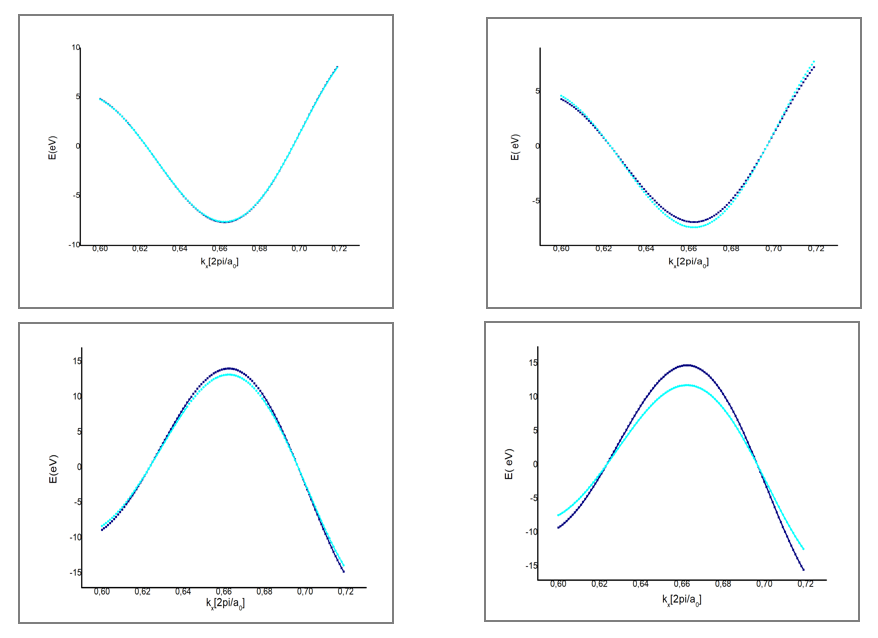}
\captionof{figure}{ Dispersion de la bande de valence ainsi de conduction au voisinage de point K de la ZB.  les résultats(a) et (c) sont pour le $MoS_2$.(b) et (d) pour $WSe_2$.}
\end{center}

\subsection{Hamiltonien à sept bandes}

En sait que le mouvement des électrons dans un solide cristallin est caractérisé par des bandes d'énergie $E_n$(k) avec n l'indice de bande et k le vecteur d'onde, mais on trouve que le couplage SO a un effet très important sur la structure de la bande d'énergie $E_n$(k).

Heureusement l'interaction spin-orbite fait décalé les bandes sans les mélanger, il suffit alors de remplacer le gap par :
\begin{equation}
E^{\uparrow (\downarrow )}_{c,v}=E_{c,v}\pm \frac{1}{2}({\Delta }^{CB}_{SO}-{\Delta }^{VB}_{SO})
\end{equation}
O\`{u} en introduit le spin-orbite splitting de la bande de valence respectivement la bande de conduction   ${\Delta }^{VB}_{SO}$,  ${\Delta }^{CB}_{SO}.$

\begin{center}
\includegraphics[scale=0.8]{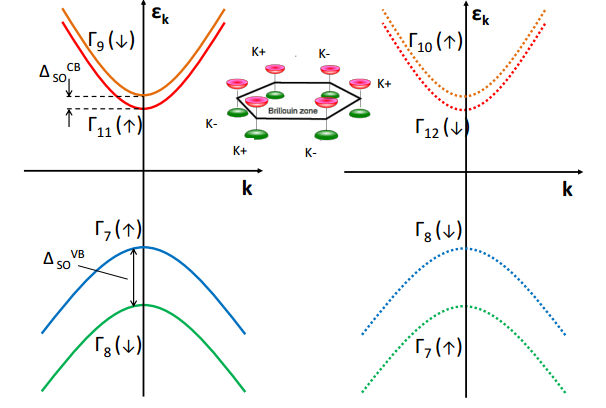}
\captionof{figure}{Illustration schématique de la structure de bande $MoS_2$ monocouche. Les représentations irréductibles sont indiquée sur les bandes ainsi l'orientation de spin électronique. Les bandes montrent un clivage de bande suite au couplage spin-orbite.}
\end{center}
La fonction d'onde de Bloch~:     ${\psi }_{n,\ k}\left(r\right)=\left\langle r\mathrel{\left|\vphantom{r {\psi }_{n,\ k}}\right.\kern-\nulldelimiterspace}{\psi }_{n,\ k}\right\rangle =e^{ik.r}U_{n,k}\left(r\right)\equiv e^{ik.r}\left\langle r\mathrel{\left|\vphantom{r U_{n,\ k}}\right.\kern-\nulldelimiterspace}U_{n,\ k}\right\rangle $      
Si en fait tenir compte de spin la fonction de Bloch s'écrit :
\[|\psi{n,k} > = \sum_{v',\sigma' = \downarrow,\uparrow} C_{n,v',\sigma'} (k) | \psi_{v',\sigma'} > \quad ; | \psi_{v',\sigma'} > =  | \psi_{v',0} > \bigotimes | \sigma' > \quad  ; v' = \left\lbrace cb,vb\right\rbrace \]

Dans le physique atomique l'interaction \acrlong{SO} s'écrit~:
\begin{equation}
H_{SO}=-\frac{\hslash }{4m^2_0c^2}\sigma .p\times \mathrm{\nabla }\mathrm{\ }V(r)
\end{equation} 
Avec, $\hslash $ la constante de planck, $m_0$ la masse de l'électron libre, p  l'opérateur impulsion, V(r) est le potentiel vu par l'électron dû aux atomes de réseaux cristallins, c célérité de lumière et $\sigma$ représente les matrices de Pauli :
\[{\widehat{\sigma }}_x=\left( \begin{array}{c}
\ 0\ \ \ \ \ \ \ \ 1 \\ 
\ 1\ \ \ \ \ \ \ \ 0 \end{array}
\ \right)\ \ \ \ ;\ \ \ \ \ \ \ {\widehat{\sigma }}_y=\left( \begin{array}{c}
\ 0\ \ \ \ \ \ \ -i \\ 
\ i\ \ \ \ \ \ \ \ \ \ \ \ 0 \end{array}
\ \right)     ; \ \ \ \ {\widehat{\sigma }}_z=\left( \begin{array}{c}
\ 1\ \ \ \ \ \ \ \ \ \ \ 0 \\ 
\ 0\ \ \ \ \ \ \ -1 \end{array}
\ \right)\] 
Si l'on considère l'interaction spin-orbite, l'Hamiltonien k.p est alors modifié :
\[\left\{ \begin{array}{c}
\left.\frac{{\left(\overrightarrow{p}+\hslash \overrightarrow{k}\right)}^2}{2m_0}+V\left(r\right)+\frac{1}{4m^2_0c^2}\left(\sigma \times \mathrm{\nabla }\mathrm{\ }V\left(r\right)\right).\left(p+\hslash k\right)\right\}U_{n,k}(r)=E_n(k)U_{n,k}(r) \end{array}
\right.\] 

Qui peut être encore écrit comme~;
\begin{equation}
\left\{ \begin{array}{c} 
\left.\frac{p^2}{2m_0}+V+\frac{\hslash k^2}{2m_0}+\frac{\hslash }{m_0}k.\tilde{p}+\frac{1}{4m^2_0c^2}p.\left(\sigma \times \mathrm{\nabla }\mathrm{\ }V\left(r\right)\right)\right\}\left|U_{n,\ k}>\right.=E_n(k)\left|U_{n,\ k}>\right. \end{array}
\right. 
\label{eqq 2.68}
\end{equation}

En posant que~;
\begin{equation}
\tilde{p}=p+\frac{1}{4m_0c^2}\left(\sigma \times \mathrm{\nabla }\mathrm{\ }V\left(r\right)\right).
\end{equation}
Maintenant, nous multiplions \ref{eqq 2.68} par le bra $\ \left\langle \left.{\psi }_{\upsilon ,\sigma \ }\right|\right.$ .On obtient une équation de valeur propre algébrique pour la dispersion $E_n$(k) qui dépend uniquement du vecteur d'onde k :

$$\sum_{v',\sigma'=\uparrow,\downarrow} \left\lbrace \left[ E_v' (0) + \frac{{\hslash k }^2}{2m_0}\right] \delta _{v,v'} , \delta_{\sigma,\sigma'} + \frac{\hslash}{m_0} k.P^{\sigma,\sigma'}_{v,v'} +\Delta^{\sigma,'}_{v,v'}\right\rbrace C_{n,v',\sigma'} (k) = E_n (k) C_{n,v,\sigma}(k)$$

\begin{equation}
P_{v,v'}^{\sigma,\sigma'} = \left\langle \psi_{v,\sigma} \left|\tilde{p}\right| \psi_{v',\sigma'}\right\rangle  = \left\langle  \psi_{v,\sigma} \left|p + \frac{1}{4m_0 c^2} \left( \sigma \times \nabla V(r)\right) \right| \psi_{v',\sigma'}\right\rangle 
\end{equation}
\begin{equation}
\Delta_{v,v'}^{\sigma,\sigma'} = \frac{1}{4m_0^2 c^2} \left\langle \psi_{v,\sigma} \left|p.(\sigma\times\nabla V(r))\right| \psi_{v',\sigma'}\right\rangle  = \Delta_{v,v'}.\delta_{\sigma,\sigma'}
\end{equation}
Avec,
\begin{equation}
\delta_{\sigma,\sigma'} = \frac{1}{4m_0^2 c^2} \left\langle \psi_{v,} \left|p.\left( \sigma\times\nabla V(r)\right) \right| \psi_{v'}\right\rangle 
\end{equation}
Généralement, les éléments de matrice de l'interaction spin-orbite~ ${\Delta }^{\sigma ,{\sigma }'}_{{\ \upsilon ,\upsilon }'}$ provoque un clivage des niveaux d'énergies dégénérée $E_{\upsilon }\left(k\right)=E_{\upsilon }\left(0\right)+\frac{\hslash k^2}{2m^*_{\upsilon }}$ 
\begin{equation}
\frac{m}{m^*_{\upsilon }}=1+\frac{2}{m}\sum_{{\ \upsilon }'}{\frac{P^2_{{\ \upsilon ,\upsilon }'}}{E_{v\ }(0)-E_{n'}(0)}}
\end{equation} 
Et dans la plupart de cas on peut négliger le couplage spin orbite dans l'équation 2.70 :
\begin{equation}
\tilde{p}=p    ;    P^{\sigma ,{\sigma }'}_{{\ \upsilon ,\upsilon }'}={\delta }_{\sigma ,{\sigma }'}P_{{\ \upsilon ,\upsilon }'}=\left\langle {\psi }_{\ \upsilon ,0\ }\mathrel{\left|\vphantom{{\psi }_{\ \upsilon ,0\ } p {\psi }_{{\ \upsilon }',0}}\right.\kern-\nulldelimiterspace}p\mathrel{\left|\vphantom{{\psi }_{\ \upsilon ,0\ } p {\psi }_{{\ \upsilon }',0}}\right.\kern-\nulldelimiterspace}{\psi }_{{\ \upsilon }',0}\right\rangle \left\langle \sigma \mathrel{\left|\vphantom{\sigma  {\sigma }'}\right.\kern-\nulldelimiterspace}{\sigma }'\right\rangle  
\end{equation}
Notons que l'Hamiltonien $ H_{SO}^K \quad \text{est écrite dans la base:}
\lbrace|{\psi }^{A'}_{vb},\ s>,\quad |{\psi }^{E'_1}_{cb},\ s> ,$ \\
$ \quad |{\psi }^{E'_2}_{vb-3},s>,\quad |{\psi }^{E'_2}_{cb+2},s>,\quad |{\psi }^{E^{''}_1}_{vb-2},s>,\quad |{\psi }^{E^{''}_2}_{vb-1},s>,\quad |{\psi }^{A''}_{cb+1},s> \rbrace$
\begin{center}
\includegraphics[scale=0.6]{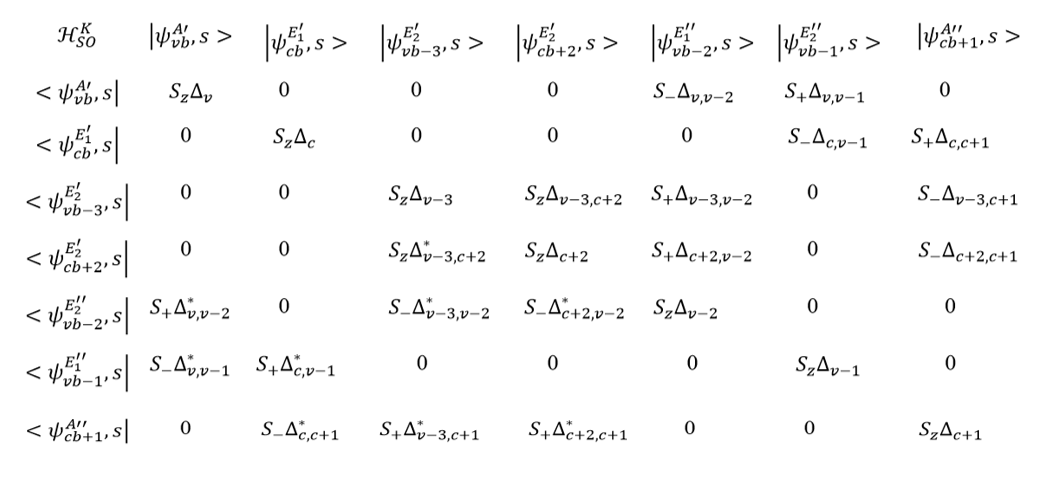}
\end{center}
Si on veut l'hamiltonien \acrshort{SO} au point $K^-$ un nombre des on doit tenir compte :
$$|{\psi }^{E'_2}_{vb-3},s> \Leftrightarrow |{\psi }^{E'_1}_{vb-3},s> \text{,} \quad |{\psi }^{E^{''}_1}_{vb-2},s> \Leftrightarrow |{\psi }^{E^{''}_2}_{vb-2},s> \text{,} \quad $$
$$ |{\psi }^{E'_1}_{cb},\ s> \Leftrightarrow |{\psi }^{E'_2}_{cb},\ s>  \text{,} \quad
|{\psi }^{E^{''}_2}_{vb-1},s> \Leftrightarrow |{\psi }^{E^{''}_1}_{vb-1},s> $$
\[{\Delta }_{{\ \upsilon ,\upsilon }'}\to {\Delta }^*_{{\ \upsilon ,\upsilon }'} , S_{\pm }\to -S_{\mp } , S_z\to -S_z\] 
 
En fait le changement de la notation de symétrie de la fonction d'onde découle de l'hypothèse que les fonctions d'ondes orbitales à $K^+$  et $K^-$ sont connectées par la symétrie d'inversion de temps.

\section*{Conclusion}
Dans ce chapitre nous présentons un hamiltonien k.p à faible énergie qui décrit la dynamique couplée du \Acrlong{BV} et du \acrlong{BC} au voisinage de point K de la \acrlong{ZB} et dont le but d'obtenir un modèle qui capture tous les caractéristiques les plus importantes des courbes de dispersion nous n'arrêtons pas à un modèle effective à deux bande mais plutôt on peut faire un développement d'un modèle à sept bandes où  on peut tenir en compte l'effet de \acrlong{CSO}.

%----------------------------------------------------------------------------------------
%	CHAPITRE 3
%---------------------------------------------------------------------------------------- 

\chapter{Gain optique pour les puits quantiques à base des métaux de transition : MoS2, MoSe2, WS2 et WSe2}  

L'amélioration des techniques expérimentales telle que l'épitaxie par jet moléculaire a permis de réaliser des systèmes de plus basse dimensionnalité qui confinent les porteurs selon plusieurs directions et qui sont aujourd'hui d'un grand intérêt expérimental et théorique, vu leurs applications dans le domaine de l'optoélectronique. Les effets les plus spectaculaires se manifestant dans ces systèmes de basses dimensionnalités sont liés aux propriétés optiques. Le traitement de ces propriétés optiques, nécessite la connaissance du gain optique. Contrairement au cas des matériaux massifs (3D), un gros travail à la fois conceptuel et numérique sera nécessaire pour obtenir le gain optique de ces systèmes à 2D. L'objectif de ce chapitre est de calculer le gain optique pour les puits quantiques à base des métaux de transition qui constituent la zone active d'une diode laser. L'évolution du gain optique en fonction de l'énergie des photons nous permet de vérifier la performance d'une telle zone active. 

\section{Gain optique : Formalisme de la matrice densité}

En utilisant la notation de Dirac, l'opérateur densité $\rho$ est défini par : 
\begin{equation}   
\rho = \sum_n f (E_n) | \Psi_n \rangle\langle \psi_n |
\end{equation}

où $f(E_n)$ représente la probabilité d'occupation de l'état d'un électron $|\Psi_n \rangle$.  On rappel que cette matrice densité permet de calculer la valeur moyenne statistique d'une observable A sur un grand ensemble de système des états $|\Psi_n \rangle$ comme suit:
\begin{equation}
\left\langle A \right\rangle = \sum_{nm} \rho_{nm} A_{nm} = Tr (\rho A)
\label{eq 3.2}
\end{equation}

L'évolution de l'opérateur dans le temps est pratiquement gouvernée par l'équation de Liouville suivante:
\begin{equation}
\frac{\partial \rho}{\partial t} = \frac{1}{i \hslash} [H, \rho]
\end{equation}

où H représente la somme de deux l'hamiltonien agissant sur un électron qui sont $H_0$  l'hamiltonien non perturbé et $H_d$ l'hamiltonien associé au champ électrique  $\vec{E}$ de l'onde et qui sera considéré comme une perturbation :
\begin{equation}
H_d = -\overrightarrow{p}_d .\overrightarrow{E}
\end{equation}

avec $p_d=er$  représente l'opérateur moment dipolaire électrique associé à la particule et e la charge élémentaire. Pour un électron de la bande de conduction dans un état $|\Psi_c\rangle$ d'énergie propre $E^c$ , la matrice densité à une particule est donnée tout simplement par:
\begin{equation}
\rho_{cc}= f \left( E^c - E^c_F\right) 
\end{equation}
où $f$ est le nombre d'occupation de Fermi-Dirac et  $E^c_F$ le quasi-niveau de Fermi de la bande de conduction. De même, pour un trou de la bande de valence sur un état $|\Psi_v\rangle$ et d'énergie propre $E^v$ on aura :
\begin{equation}
\rho_{vv} = f \left( E^v - E^v_F \right) 
\end{equation}
où $E^v_F$  représente le quasi-niveau de Fermi de la bande de valence.\\
D'après la relation \ref{eq 3.2} , la polarisation électrique moyenne par unité de volume est donnée par:   
\begin{equation}
<P>=\frac{1}{V} Tr \left( \rho_{cv} M_{cv}\right) 
\end{equation}
avec $M_{cv}$ la matrice associée au moment dipolaire.
On introduit alors la susceptibilité électrique complexe $\tilde{\chi}_e$ définie par :
\begin{equation}
<P> = \varepsilon_0 \tilde{\chi}_e (\omega) \tilde{E} e^{-i\omega t} + \varepsilon_0 \tilde{\chi}_e (-\omega) \tilde{E}^* e^{i\omega t}
\end{equation}
La susceptibilité électrique complexe sera alors :
\begin{equation}
\tilde{\chi}_e(\omega) = \chi'(\omega)+i \chi^{''} (\omega) = \frac{1}{V} \frac{{|R_{cv}|}^2}{\varepsilon_0}\frac{ \left[ f \left( E^c - E^c_F\right)  - f \left( E^v_F-E^v \right) \right] }{\left( \hslash\omega - E^c - E^v - i\hslash\gamma_{int}\right) }
\end{equation}
où V est le volume de la zone active du matériau.\\
Cette relation peut être généralisée au cas où les transitions impliquent plusieurs états de conduction et de valence. On arrive alors à l'expression générale :
\begin{equation}
\tilde{\chi}_e(\omega) = \frac{1}{V\varepsilon_0} \sum_{cv} {|R_{cv}|}^2 \frac{\left[ f \left( E^c - E^c_F\right)  - f \left( E^v_F-E^v \right) \right]}{\left( \hslash\omega - E^c - E^v - i\hslash\gamma_{int}\right)}
\end{equation}

A partir de la susceptibilité électrique complexe, il est possible de définir plusieurs autres grandeurs fondamentales, telles que le gain intrinsèque G Il vient alors :
$$G^{3D} = \frac{\omega}{nc} Im (\tilde{\chi}_e) = \frac{\omega}{nc\varepsilon_0} \frac{1}{V} \sum_{cv} {|R_{cv}|}^2 \left[ f \left( E^c - E^c_F\right)  - f \left( E^v_F-E^v \right) \right]\times $$
\begin{equation}
\frac{\hslash \gamma_{int}}{\left( \hslash\omega - E^c - E^v\right) ^2 + \left( \hslash \gamma_{int}\right) ^2}
\end{equation}

Dans un puits, il faut tenir compte du fait qu'il s'agit d'un système à 2D. En sommant sur toutes les sous-bandes (m) de valence et de conduction (n), le gain optique s'écrit pour un puits quantique : 
$$G^{2D} (\omega) = \frac{e^2}{m_0^2 n_r c\varepsilon_0 \omega} \frac{1}{L_{eff}} \sum_{nm} J_{cv}^{nm} \int_0^{+\infty} \rho_{DOS}^{2D}{| M_{cv} (K_p) }^2| \times $$
\begin{equation}
 \left[ f \left( E^c_{n,k_p} - E^c_F\right)  - f \left( E^v_F - E^v_{m,k_p} \right) \right] \times 
\frac{{\hslash}/{\tau}}{{\left( \hslash \omega \left( E^c_{n,k_p} -  E^v_{m,k_p} \right) - E \right) }^2 + {\left( {\hslash}/{\tau} \right)}^2  } dE
\end{equation}

où $\omega$ est la fréquence du photon incident, c et $\varepsilon_0$  sont respectivement la vitesse de la lumière et la permittivité du vide. $n_r,L_{eff},J^{nm}_{cv}$  sont respectivement l'indice de réfraction, la largeur effective de la zone active et le recouvrement des fonctions d'onde.  $\tau$ est le temps de relaxation intra-bande dont la valeur est de l'ordre de $10^{-3}$ s . $\rho^{2D}_{DOS}$ est la densité d'état par unité de surface d'un système d'électrons à 2D donnée par :  
$\rho^{2D}_{DOS} (E) = (m^* /\pi \hslash^2 )$.\\

Notons ici que, si le système a plus d'un état quantique, la densité totale des états à 2D peut être écrite sous la forme $\rho^{2D}_{DOS} (E) = \frac{m^*}{\pi \hslash^2} \sum_n \Theta (E-E_n)$

où $E_n$ et  représente l'énergie des états quantifiés et $\Theta (E-E_n)$ est la fonction échelon de Heaviside.
\begin{center}
\includegraphics[scale=0.5]{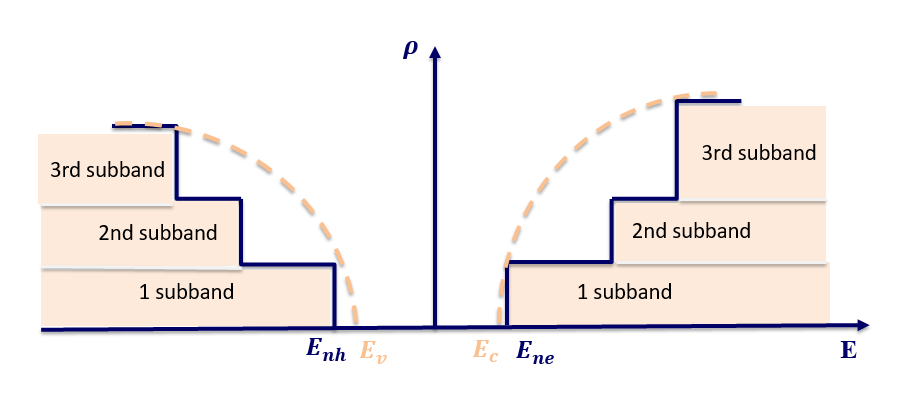}
\captionof{figure}{Schéma du changement de la densité d'état massif suite à un confinement quantique.}
\end{center}

Il est important de signaler que l'utilisation de la distribution de Lorentz qui relaxe la loi de conservation de l'énergie dans la formule du gain induit deux artefacts mathématiques à savoir :
\renewcommand{\labelitemi}{$\bullet$}
\begin{itemize}
\item
une absorption artificielle sous le gap.
\item
une sous-estimation de la condition de Bernard et Duraffourg concernant l'inversion de population: $E^c_F - E^v_F > E^c_n - E^v_m = \hslash \omega$.
\end{itemize}

\section{Semi-conducteurs à basse dimension et confinement quantique}
\subsection{Définition et réalisations}

La première réalisation de confinement quantique a été effectuée par \textit{Esaki} en \textbf{1970} dans le cadre de ses travaux pionniers sur l'effet tunnel. Ainsi le développement des techniques d'épitaxie par jets moléculaires ce qui montre le figure \ref{fig 3.1} ci dissous ou par dépôt chimique de composés organométalliques en phase vapeur (\acrshort{MOCVD}) \acrlong{MOCVD} au milieu des années 70 il devient possible de synthétiser des empilements de différents matériaux semiconducteurs en contrôlant l'épaisseur des couches déposées avec une précision voisine de la monocouche atomique \cite{François H.François H.}. En jouant sur la composition des semi-conducteurs lors de l'épitaxie, un puits quantique artificiel peut être crée dans la bande de conduction ou de valence où les porteurs de charge n'auront pas la même énergie potentielle, et ils sont confinées par une barrière de potentiel à l'interface et dont l'épaisseur est de l'ordre nanométrique (comprises entre 10 et 300 A${}^\circ$ environ $\lambda_D$: longueur d'onde de De Broglie), mais ils restent libres de se déplacer dans le plan des couches. Cette faible dimension à des conséquences sur la quantification du mouvement électronique tel que de sous-bandes d'énergie faisant apparaitre qui modifient radicalement les propriétés optiques et électriques du matériau semi-conducteur synthétisé. Ceci entraine l'observation de propriétés inhabituelles qui nous permet de renforcer la performance des optoélectroniques.

L'énergie totale du porteur s'exprime alors comme la somme de son énergie de confinement (quantifiée) suivant la direction de croissance et de son énergie cinétique (continue) dans le plan des couches. Si le puits quantique sera orienté de fa\c{c}on à ce que le confinement soit dans la direction \textit{z }et que les interfaces soient parallèles au plan de coordonnées (\textit{xy)}. L'énergie d'un porteur dans une bande $b=\left\{c,v\right\}$ donnée peut s'écrire sous la forme~:

$E_n+(\hslash k_{\parallel })^2/2m^*_b$ \quad o\`{u} \quad $k_{\parallel }=k_x+k_y$ \quad le vecteur d'onde des porteurs dans la direction (\textit{xy}) et l'énergie $E_n$ est celle de la particule confinée de masse $m^*_b$ .
\begin{center}
\includegraphics[scale=0.4]{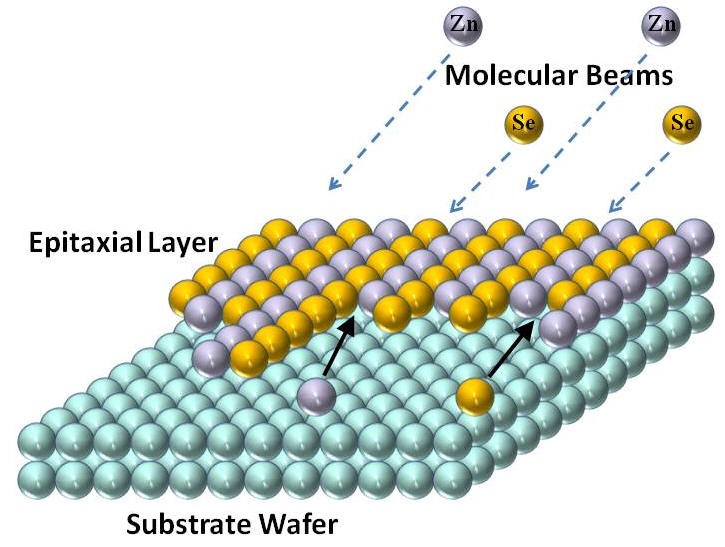}
\captionof{figure}{Principe de technique MBE sur le TMDC $WSe_2$}
\label{fig 3.1}
\end{center}
\subsection{Effet de confinement sur la structure de bande}

Le confinement restreigne la structure de bande aux valeurs $\left|\overrightarrow{k}\right|>\frac{\pi }{a}$ .  La bande d'énergie interdite $E_g$ voit sa valeur augmenter ce qui influe directement sur les propriétés optiques du matériau. Plus celui-ci est confiné plus, plus l'énergie de bord de bande est donc élevée. C'est l'intérêt majeur des semi-conducteurs à basse dimension. 
\begin{center}
\includegraphics[scale=0.5]{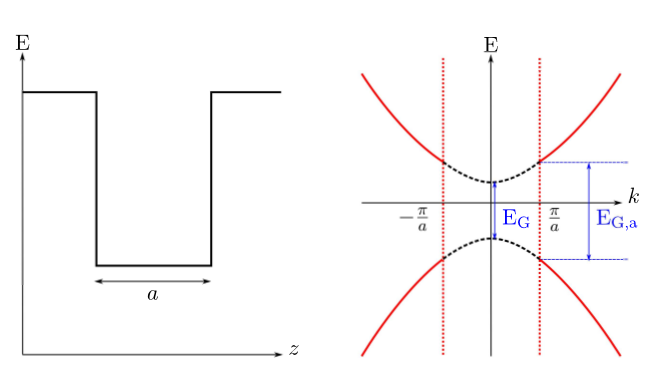}
\captionof{figure}{Représentation schématique de diagramme énergétique d'un puits quantique d'épaisseur a, et l'effet de confinement sur la structure de bande d'un semi-conducteur}
\end{center}

La température infule de même sur la largeur de la bande interdite comme le montre le figure ci dissous pour les TMDCs MoS2 et WSe2.

\begin{minipage}{\textwidth} 
\begin{minipage}{0.5\textwidth}
  \centering
  \includegraphics[scale=0.28]{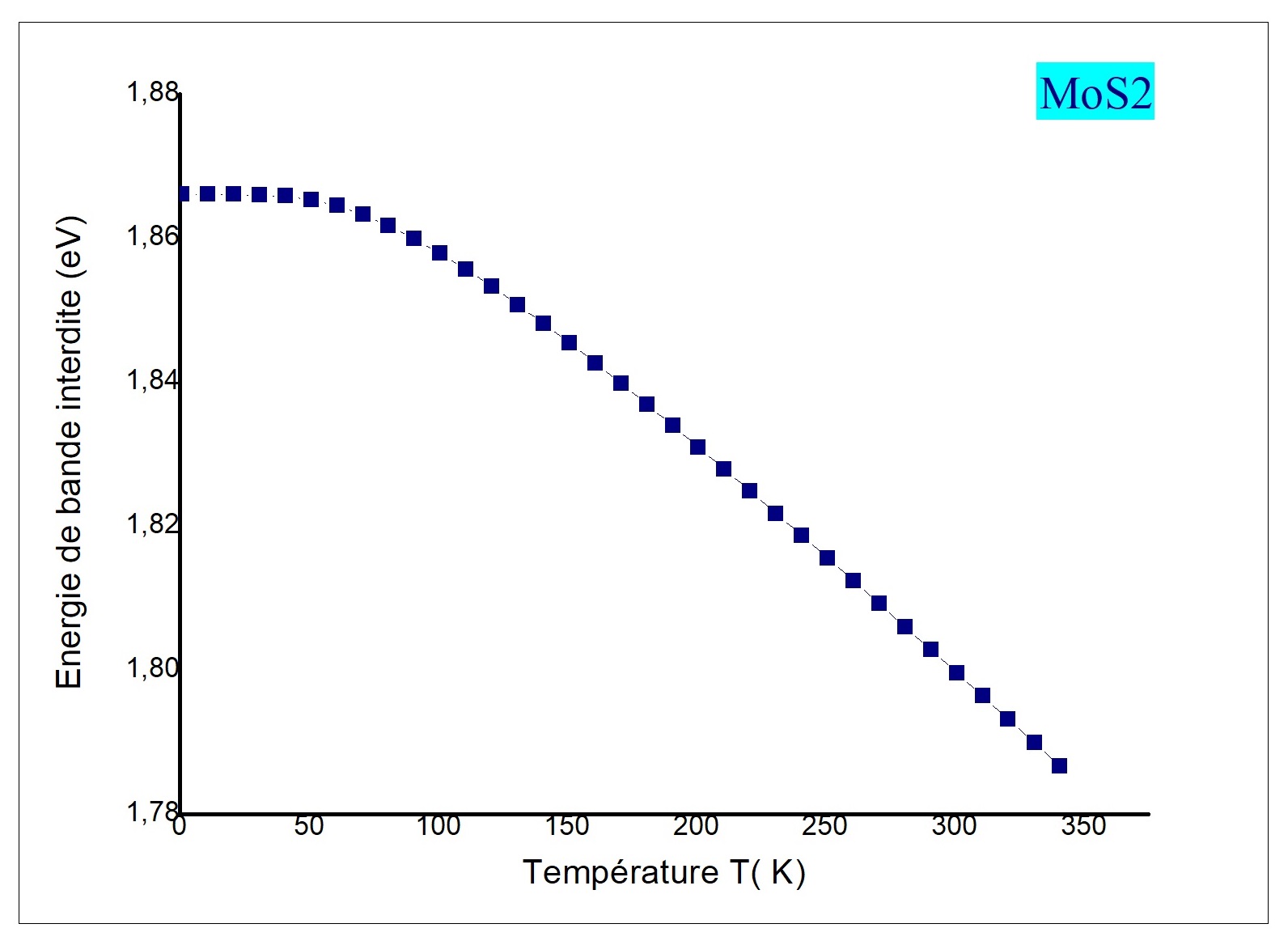}
\end{minipage}%
\begin{minipage}{0.5\textwidth}
  \centering 
  \includegraphics[scale=0.28]{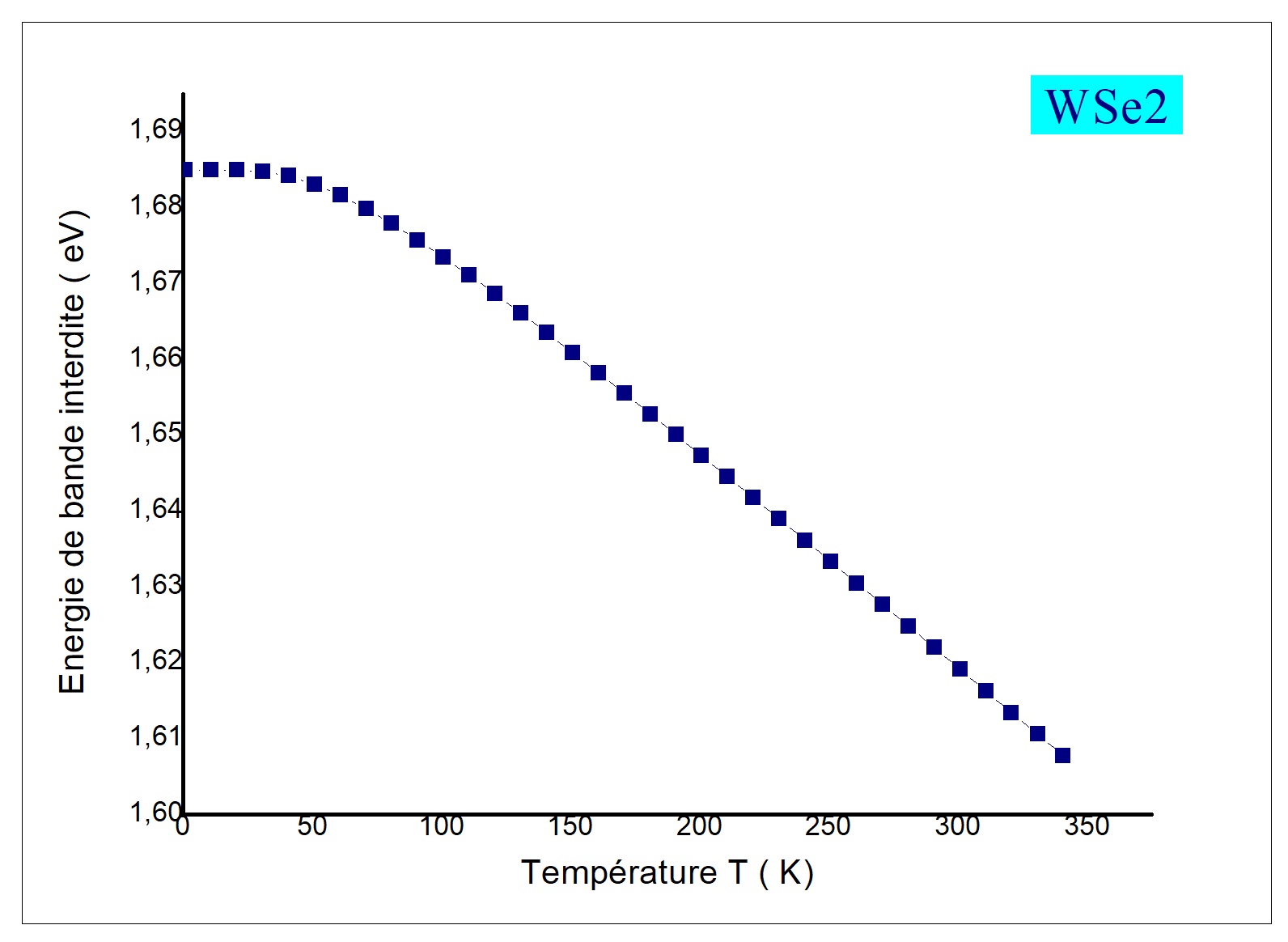}
\end{minipage} 
\captionof{figure}{Variation de gap en fonction de la température pour les semi-conducteurs MoS2 et WSe2.} 
\end{minipage}

\subsection{Stabilité des hétérostructures }

La stabilité des hétérostructures nécessite que les deux matériaux hétéro-épitaxié aient la même structure cristalline, des paramètres de mailles voisins et des coefficients de dilatation proches puisque l'épitaxie se réalise à des températures élevées, sinon le matériau constituant la couche de plus grand épaisseur impose sa maille à l'autre induisant un effet de pression sur le paramètre de maille faible. Ceci entraine l'existence d'une contrainte biaxial (pression hydrostatique) dans le plan des couches de matériau de faible épaisseur qui peuvent modifier la largeur de la bande interdite et lever la dégénérescence des bandes.

Le désaccord de maille est défini par le paramètre sans dimension $\varepsilon$.
\begin{equation} 
\varepsilon =\frac{a_{barrière} -a_{puits} }{a_{puits} }  
\end{equation} 

où $a_{barriére}$, $a_{puits}$ représente respectivement le paramètre de matériaux barrière et puits.
Le signe de $\varepsilon$ défini l'état de la couche déposé par rapport au substrat, soit en compression pour une valeur négatif de $\varepsilon$ soit en tension si $\varepsilon$ est positif.

La figure \ref{fig 3.3} ci dissous décrit l'état d'une 1L-TMDCs dans une hétérostructure type $MoS_2$/$MX_2$ tel que la monocouche $WS_2$ est en tension lorsqu'il est déposé sur $MoS_2$ alors que pour les restes monocouche des TMDCs sont dans un état de compression.
\begin{center}
\includegraphics[scale=0.6]{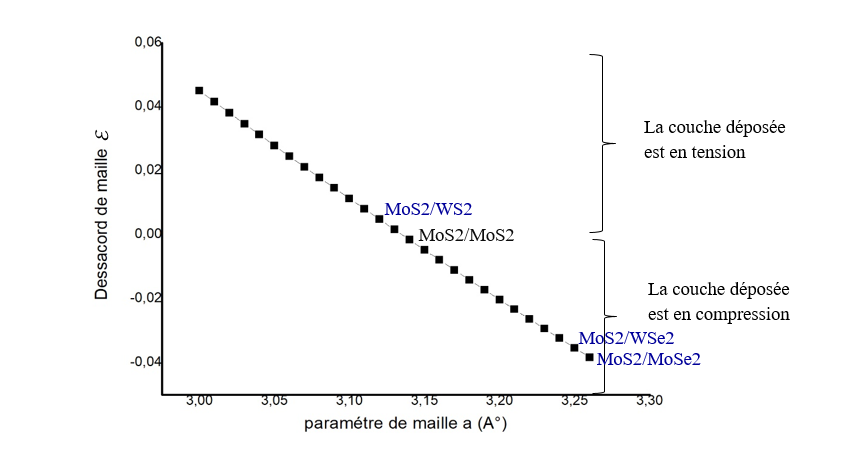}
\captionof{figure}{Courbe qui représente le désaccord de maille dans les hétérostructures $MoS_2$/$MX_2$.}
\label{fig 3.3}
\end{center}
\section{Conditions de formation d'un puits quantique }
\subsection{Diagrammes de bandes d'énergies }

Lorsque les SC (1) et (2) ne sont pas en contacte on notera l’apparition de deux paramètres importante un décalage dans la bande de conduction et un autre dans la bande de valence qui sont défini par :
\begin{equation} 
\label{eq:3.2} 
{\Delta E}_{c0}=E_{c2}-E_{c1}={\chi }_1-{\chi }_2\ \ \ \ \ \ \ \ ;\ \ \ \ \ \ {\Delta E}_{v0}=E_{v2}-E_{v1} 
\end{equation} 

\begin{center}
\includegraphics[scale=0.7]{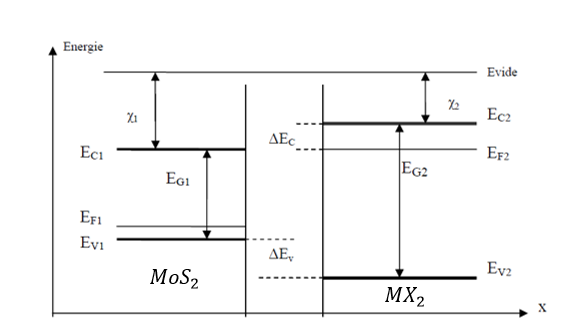}
\captionof{figure}{Les diagrammes énergétiques du chacun de semi-conducteurs en l'absence de tout contact \protect\cite{Xiangying Su}.}
\end{center}

Lorsque les deux semi-conducteurs sont mis en contact, ils échangent des électrons de manière à aligner leurs niveaux de Fermi $E_{F1}$=$E_{F2}$. Le retour à l'équilibre implique l'apparition d'une ddp interne $V_d$(la tension de diffusion) provenant de l'échange de charge entre les deux SC dans une zone d'échange au voisinage de la jonction apparaissent dans chacun des SC de part et d'autres provoquant des courbures des bandes d'énergies.

\begin{center}
\includegraphics[scale=0.6]{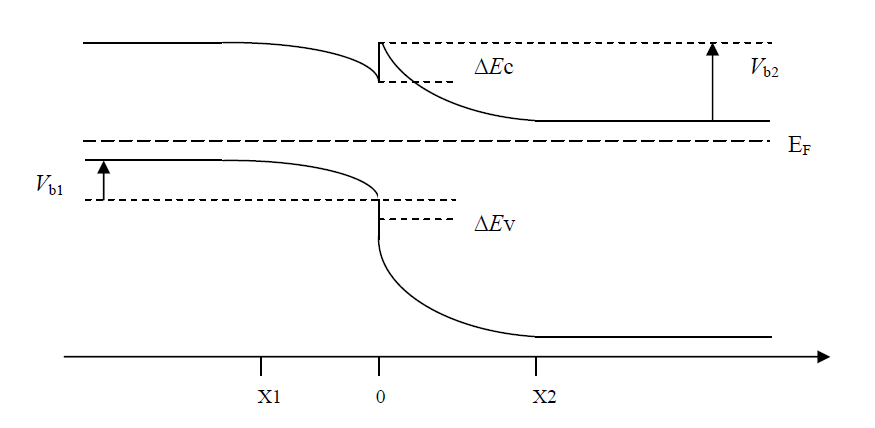}
\captionof{figure}{Le diagramme énergétique de l'hétérojonction lorsque le contacte est établit \protect\cite{Xiangying Su}.}
\end{center}
\subsection{Configuration de bandes et confinement de porteurs }

Il existe plusieurs types de contacts semi-conducteur/semi-conducteur, schématisés en figure \ref{fig 3.6} ainsi l'utilisation de ces types d'hétérojonctions tout dépend de façon dont on souhaite profiter de l'existence des barrières et de puits de potentiel.
\begin{center}
\includegraphics[scale=0.7]{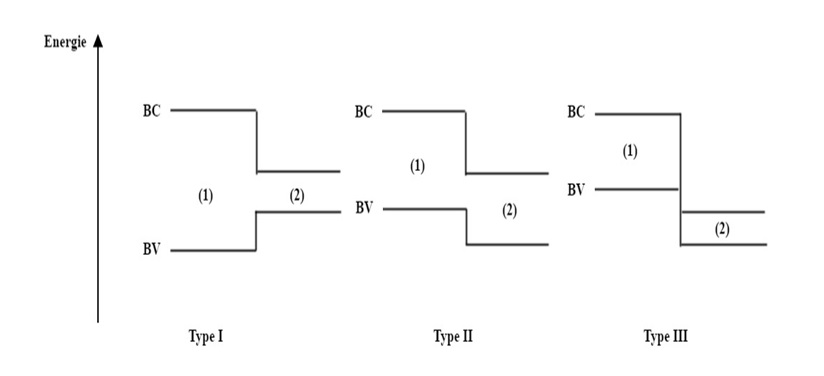}
\captionof{figure}{Représentation schématique des différents types d'alignement possibles pour une structure quantique.}
\label{fig 3.6}
\end{center}

Dans le type I, le gap le plus faible se positionne « à l'intérieur » du gap le plus large : $E_{c1}<E_{c2}\ $ et $E_{v1}>E_{v2},$ si$\ \ E_{g2}>E_{g1}.$ Dans ce cas le gap de matériau (1) est inclus dans le gap du matériau (2), les offsets de bande pour la bande de conduction et de valence agissent comme des barrières de potentiel gardent les électrons et les trous confinés dans le matériau de gap inférieur (1). 

Dans le type II, les énergies de bande interdites des matériaux sont décalé les uns par rapport à l'autre c. à d : si   $E_{c1}<E_{c2}$ et $E_{v1}<E_{v2}\ \ $   CB1 est incluse dans le gap de SC2 se qui conduit les électrons à s'accumuler dans le SC1 et les trous dans le SC2.

L'hétérojonction type III : est un cas particulier de type II, où les énergies de bande interdite des deux matériaux sont totalement décalées les unes par rapport aux autres.

\section{Les hétérostructures MoS2/MX2}

La couche mince le plus étudié a été le graphène depuis son isolement en 2004 alors que le graphène n'est pas adapté à la construction des puits quantique en raison de son gap nul.
Cependant les matériaux \acrshort{TMDCs} représente un bon candidat et comme l'effet de confinement quantique dans le $MoS_2$ a été montré qu'il est efficace dans l'amélioration de performance des propriétés optoélectronique \cite{A. Kuc,K.F.Mak}, en limite le confinement quantique dans le $MoS_2$ alors que les $MX_2$ : $\{$$WS_2$, $WSe_2$ $MoSe_2$$\}$ agit comme une barrière de potentiel.  
\begin{center}
\includegraphics[scale=0.45]{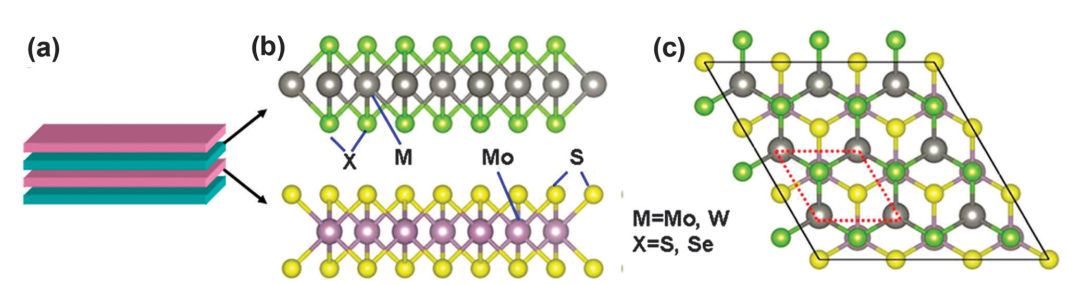}
\captionof{figure}{(a) Empilement des monocouches des \acrshort{TMDCs} formant un super-réseau. (b) vue de dessus d'un super-réseau $MoS_2$/ $MX_2$  où l'atome Mo est juste au-dessus du atome X .(c)vue de dessus de plan des couches \protect\cite{Xiangying Su}.}
\end{center}

L'empilement d'un monocouches de matériaux SC-$MoS_2$ d'énergie de gap faible deux régions d'énergie de gap plus élevée d'un SC-$MX_2$ provoque un changement de structure de bandes à l'interface. Ce changement conduit à l'apparition de discontinuités de bandes de conduction ou/et de valence (offsets), c'est-à-dire des sauts de potentiel.

On se propose alors de déterminer les énergies de confinement dans un puits quantique de hauteur   \acrlong{BCO} (\acrlong{BVO}) de largeur L se forme dans la bande de conduction(valence) dans un hétérostructure type $MoS_2$ / $MX_2$ avec $MX_2$ : $\left\lbrace {WS_2, WSe_2, MoSe_2}\right\rbrace $. 

Pour la détermination de l'énergie de quantification dans le puits quantique de potentiel fini d'une structure à base de $MoS_2$/$MX_2$ le formalisme le plus simple est l'utilisation de l'approximation de la fonction enveloppe.

\subsection{Hétérostructures et quantification d'énergie }
Pour simplifier les études théoriques, on adopte un schéma plat des bandes. 
\begin{center}
\includegraphics[scale=0.5]{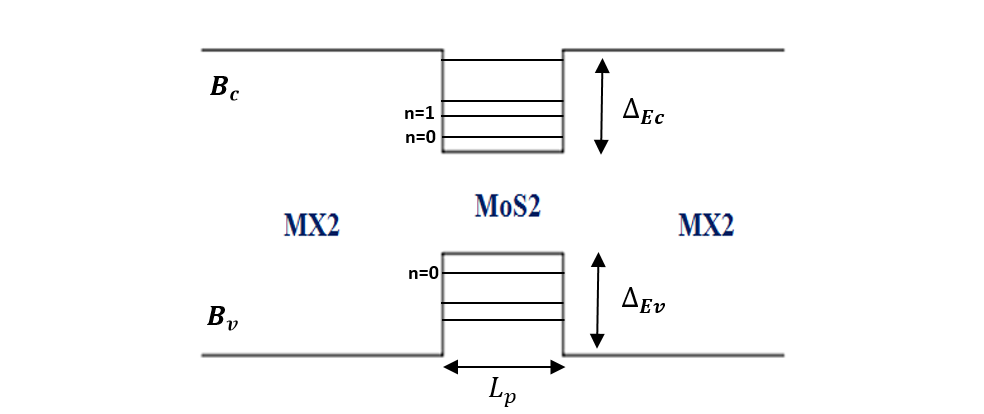}
\captionof{figure}{Schéma illustratif montre la formation d'un puits quantique de largeur $L_p$.}
\end{center}
Le puits formé fait confiner les particules dans une région de l'espace tel que leurs probabilités de les trouver à l'infini devient nulle à tout moment, la fonction d'onde est donc normalisable et les valeurs de leur énergie sont quantifiées, le spectre est discontinu et on dit que les particules se trouvent dans des états liés. Tandis qu'à l'extérieur du puits l'énergie de la bande de valence et de la bande de conduction est un continuum vu que les électrons et les trous sont libres de se déplacer, la fonction d'onde n'est plus normalisable et le spectre en énergie est continu et on dit que les particules se trouvent dans des états non liés.

\subsection{Energie de confinement dans un puits quantique}

Notons que le puits quantique sera orienté de façon à ce que le confinement soit dans la direction z et que les interfaces soient parallèles au plan de coordonnées (xy).

Un terme s'ajoute à l'énergie de bande qui n'est autre l'énergie de celle de la particule de masse $m_{bw}^*$ confinée entre les deux couche $MX_2$ séparé par la distance a. L'énergie totale est donnée par :
\[E_{bn}+\frac{(\hslash k_b)^2}{2m_{bw}} \quad \text{;} b= \left\lbrace c,v\right\rbrace \] 
O\`{u}  $m_{bw}$ la masse effective de porteur de charge pour une bande b donnée à l'intérieur de puits.
La bande se décompose en sous bandes qui correspondant aux différentes valeurs du nombre quantique n. Notons qu'il y'a au moins un niveau d'énergie confiné quel que soit la largeur et la hauteur de barrière de potentiel.

Dont le but de trouver ces niveaux d'énergies on limite notre étude au bande de conduction, on va s'intéresser à un confinement dans un potentiel carré fini pour une hétérostructure de type I.
\begin{center}
\includegraphics[scale=0.8]{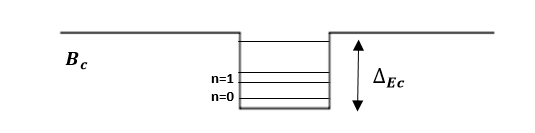}
\captionof{figure}{Les niveaux de confinement pour la bande de conduction.}
\end{center}
En résolvant l'équation aux valeurs propres de Schrödinger donné par :
\begin{equation} 
\label{eq:3.3} 
\left\{-\frac{{\hslash }^2}{2m_*(z)}\frac{{\partial }^2}{\partial z^2}+V\left(z\right)\right\}{\psi }_n(z)=E_n(z){\psi }_n(z) 
\end{equation} 
$m_*:\ $Masse effective de l'électron et $V\left(z\right)$ l'énergie potentielle. \\
On écrit alors l'équation de Schr\"{o}dinger dans les trois régions ou agit le potentiel~:
\begin{equation} 
\label{eq:3.4} 
V\left(z\right)=\left\{ \begin{array}{c}
V_0\ \ \ \ \ \ \ \ \ \ \ \ \ \ \ \ \ \ \ \ z<-a \\ 
\ \ 0\ \ \ \ \ \ \ \ \ \ \ \ -a<z<a \\ 
V_0\ \ \ \ \ \ \ \ \ \ \ \ \ \ \ \ \ \ \ \ \ \ z>a \end{array}
\right. 
\end{equation} 
Rappelons que les conditions aux limites qui permettent de résoudre complètement l'équation différentielle sont les suivantes : 

\begin{itemize}
\item  Probabilité de trouver la particule à l'infini est nulle~: \begin{equation}{lim}_{z\to \infty }\left|\psi \left(z\right)\right|=0 \end{equation}     
\item  Continuité à l'interface pour~: \begin{equation}\psi \left(z\right)\ \ \mathrm{\ et\ }\ \frac{1}{m_*}\frac{\partial {\psi }_n(z)}{\partial z}\ \ \end{equation}                                          
\end{itemize}
\underbar{\textbf{A l'extérieur du puits}~}:\textbf{  }$z>\left|a\right|$

L'équation de Schr\"{o}dinger s'écrit~:
\begin{equation} 
\label{eq:3.7} 
\frac{{\partial }^2}{\partial z^2}{\psi }_n(z)-{\rho }^2{\psi }_n(z)=0\ \ \ \ \ \ \ \ ;\ \ \ \ \ \ \ {\rho }^2=\frac{2m_{cb}}{{\hslash }^2}(V_0-E_n(z)) 
\end{equation} 

$m_{cb}:\ $La masse effective des électrons dans la barrière de potentiel.

Les solutions sont~:
\begin{equation}
Région~1~: {\psi }_1(z)=A_1e^{\rho z}+{A'}_1e^{-\rho z}
\end{equation}
\begin{equation}
Région~3~: {\psi }_3(z)=A_2e^{\rho z}+{A'}_2e^{-\rho z}                                       
\end{equation}
 Comme ${\psi }_n(z)$ doit \^{e}tre bornée dans les régions (1) et (2) , forcement~: ${A'}_1=A_2=0.$

\underbar{\textbf{A l'intérieur du puits}~}:  $z<\left|a\right|$.

L'équation de Schr\"{o}dinger est~:
\begin{equation} 
\frac{\partial ^2{\psi }_n(z)}{\partial z^2}+k^2{\psi }_n(z)=0\ \ \ \ \ \ ;\ \ \ \ \ \ k^2=\frac{2m_{cw}}{{\hslash }^2}E_n(z) 
\end{equation} 

$m_{cw}:\ $La masse effective des électrons dans le puits.\\
La solution s'écrit alors comme suit~:
\begin{equation}
Région~2~: {\psi }_n(z)=B_1e^{ikz}+{B'}_1e^{-ikz}
\end{equation}
Vu que la potentiel est pair, les fonctions d'ondes à l'intérieur du puits sont soit paires (symétriques) soit impaires (antisymétriques), c'est-à-dire qu'on a~:
\begin{equation}
{\psi }^P_2\left(z\right)=Bcos(kz) \quad  \text{et} \quad {\psi }^I_2\left(x\right)=Bsin(kz)                          
\end{equation}

Le problème nous donnant deux ensembles de solutions, soit paire soit impaire et qui sont données respectivement par~:\\
\begin{equation}
Etats~paires~:\left\{ \begin{array}{c}
{\psi }_1\left(z\right)=A_1e^{\rho z}\ \ \ \ \ \ \  \\ 
{\psi }^P_2\left(z\right)=Bcos(kz) \\ 
{\psi }_3\left(z\right)={A'}_2e^{-\rho z}\ \ \ \  \end{array}
\right.
\end{equation}

\begin{equation}
Etas~impaires~: \left\{ \begin{array}{c}
{\psi }_1\left(z\right)=A_1e^{\rho z}\ \ \ \ \ \ \  \\ 
{\psi }^I_2\left(z\right)=Bsin(kz) \\ 
{\psi }_3\left(z\right)={A'}_2e^{-\rho z}\ \ \ \  \end{array}
\right.        
\end{equation}

Afin de déterminer les relations donnant la quantification de l'énergie de la particule, on va imposer aux fonctions d'ondes et de ses dérivées premiers d'\^{e}tre continus~au point \textit{a} et \textit{-a }:

\begin{equation} 
\left\{ \begin{array}{c}
{\psi }_1\left(-a\right)={\psi }_2\left(-a\right)\  \\ 
{\psi '}_1\left(-a\right)=\frac{m_{cb}}{m_{cw}}{\psi '}_2\left(-a\right)\  \\ 
\ \ \ \ \  \end{array}
\ \ \ \ ;\ \ \ \left\{ \begin{array}{c}
{\psi }_2\left(a\right)={\psi }_3\left(a\right)\  \\ 
{\psi '}_2\left(a\right)=\frac{m_{cw}}{m_{cb}}{\psi '}_3\left(a\right)\  \\ 
\ \ \ \ \ \  \end{array}
\right.\right. 
\end{equation}
 
Ainsi, ces deux conditions aux limites conduisant pour les deux ensembles (paires et impairs) à  deux conditions de quantifications~:
  
\begin{equation} 
tg\left(ka\right)=\frac{\rho }{k} .\frac{m_{cw} }{m_{cb} } =\frac{m_{cw} }{m_{cb} } .\sqrt{\frac{m_{cb} \left(V_{0} -E_{n} \left(z\right)\right)}{m_{cw} E_{n} } }  
\label{eq 3.16}
\end{equation} 

et,

\begin{equation} 
cotg\left(ka\right)=-\frac{\rho }{k} .\frac{m_{cw} }{m_{cb} } =-\frac{m_{cw} }{m_{cb} } .\sqrt{\frac{m_{cb} \left(V_{0} -E_{n} \left(z\right)\right)}{m_{cw} E_{n} } } 
\label{eq 3.17} 
\end{equation} 

Pour les états pairs nous pouvons écrire~:
\begin{equation} 
f(E_{n} )=tg\left(\sqrt{a^{2} .\frac{2m_{cw} }{\hbar ^{2} } E_{n} (z)} \right)-\frac{m_{cw} }{m_{cb} } .\sqrt{\frac{(m_{cb} (V_{0} -E_{n} (z))}{m_{cw} E_{n} (z)} } =0 
\label{eq 3.18}
\end{equation} 

Pour les états impairs~:

\begin{equation} 
f(E_{n} )=cotg\left(\sqrt{a^{2} .\frac{2m_{cw} }{\hbar ^{2} } E_{n} (z)} \right)+\frac{m_{cw} }{m_{cb} } .\sqrt{\frac{(m_{cb} (V_{0} -E_{n} (z))}{m_{cw} E_{n} (z)} } =0 
\label{eq:3.19} 
\end{equation} 
Notons que les deux équations \ref{eq 3.16} et \ref{eq 3.17} peut prendre un forme plus simple dans le cas ou $m_{cw}=m_{cb}=m_0:$

\begin{equation} 
\left\{\begin{array}{c} {tg} \\ {-cotg} \end{array}\right\}\left(\sqrt{\frac{a^2}{E_{0} } E_{n} \left(z\right)} \right)=\sqrt{\frac{\left(V_{0} -E_{n} \left(z\right)\right)}{E_{n} \left(z\right)} }  
\end{equation} 

Posons que~;$\ E_0=\frac{{\hslash }^2}{2m_0[A{}^\circ ]^2}\approx 3.78\ eV$\\
On sort de général et on essaye d'écrire les équations \ref{eq 3.18} et \ref{eq:3.19} pour notre système o\`{u} $V_0$ représente$\ \Delta E_c\ $et $\Delta E_v$ respectivement les bandes offset de conduction et de valence~et \textit{a} la largeur de puits noté \textit{W }:

\begin{equation} 
\left\{\begin{array}{l} {tg} \\ {-cotg} \end{array}\right\}\left(\sqrt{W^{2} .\frac{2m_{cw} }{\hbar ^{2} } E_{c,n} (z)} \right)=\sqrt{\frac{m_{cw} }{m_{cb} } .\frac{(\Delta E_{c} -E_{c,n} (z))}{E_{c,n} (z)} } 
\label{eq 3.21} 
\end{equation} 

et,
\begin{equation} 
\left\{\begin{array}{l} {tg} \\ {-cotg} \end{array}\right\}\left(\sqrt{W^{2} .\frac{2m_{vw} }{\hbar ^{2} } E_{v,n} (z)} \right)=\sqrt{\frac{m_{vw} }{m_{vb} } .\frac{(\Delta E_{v} -E_{v,n} (z))}{E_{v,n} (z)} } 
\label{eq 3.22} 
\end{equation} 

O\`{u}  $m_{cb}\ $ est la masse effective des électrons de conduction dans la barrière de potentiel, $m_{cw}\ $est la masse effective des électrons de conduction dans le puits $m_{vb}\ $est la masse effective des trous dans la barrière de potentiel, $m_{v,w}$ est la masse effective des trous dans le puits. $\Delta E_c$ représente la différence entre l'énergie de conduction des barrières et celle du puits ou ce qu'on appelle la bande de conduction offset ainsi  $\Delta E_v\ $est la  différence entre l'énergie de valence des barrières et celle du puits ou la bande de valence offset.  Graphiquement, il est possible de résoudre ces équations implicites en \textit{E.}

Une comparaison des valeurs offset de bande conduction ainsi de valence pour les différentes TMDCs nous permet de limité notre problème de simulation seulement à l'héterostructure $MoS_2$/$WSe_2$ vu que ce dernière possède la plus grande valeur \acrlong{BVO} et \acrlong{BCO} par rapport autres type d'héterostructures.
\begin{center}
\includegraphics[scale=0.55]{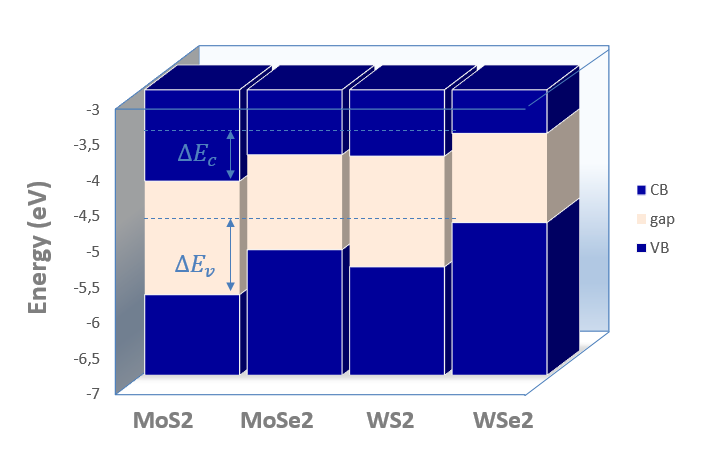}
\captionof{figure}{L'alignement de bande pour une monocouche des matériaux TMDCs monocouche.}
\end{center}

\begin{center}
{\renewcommand{\arraystretch}{1.5} %donne la distance entre les lignes%
{\setlength{\tabcolsep}{0.55cm} %donne la distance entre les collones%
\begin{tabular}{|p{3.3in}|p{1.7in}|} \hline 
Température & T=300K \\ \hline 
Masse effective des électrons pour le $MoS_2$ & $m_{cw}=0.56 m_0$ \\ \hline 
Masse effective des trous pour le $MoS_2$ & $m_{vw}=0.64 m_0$ \\ \hline 
Masse effective des électrons pour le $WSe_2$ & $m_{cb}=0.35 m_0$ \\ \hline 
Masse effective des trous pour le $WSe_2$ & $m_{vb}=0.46 m_0$ \\ \hline 
Energie de bande interdite du $MoS_2$ & $E_g=1.8\ eV$ \\ \hline 
Energie d'interaction spin-orbite  & ${\Delta }_{SO}=0.20\ eV$ \\ \hline 
Temps de relaxation intra-bande & $\tau ={10}^{-13}\ s$ \\ \hline 
Largeur du puits quantique & W=10 \textit{nm} \\ \hline 
Constante diélectrique du $MoS_2$ & ${\varepsilon }_r=4.8$ \\ \hline 
Bande de valence offset  & ${\Delta E}_v=0.35\pm 0.07\ eV$ \\ \hline 
Bande de conduction offset & ${\Delta E}_c=0.17\pm 0.12\ eV$ \\ \hline 
\end{tabular}}}\\
\captionof{table}{Constantes utilisées pour les simulations \protect\cite{Jiwon C} \protect\cite{Ming-Hui C,M. Asada}.}
\end{center}

Il est possible de connaitre à l'avance le nombre des racines recherchées(le nombre des niveaux d'énergies confinées) tel que cette information facilite énormément le code~:
\begin{equation} 
n_{max}=anint(\frac{W\sqrt{2m_{bw}{\Delta E}_b}}{\pi \hslash }~); b=\left\{v,c\right\} 
\end{equation} 

o\`{u} \textit{anint }signifie qu'on doit arrondir à l'entier supérieur et \textit{b }l'indice de bande alors que ${\Delta E}_b$  est la largeur de la bande offset de conduction ou de valence. 

La méthode la plus simple pour trouver les énergies de confinements de la bande de valence et de conduction est de faire représenter séparément la partie gauche et la partie droite de l'équation de l'équation de \ref{eq 3.21} et \ref{eq 3.22} à résoudre. On soustrait les deux parties et on recherche les points d'intersections où la différence est presque nulle.

\begin{minipage}{\textwidth} 
\begin{minipage}{0.5\textwidth}
  \centering
  \includegraphics[scale=0.3]{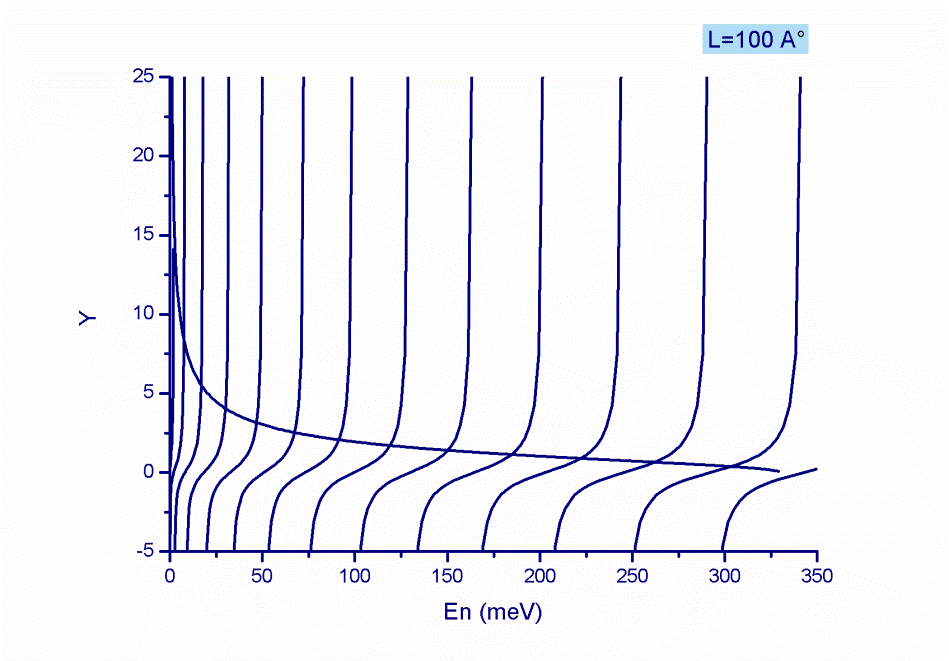}
\end{minipage}%
\begin{minipage}{0.5\textwidth}
  \centering 
  \includegraphics[scale=0.3]{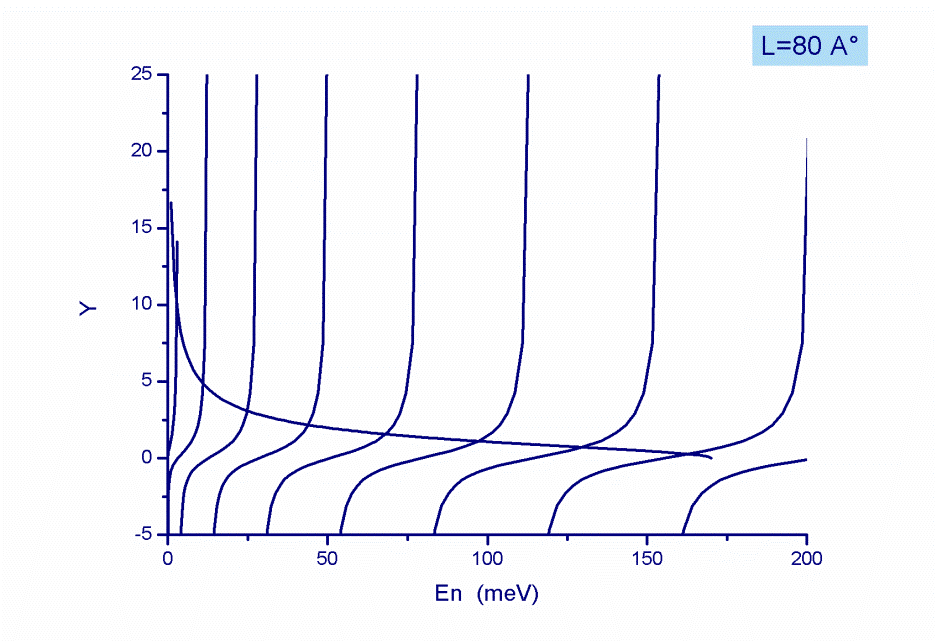}
\end{minipage} 
\begin{minipage}{0.5\textwidth}
  \centering
  \includegraphics[scale=0.3]{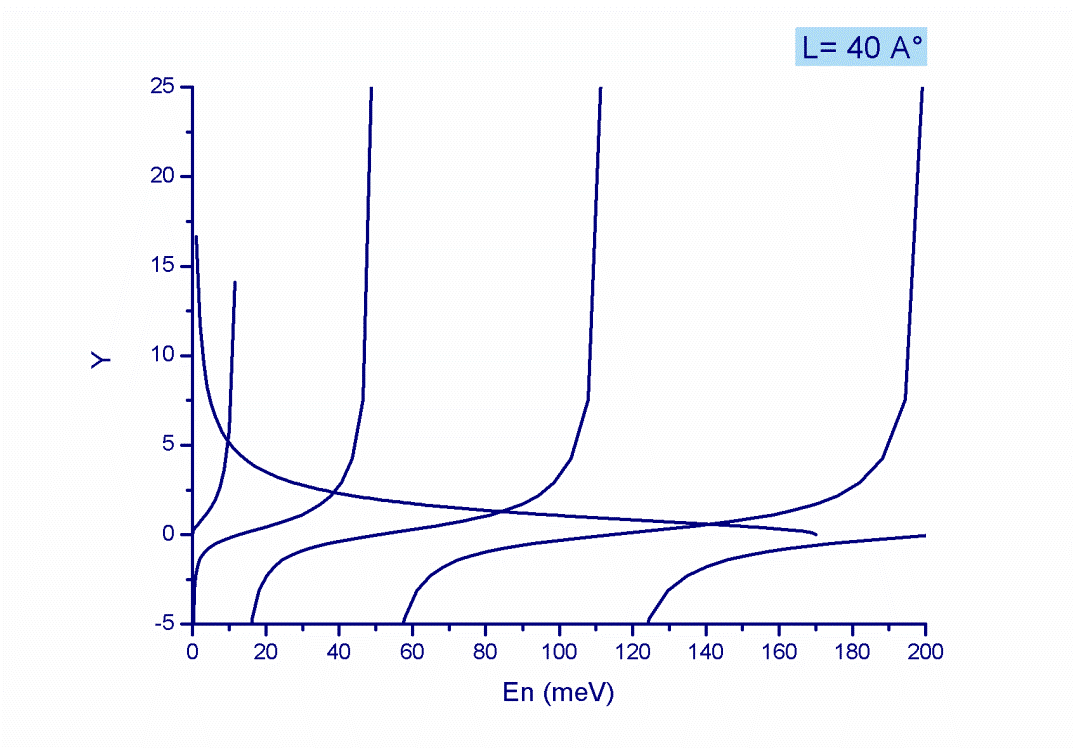}
\end{minipage} 
\captionof{figure}{Détermination des énergies de confinement par une représentation graphique des équations transcendantales pour différentes valeurs de largeur de puits.} 
\label{fig 3.13}
\end{minipage}\\

La figure \ref{fig 3.13} ci-dessus montre le grand dépendance de la largeur de puits avec les niveaux de confinement, plus le puits est large plus on'a des niveaux confinés. 
\subsection{Transitions inter- et intra-bandes }

Ils existent plusieurs types des transitions électroniques ; soit des transitions intra-bandes si elles sont internes pour une même bande d'énergie, soit des transitions inter-bandes pour des transitions d'une bande de valence vers une autre de conduction ou les transitions inter-sous-bandes. On suppose que la probabilité qu'un électron d'un niveau d'énergie n se recombine avec un trou d'un niveau $m\neq n$ sont très faibles. On peut négliger ces transitions interdites et toujours supposer que m=n. De plus, les probabilités pour des transitions où $k_c\neq k_v$ sont nulles.

\begin{center}
\includegraphics[scale=0.3]{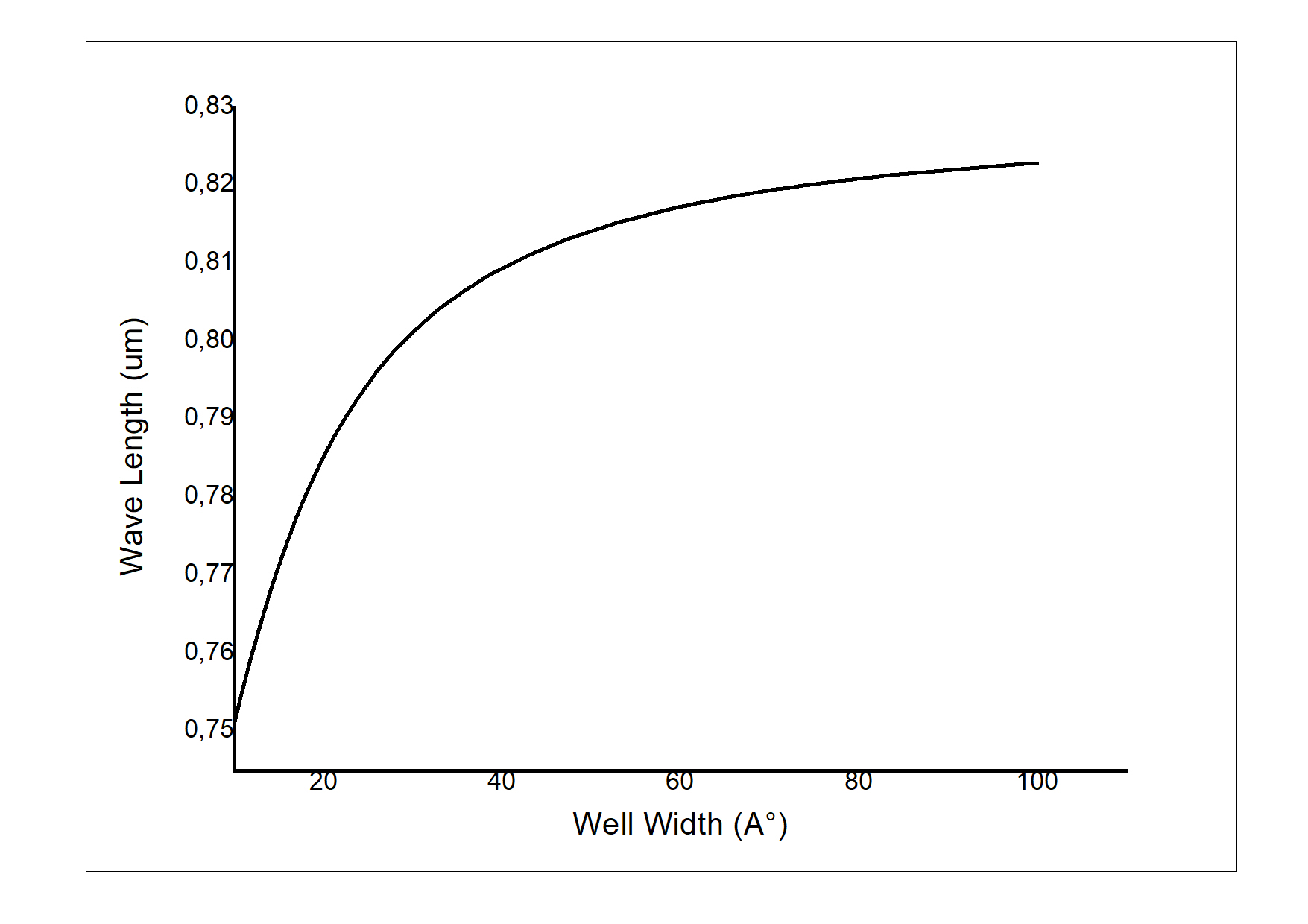}
\captionof{figure}{Longueur d'onde correspondant à la première transition entre les niveaux d'énergie quantifiés des électrons de conduction et des trous lourds en fonction de la largeur du puits calculée pour le puits $MoS_2$ et la barrière de $WSe_2$.}
\label{fig 3.14}
\end{center}

La longueur d'onde correspondant à la transition entre les niveaux confinés de la bande de valence et de conduction peut être exprimé par :
\[\lambda_n = 1.24 / \left( E_g + E_{cn} + E_{vn}\right)  \left( \mu m\right)\] 

La figure \ref{fig 3.14} ci-dessus représente la longueur d'onde  $\lambda_0$ pour la première transition (n=0) entre les trous lourds de la bande de valence et les électrons de la bande de conduction en fonction de la largeur de puits.

\begin{center}
\includegraphics[scale=0.3]{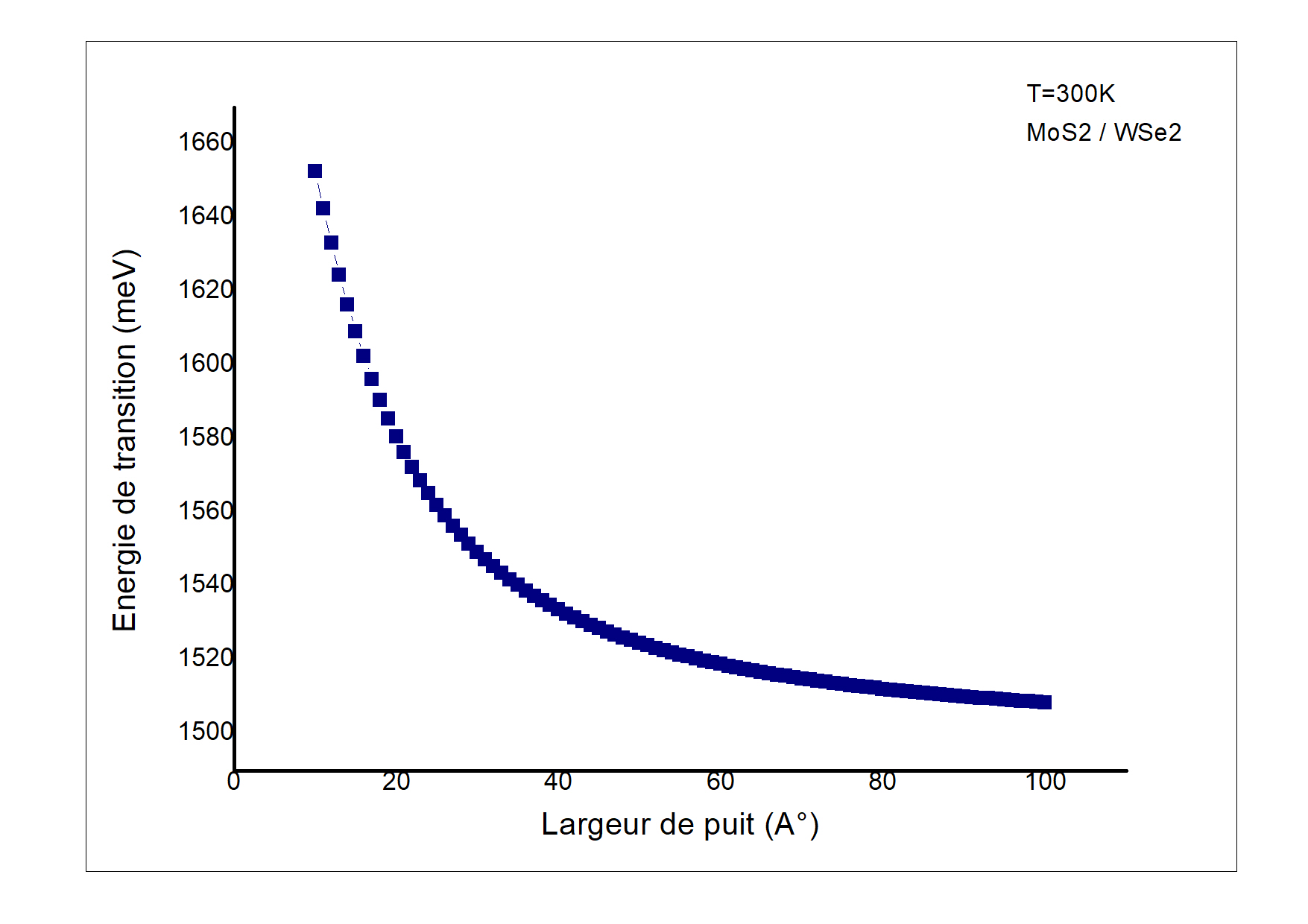}
\captionof{figure}{Energie de transition en fonction de la largeur de puits à T=300K}
\end{center}

\section{Gain optique: Résultats et interprétations} 

Nous avons calculé le gain optique pour une zone active formée pour le puits quantique $MoS_2$/$WSe_2$. Les figures ci-dissous donnant l'évolution du gain optique en fonction de l'énergie des photons pour différents niveaux d'injection  $N_{3D} \left( \times 10^{19} cm^{-3} \right)$ en mode TE pour un temps de relaxation intra-bandes $\tau_{int}$ et pour une largeur de puits 2 nm $\leq$ L $\leq$ 8 nm à T=100K, T=200K et T=300K. Ce qui nous permet par la suite de vérifier la performance de la zone active.

\begin{center}
\includegraphics[scale=0.7]{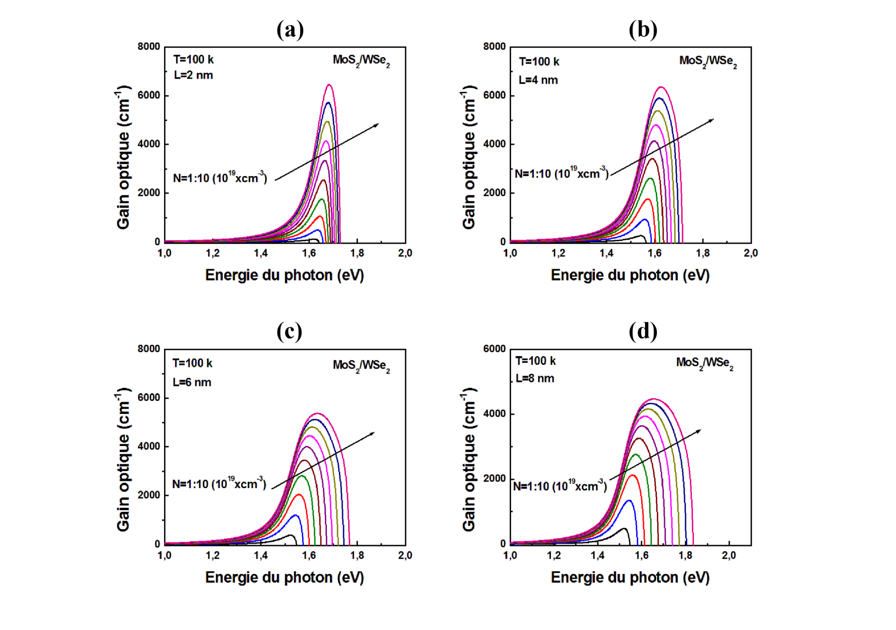}
\captionof{figure}{Evolution de gain optique en fonction de la largeur de puits à température T=100K}
\label{fig 3.16}
\end{center}
\begin{center}
\includegraphics[scale=0.55]{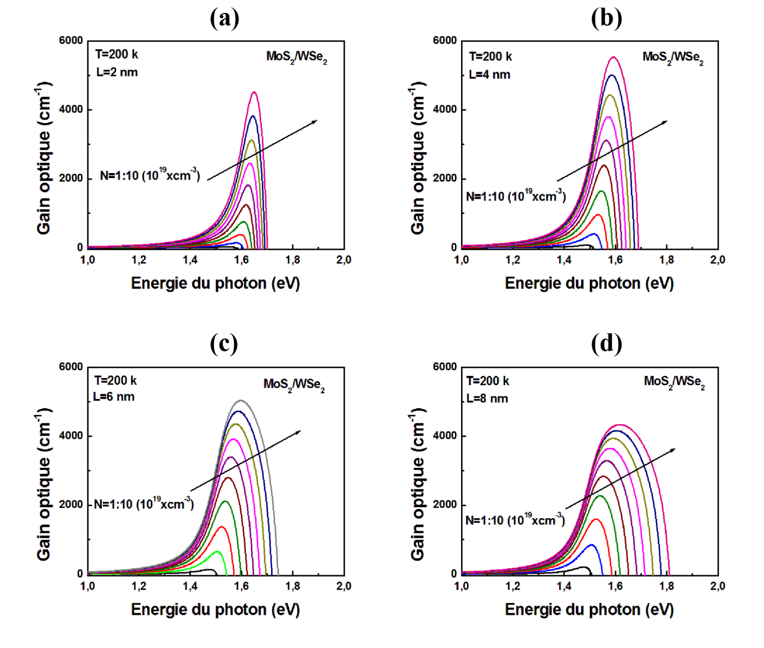}
\captionof{figure}{Evolution de gain optique en fonction de la largeur du puits à température T=200K}
\label{fig 3.17}
\end{center}
\begin{center}
\includegraphics[scale=0.55]{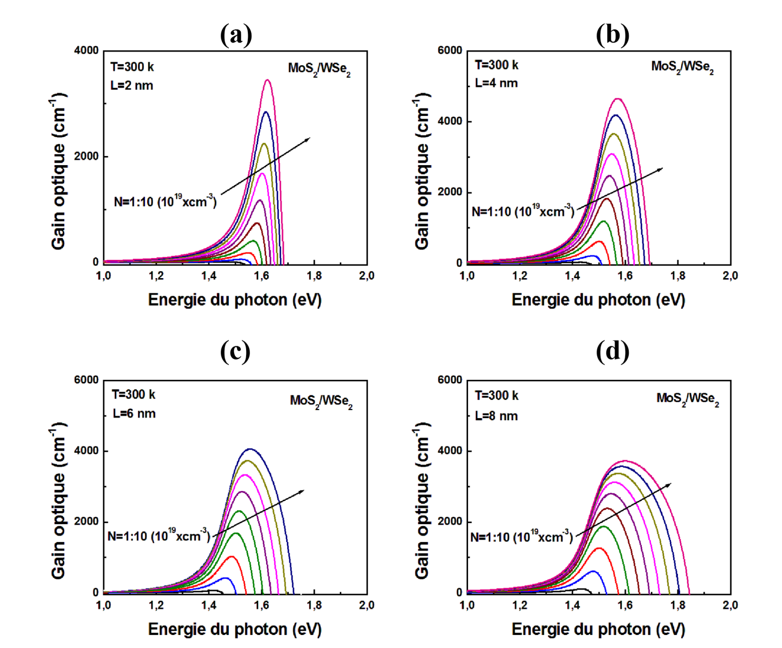}
\captionof{figure}{Evolution du gain optique en fonction de la largeur du puits à température ambiante.}
\label{fig 3.18}
\end{center}

Les figures \ref{fig 3.16}, \ref{fig 3.17} et \ref{fig 3.18} montrent qu'une augmentation de la largeur du puits entraîne une diminution du gain provoquent aussi un élargissement du spectre et un décalage vers les faibles longueurs d'onde. Cela est dû au faite que l'énergie de confinement est inversement proportionnelle à la largeur de puits. 

\subsection{Gain maximal}
\begin{center}
\includegraphics[scale=0.7]{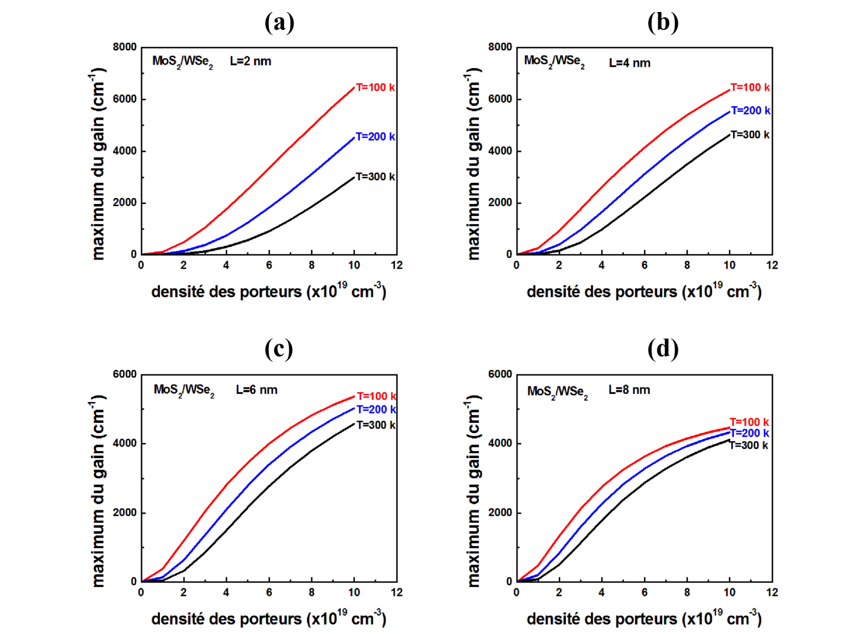}
\captionof{figure}{Variation du gain maximal en fonction de la densité de porteurs injectés $N_{3D}$ à 100K, 200K, 300K et différentes largeur du puits pour les structures lasers $MoS_2$/$WSe_2$.}
\end{center}
Au dessus du seuil de transparence, le gain maximal $G_{max}$ augmente rapidement avec la densité des porteurs injectés. A température ambiante T=300 K, autour d'une concentration $N_{3D} = 5 \times 10^{19}    cm^{-3}$, le système quantique, c'est-à-dire le milieu amplificateur de photons satisfait la condition de Bernard-Durrafourg $\left( E_{Fc} - E_{Fv} > \hslash\omega \right)$. Ceci permet de définir le seuil de transparence du système, c'est-à-dire la concentration de porteurs nécessaire pour satisfaire la condition d'inversion de population. Notons que au-dessus de ce seuil, plus la densité de porteurs est importante, plus la courbe du gain est large.
Ce comportement peut être expliqué par le fait que la dispersion des niveaux d'électrons et de trous change de manière significative avec l'augmentation de la température et de la largeur du puits. 
\section*{Conclusion}
Dans cette partie nous avons vu l'importance de la bande offset qui affirme la possibilité de formation de puits quantique dans l'hétérostructure à base des TMDCs telle que le comportement d'une jonction semi-conductrice dépend bien de l'alignement des bandes d'énergie sur l'interface. 

	La largeur de puits influe largement sur le nombre des niveaux d'énergie confinée à l'intérieure de ce dernière. De plus le contrôle de la bande interdite via le nombre des couches rendre ces matériaux très promoteur d'être intégré dans divers application opto-électronique où il devient possible de contrôler l'émission émise suite à un contrôle de gap.
	
	Enfin, on s'attend également à ce que les 2D-TMDCs possèdent les avantages des lasers à puits quantique à savoir leur gain optique et leur densité d'état bidimensionnelle associé après l'étude de la performance de la zone active de la structure laser à puits quantique $MoS_2$/$WSe_2$ suite à un calcul du gain optique en fonction de l'énergie des photons pour différents niveaux d'injection et à différente température qui met en évidence la possibilité que ces dernières peuvent être intégrés dans diverses applications opto-électroniques notamment les diodes laser.

%- - - - - - -- - -- - -- - - - -- -- - - -- - - -- - 
%CONCLUSION Générale
%- - - - -- - - - - - - - - - -- - - -- - - -- - - - 

\shorttableofcontents{Conclusion générale}{1}

Durant ces dernières années, la physique des semi-conducteurs à 2D à base de \acrlong{TMDCs} a révolutionné la technologie grâce à la maitrise de l'élaboration des puits quantiques et des structures lasers à puits quantiques. L'intérêt pour interpréter les résultats des expériences de transport et d'optique effectuées sur de telles structures, a motivé le développement de nouveaux aspects et modèles théoriques appropriés aux propriétés spécifiques des puits quantiques et des diodes lasers.
Le travail exposé dans ce manuscrit avait comme objectif principal de contribuer à une meilleure description de la zone active des structures lasers à base de $MX_2$ ($MoS_2$, $MoSe_2$, $WS_2$ et $WSe_2$), émettant dans le domaine du moyen-infrarouge, a été axée essentiellement sur trois points :\\
$\bullet$ Nous avons d'abord présenté le modèle théorique utilisé pour calculer le gain optique des structures dont les zones actives sont formées par des puits quantiques contraints à base de métaux de transition $MX_2$ ($MoS_2$, $MoSe_2$, $WS_2$ et $WSe_2$).\\
$\bullet$ Nous avons analysé les performances de la zone active des structures lasers telles que le gain optique, ainsi que les outils de calcul. En effet la spécificité du système à étudier, nous a poussé à considérer une méthode qui permet d'une part d'expliciter les différentes expressions de gain optique pour les puits quantiques, et d'autre part, de préciser les hypothèses de calcul de gain optique.\\

 L'étude de ces performances montre que les structures lasers à puits quantique sont capables de fonctionner à température ambiante. Tous ces arguments nous permettent de dire que les puits quantiques jouent un rôle très important dans le fonctionnement des diodes lasers à base de ces matériaux. Il est donc conseillé de fabriquer dans le futur des diodes lasers dont la zone active est formée par des puits quantiques.

\addcontentsline{toc}{chapter}{Conclusion générale}%ajout de la conclusion à la table de matière

En perspectives, nous envisageons dans la suite d'élaborer des couches minces à base des MX2 et en particulier les héterostructures  $MoS_2$/$WSe_2$. De point de vue théorique, nous envisageons d'étudier les performances d'une zone active fabriquée par des puits quantiques $MoS_2$/$WSe_2$ en cascades et d'étudier les systèmes de plus basse dimensionnalité tels que les fils quantiques et les boites quantiques à base de $MoS_2$/$WSe_2$ qui sont aujourd'hui d'un grand intérêt expérimental et théorique, vu leur application dans le domaine de l'optoélectronique.

%- - - - - - -- - -- - -- - - - -- -- - - -- - - -- - 
%ANNEXE
%- - - - -- - - - - - - - - - -- - - -- - - -- - - - 

\addcontentsline{toc}{chapter}{Bibliographie}

\end{document}